\documentclass{pasj01}
\usepackage{url}
\usepackage{subfigure}
\usepackage{multirow}
\usepackage{lscape}
\usepackage{breqn}

\Received{$\langle$reception date$\rangle$}
\Accepted{$\langle$acception date$\rangle$}
\Published{$\langle$publication date$\rangle$}

\begin{document}

\title{A systematic study of Galactic infrared bubbles along the Galactic plane with AKARI and Herschel. II. Spatial distributions of dust components around the bubbles.}
\author{Misaki \textsc{Hanaoka}\altaffilmark{1}}%
\author{Hidehiro \textsc{Kaneda}\altaffilmark{1}}%
\author{Toyoaki \textsc{Suzuki}\altaffilmark{1}}%
\author{Takuma \textsc{Kokusho}\altaffilmark{1}}%
\author{Shinki \textsc{Oyabu}\altaffilmark{1}}%
\author{Daisuke \textsc{Ishihara}\altaffilmark{1}}%
\author{Mikito \textsc{Kohno}\altaffilmark{1}}%
\author{Takuya \textsc{Furuta}\altaffilmark{1}}%
\author{Takuro \textsc{Tsuchikawa}\altaffilmark{1}}%
\author{Futoshi \textsc{Saito}\altaffilmark{1}}%

\altaffiltext{1}{Graduate School of Science, Nagoya University, Furo-cho, Chikusa-ku, Nagoya 464-8602, Japan}
\email{hanaoka@u.phys.nagoya-u.ac.jp, kaneda@u.phys.nagoya-u.ac.jp}




\maketitle

\begin{abstract}
  Galactic infrared (IR) bubbles, which can be seen as shell-like structures at mid-IR wavelengths, are known to possess massive stars within their shell boundaries.
  In our previous study, \citet{Hanaoka2019} expanded the research area to the whole Galactic plane (0$^{\circ}$~$\leq l \leq$~360$^{\circ}$, $|b| \leq$~5$^{\circ}$) and studied systematic differences in the shell morphology and the IR luminosity of the IR bubbles between inner and outer Galactic regions.
  In this study, utilizing high spatial-resolution data of AKARI and WISE in the mid-IR and Herschel in the far-IR, we investigate the spatial distributions of dust components around each IR bubble to discuss the relation between the star-formation activity and the dust properties of the IR bubbles.
  For the 247 IR bubbles studied in \citet{Hanaoka2019}, 165 IR bubbles are investigated in this study, which have the Herschel data ($|b|\leq 1^{\circ}$) and known distances.
  We created their spectral energy distributions on a pixel-by-pixel basis around each IR bubble, and decomposed them with a dust model consisting of polycyclic aromatic hydrocarbons (PAHs), hot dust, warm dust and cold dust. 
  As a result, we find that the offsets of dust heating sources from the shell centers in inner Galactic regions are systematically larger than those in outer Galactic regions.
  Many of the broken bubbles in inner Galactic regions show large angles between the offset and the broken shell directions from the center.
  Moreover, the spatial variations of the PAH intensity and cold dust emissivity around the IR bubbles in inner Galactic regions are larger than those in outer Galactic regions.
  We discuss these results in light of the interstellar environments and the formation mechanism of the massive stars associated with the IR bubbles.
\end{abstract}

\section{Introduction}
There are a large number of Galactic infrared (IR) bubbles existing along the Galactic plane.
They have shell-like structures, which can be clearly seen at mid-IR wavelengths.
Churchwell et al. (2006, 2007) cataloged 591 IR bubbles which are located in inner Galactic regions ($|l|\leq$~65$^{\circ}$, $|b|\leq$~1$^{\circ}$), using the 8~$\mu$m band images of the Galactic Legacy Infrared Mid-plane Survey Extraordinaire (GLIMPSE: \cite{Benjamin2003}; \cite{Churchwell2009}) program with Spitzer.
In our previous study, \citet{Hanaoka2019} expanded the research area to the whole Galactic plane (0$^{\circ}$~$\leq l \leq$~360$^{\circ}$, $|b| \leq$~5$^{\circ}$) and newly found 175 IR bubbles with large angular radii $R>1'$ using the 9~$\mu$m band images of the all-sky surveys with AKARI (\cite{Murakami2007}; \cite{Ishihara2010}).

The shell-like structure associated with the IR bubbles can be seen clearly in the Spitzer 8~$\mu$m and AKARI 9~$\mu$m band images.
These band intensities are expected to be dominated by polycyclic aromatic hydrocarbon (PAH) emission from photodissociation regions (PDRs).
PAHs are easily photodissociated by strong ultraviolet (UV) radiation ($\geq13.6$~eV), thus they can hardly exist in H\emissiontype{II} regions.
The IR bubbles are expected to possess the spherical shell structure formed on the edge of the H\emissiontype{II} region.
Ionizing gases are distributed within the shell structures, producing the mid-IR continuum emission which mostly traces hot dust in H\emissiontype{II} regions (\cite{Deharveng2010}).
\citet{Hanaoka2019} confirmed that 97\% of the 247 IR bubbles along the Galactic plane show diffuse hot dust emission within the shell boundaries using the AKARI 18~$\mu$m images.
Thus, most of the IR bubbles possess massive stars within their shell structures.

In order to understand the massive star formation, triggered star-formation mechanisms have been discussed. 
Typical massive star-formation mechanisms are triggered by expansion of the H\emissiontype{II} regions and cloud-cloud collision (CCC) (e.g., \cite{Elmegreen1998}; Zinnecker \& Yorke 2007; \cite{Deharveng2010}).
The former mechanism compresses the interstellar media (ISM) at the edges of the H\emissiontype{II} regions and dense clumps by radiation from the pre-existing massive stars (e.g., \cite{Deharveng2010}; Dale et al. 2007; \cite{Gritschneder2009}).
The CCC mechanism triggers a massive star formation on the collision surface between two molecular clouds (Habe \& Ohta 1992).
A collision of two molecular clouds having different velocities makes an intermediate velocity component at the collision surface, which is seen as a broad bridge feature on the position-velocity map (e.g., \cite{Hawarth2015}; \cite{Takahira2018}).

In the previous study, \citet{Hattori2016} classified the IR bubbles cataloged in Churchwell et al. (2006, 2007) by the shell morphologies according to the quantitative criteria established in their study.
They also estimated the IR flux densities and the spatial distributions of dust emission in and around each IR bubble using the 9, 18, 65, 90, 140 and 160~$\mu$m band images of the AKARI all-sky surveys.
Then, they found that large broken bubbles tend to have higher total IR luminosities, lower fractional luminosities of the PAH emission and dust heating sources located nearer to the shells.
Based on these results, \citet{Hattori2016} suggested that many of the large broken bubbles might have been formed by the CCC mechanism.
\citet{Hanaoka2019} expanded the study of \citet{Hattori2016} to the whole Galactic regions (0$^{\circ}$~$\leq l \leq$~360$^{\circ}$, $|b| \leq$~5$^{\circ}$) and newly found 175 IR bubbles.
\citet{Hanaoka2019} obtained the shell radii, the covering fractions (CFs) of the shells and the IR luminosities of 247 IR bubbles with large angular radii $R>1'$ including the IR bubbles studied by \citet{Hattori2016}.
They found that there are systematic differences in the global IR properties of the IR bubbles between inner and outer Galactic regions and suggested that the results are interpreted by the interstellar environments around the IR bubbles and the star-formation mechanisms of the central massive stars.
However, \citet{Hanaoka2019} did not perform a spatially-resolved study of the IR properties.

In this study, we investigate the spatial distributions of dust components of the IR bubbles to discuss the relation between the star-formation activity and the dust properties.
By using high spatial resolution data of Herschel Space Observatory (\cite{Pilbratt2010}) whose resolution is comparable to that of the AKARI 9~$\mu$m band image tracing the PAH emission, the spatial variations of the dust properties in the IR bubbles are discussed in more detail.
The mid-IR wavelength data observed by the Wide-field Infrared Survey Explorer (WISE: \cite{Wright2010}) are also added to estimate a hot dust ($>100$~K) component, which traces the environments closer to massive stars.
Then, we obtain the overall properties of the IR bubbles along the whole Galactic plane and discuss the effects of the massive stars and interstellar environments on the properties of the IR bubbles in inner and outer Galactic regions.

\section{Observation and data analysis}
\begin{table}
  \caption{Observation data properties}
  \label{table:observation_data}
  \begin{center}
    \begin{tabular}{cccc}
      \hline
      \multirow{2}{*}{Satellite} & Center & Pixel & FWHM \\
      & wavelength & scale & of PSF \\\hline
      \multirow{2}{*}{AKARI\footnotemark[$*$]} & 9 $\mu$m & \multirow{2}{*}{4$''$.68} & 5$''$.5\\
      & 18 $\mu$m & & 5$''$.7 \\\hline
      \multirow{5}{*}{Herschel\footnotemark[$\dag$]} & 70 $\mu$m & 3$''$.2 & 6$''$.0 \\
      & 160 $\mu$m & 4$''$.5 & 12$''$.0\\
      & 250 $\mu$m & 6$''$.0 & 18$''$.0\\
      & 350 $\mu$m & 8$''$.0 & 24$''$.0\\
      & 500 $\mu$m & 11$''$.5 & 35$''$.0\\\hline
      WISE\footnotemark[$\ddag$] & 22 $\mu$m & 1$''$.375 & 12$''$.0\\\hline
    \end{tabular}
  \end{center}
  \begin{tabnote}
    \footnotemark[$*$] \citet{Onaka2007}\\
    \footnotemark[$\dag$] \citet{Molinari2016}\\
    \footnotemark[$\ddag$] \citet{Cutri2013}
  \end{tabnote}
\end{table}

\begin{figure*}
  \begin{center}
    \includegraphics[width=0.85\linewidth,clip]{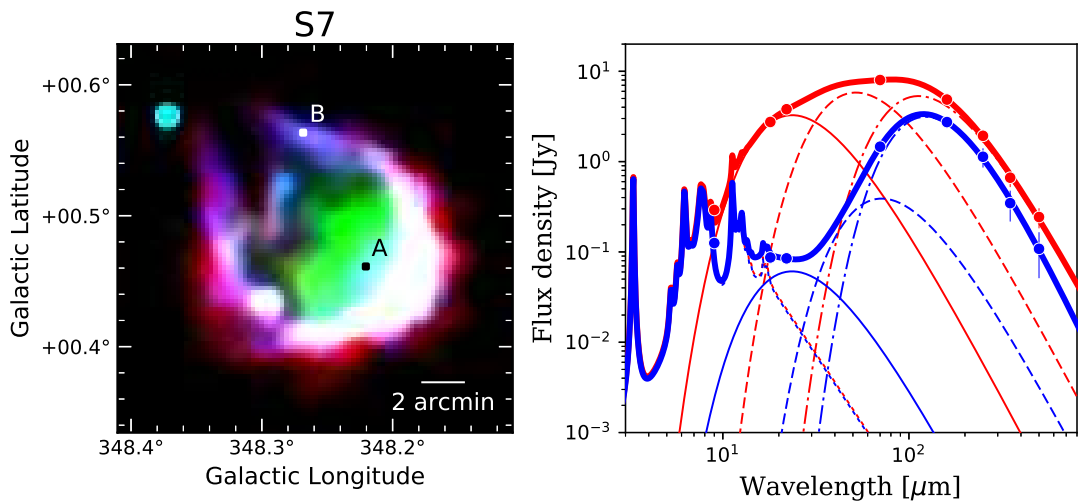}
  \end{center}
  \caption{
    Examples of the local SED-fitting results for one of our sample IR bubbles, S7 (named ``RCW 120'' by \cite{Rodgers1960}).
    The left panel shows the composite image of S7, consisting of AKARI 9~$\mu$m, 18~$\mu$m and the Herschel 160~$\mu$m band images in blue, green and red, respectively.
    The right panel shows the fitting results of the local SED at the positions denoted in the left panel, the red and blue curves for ``A'' and ``B'', respectively.
    The local SEDs consist of the AKARI 9~$\mu$m, 18~$\mu$m, the WISE 22~$\mu$m, the Herschel 70~$\mu$m, 160~$\mu$m, 250~$\mu$m, 350~$\mu$m and 500~$\mu$m data points.
    The thick solid line shows the best-fit model, while the dotted, thin solid, dashed and dash-dotted lines correspond to the PAH, hot, warm and cold dust components, respectively.
    }
  \label{fig:Local_SED_fit}
\end{figure*}

To investigate the local infrared properties of the IR bubbles studied by \citet{Hanaoka2019}, we used AKARI, Herschel and WISE photometric data.
The AKARI mid-IR all-sky surveys were carried out in two photometric bands (the central wavelengths are 9, 18~$\mu$m; \cite{Onaka2007}), which covered more than 90\% of the whole sky (\cite{Ishihara2010}).
The Herschel infrared Galactic Plane Survey (Hi-GAL: Molinari et al. 2010, 2016) is an unbiased photometric survey of the Galactic plane ($0^{\circ} \leq l < 360^{\circ}$, $|b|<1^{\circ}$) at 70, 160, 250, 350 and 500~$\mu$m wavelengths (\cite{Poglitsch2010}; \cite{Griffin2010}).
All the Hi-GAL images are publicly available in the Herschel Science Archive\footnote{Herschel Science Archive \url{http://archives.esac.esa.int/hsa/whsa/}}, and the high-quality data of Hi-GAL (Hi-GAL DR1 images: covering $-70^{\circ}\lesssim l \lesssim68^{\circ}$, $|b|\leq 1^{\circ}$), which had been better calibrated with the ROMAGAL pipeline, were also released on the online site of the VIALACTEA project\footnote{VIALACTEA project \url{http://vialactea.iaps.inaf.it}} (\cite{Molinari2016}).
Moreover, we used the WISE 22~$\mu$m band images to investigate the distribution of a hot dust ($\sim$120~K) component.
The AllWISE data had been improved with enhanced sensitivity and accuracy compared with earlier WISE data releases (\cite{Cutri2013}).
The WISE data are given in units of DN, and the DN-to-Jy conversion factor is $5.2269 \times 10^{-5}$ (\cite{Cutri2013}; \url{http://wise2.ipac.caltech.edu/docs/release/allwise/expsup/sec4_3a.html}).
The specifications of the observation data, i.e., the central wavelengths, the pixel scales and the sizes of the point spread functions (PSFs) are summarized in table~\ref{table:observation_data}.

In this study, we investigate the IR bubbles having the Herschel Hi-GAL data ($|b|\leq 1^{\circ}$) and known distances out of the IR bubbles studied by \citet{Hanaoka2019}. 
165 IR bubbles remain as our sample, which consist of 29 IR bubbles found by \citet{Hanaoka2019} and 136 IR bubbles studied by \citet{Hattori2016} (i.e., cataloged by Churchwell et al. 2006, 2007).

\begin{figure*}
  \begin{center}
    \subfigure{
      \mbox{\raisebox{3mm}{\rotatebox{90}{\small{Galactic Latitude}}}}
      \mbox{\raisebox{0mm}{\includegraphics[width=150mm, bb=0 15 850 185, clip]{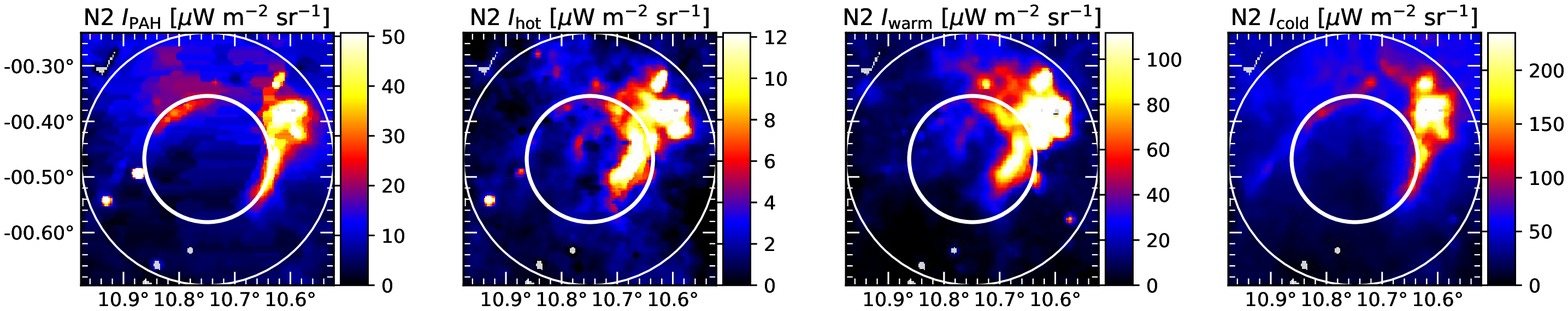}}}
    }
    \subfigure{
      \mbox{\raisebox{3mm}{\rotatebox{90}{\small{Galactic Latitude}}}}
      \mbox{\raisebox{0mm}{\includegraphics[width=150mm, bb=0 15 850 185, clip]{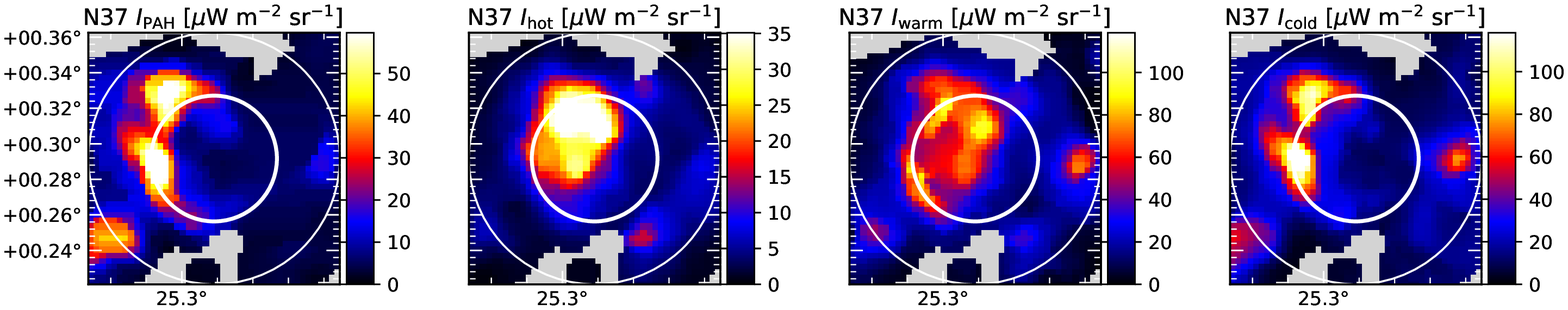}}}
    }
    \subfigure{
      \mbox{\raisebox{3mm}{\rotatebox{90}{\small{Galactic Latitude}}}}
      \mbox{\raisebox{0mm}{\includegraphics[width=150mm, bb=0 15 850 185, clip]{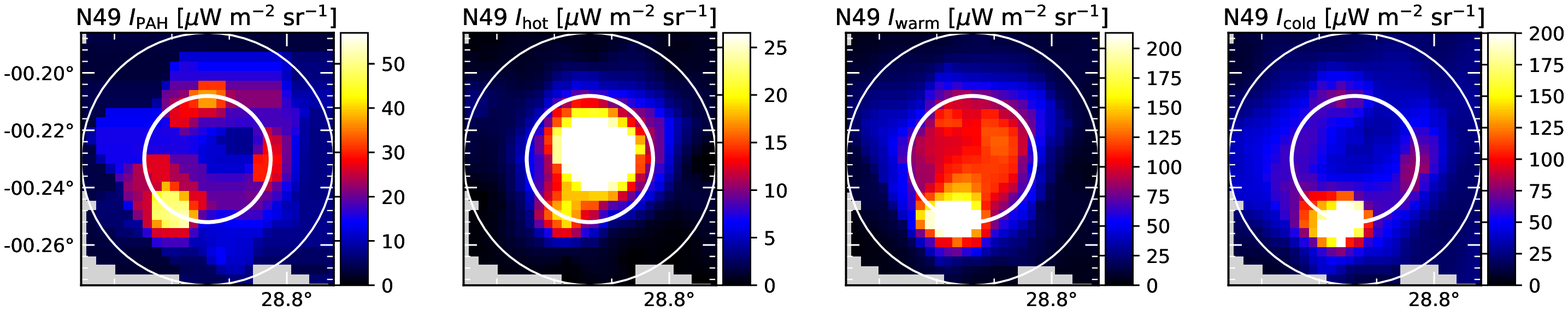}}}
    }
    \subfigure{
      \mbox{\raisebox{3mm}{\rotatebox{90}{\small{Galactic Latitude}}}}
      \mbox{\raisebox{0mm}{\includegraphics[width=150mm, bb=0 15 850 185, clip]{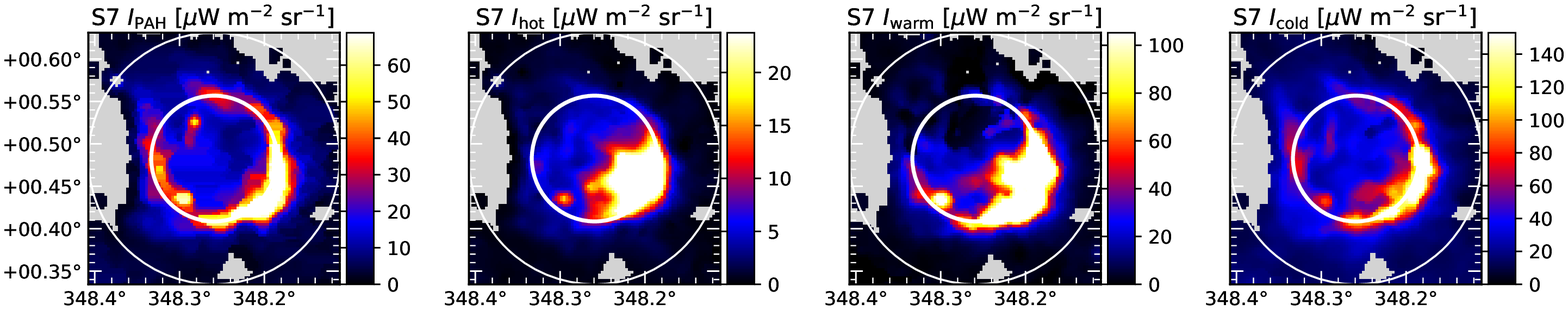}}}
    }
    \subfigure{
      \mbox{\raisebox{3mm}{\rotatebox{90}{\small{Galactic Latitude}}}}
      \mbox{\raisebox{0mm}{\includegraphics[width=150mm, bb=0 15 850 185, clip]{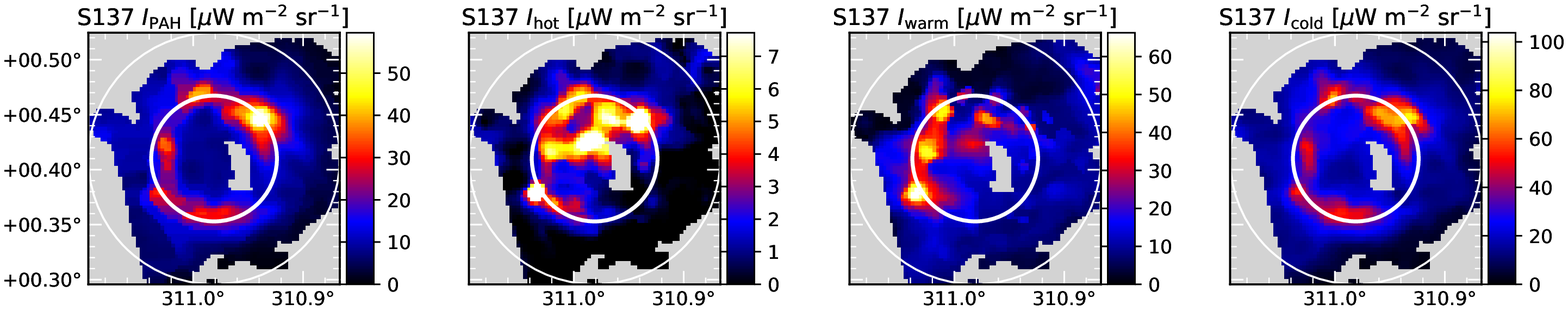}}}
    }
    \subfigure{
      \mbox{\raisebox{3mm}{\rotatebox{90}{\small{Galactic Latitude}}}}
      \mbox{\raisebox{0mm}{\includegraphics[width=150mm, bb=0 15 850 185, clip]{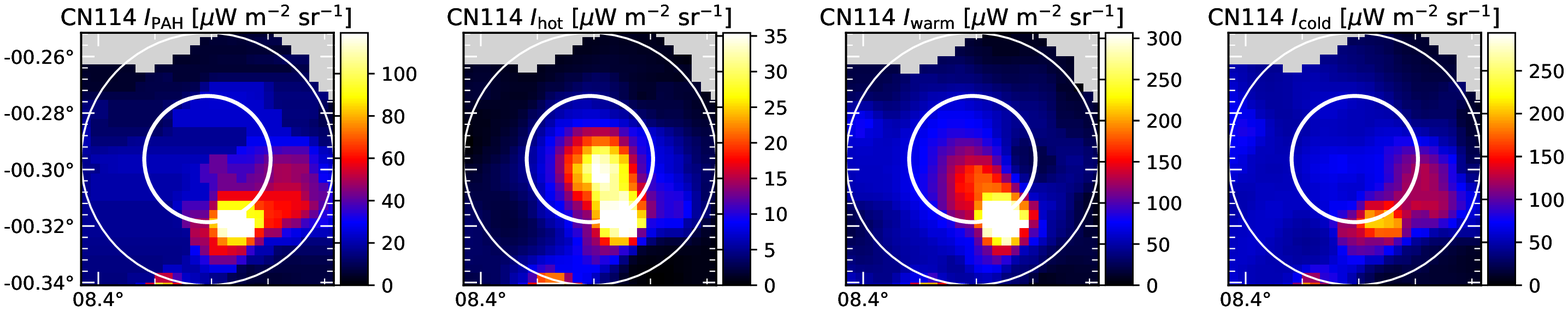}}}
    }
    \subfigure{\mbox{\raisebox{0mm}{\hspace{13mm}\small{Galactic Longitude}\hspace{8mm}\small{Galactic Longitude}\hspace{8mm}\small{Galactic Longitude}\hspace{8mm}\small{Galactic Longitude}\hspace{10mm}}}}
  \end{center}
  \caption{
    Examples of the spatial distributions of the PAH, hot dust, warm dust and cold dust from left to right.
    The thick and thin white circles correspond to the $R$ and $2R$ circular regions, respectively.
  }
  \label{fig:intensity_map}
\end{figure*}
\addtocounter{figure}{-1}
\begin{figure*}
  \begin{center}
    \subfigure{
      \mbox{\raisebox{3mm}{\rotatebox{90}{\small{Galactic Latitude}}}}
      \mbox{\raisebox{0mm}{\includegraphics[width=150mm, bb=0 15 850 185, clip]{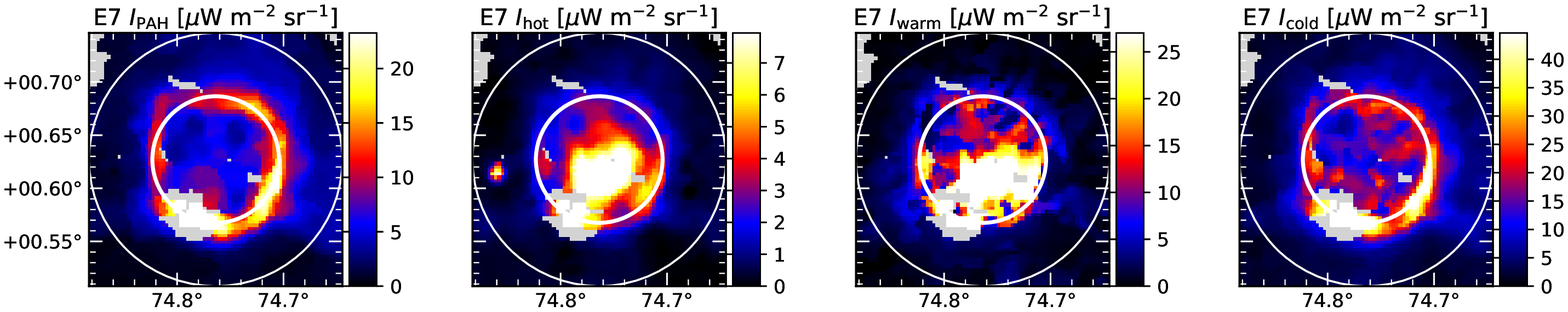}}}
    }
    \subfigure{
      \mbox{\raisebox{3mm}{\rotatebox{90}{\small{Galactic Latitude}}}}
      \mbox{\raisebox{0mm}{\includegraphics[width=150mm, bb=0 15 850 185, clip]{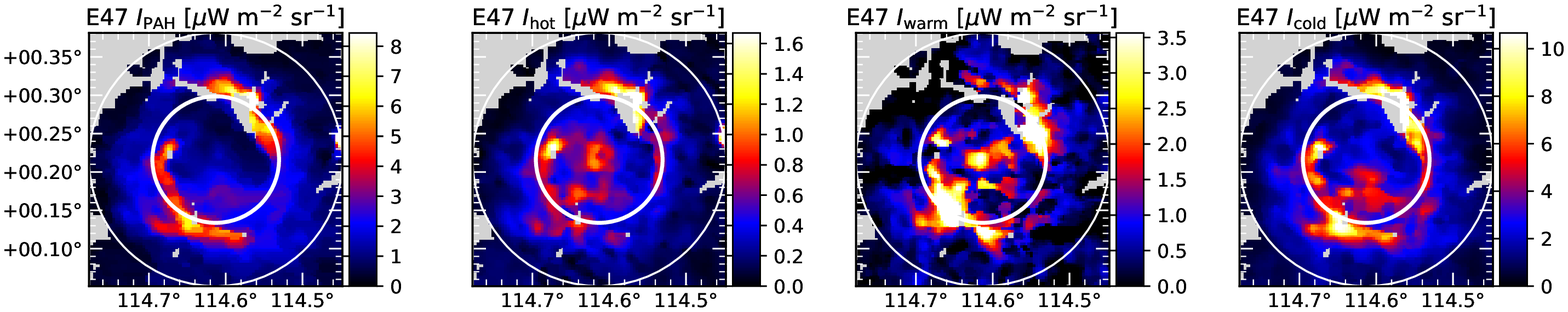}}}
    }
    \subfigure{
      \mbox{\raisebox{3mm}{\rotatebox{90}{\small{Galactic Latitude}}}}
      \mbox{\raisebox{0mm}{\includegraphics[width=150mm, bb=0 15 850 185, clip]{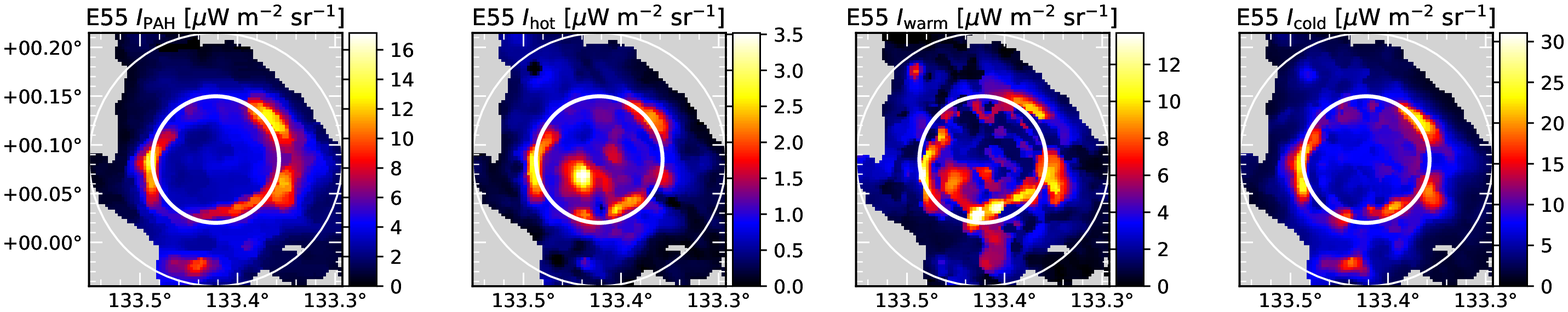}}}
    }
    \subfigure{
      \mbox{\raisebox{3mm}{\rotatebox{90}{\small{Galactic Latitude}}}}
      \mbox{\raisebox{0mm}{\includegraphics[width=150mm, bb=0 15 850 185, clip]{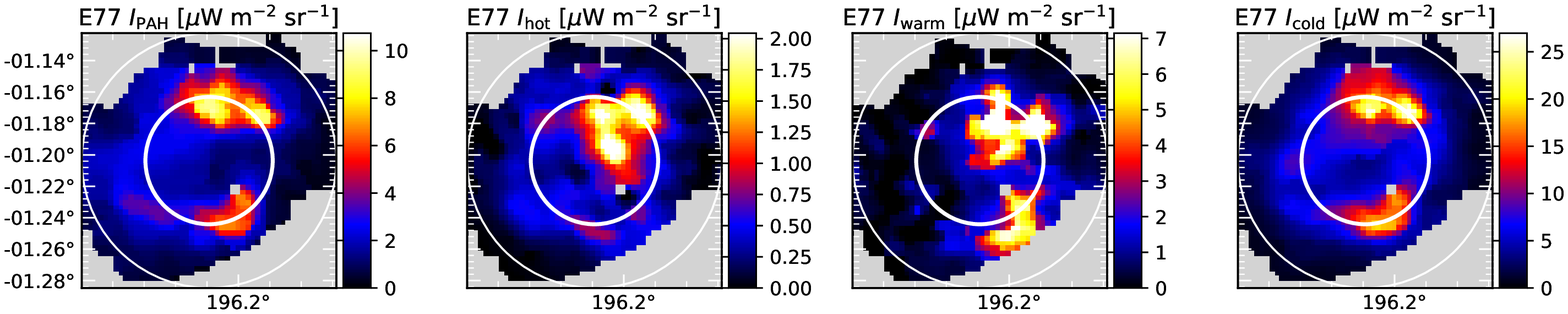}}}
    }
    \subfigure{
      \mbox{\raisebox{3mm}{\rotatebox{90}{\small{Galactic Latitude}}}}
      \mbox{\raisebox{0mm}{\includegraphics[width=150mm, bb=0 15 850 185, clip]{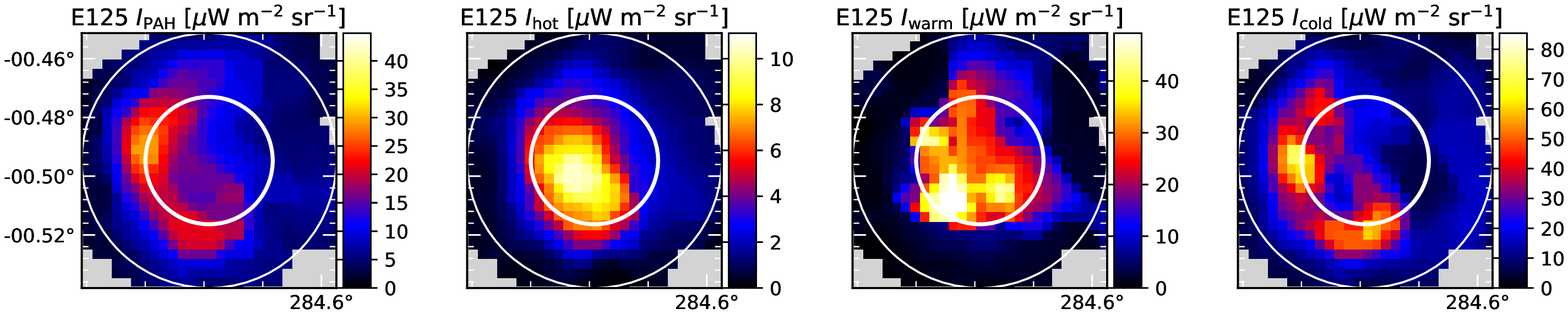}}}
    }
    \subfigure{
      \mbox{\raisebox{3mm}{\rotatebox{90}{\small{Galactic Latitude}}}}
      \mbox{\raisebox{0mm}{\includegraphics[width=150mm, bb=0 15 850 185, clip]{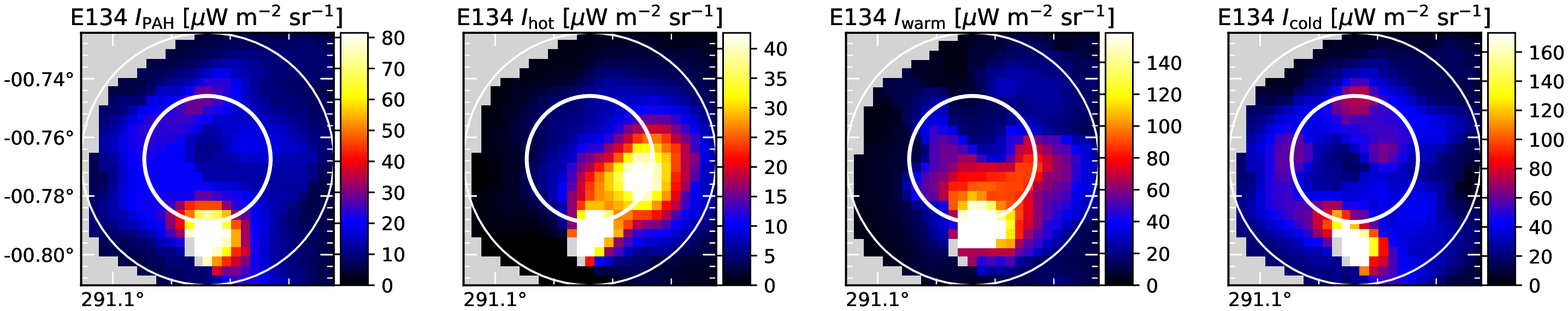}}}
    }
    \subfigure{\mbox{\raisebox{0mm}{\hspace{13mm}\small{Galactic Longitude}\hspace{8mm}\small{Galactic Longitude}\hspace{8mm}\small{Galactic Longitude}\hspace{8mm}\small{Galactic Longitude}\hspace{10mm}}}}
  \end{center}
  \caption{Continued.}
\end{figure*}

\begin{figure*}
  \begin{center}
    \includegraphics[width=0.8\linewidth,bb=5 0 700 605,clip]{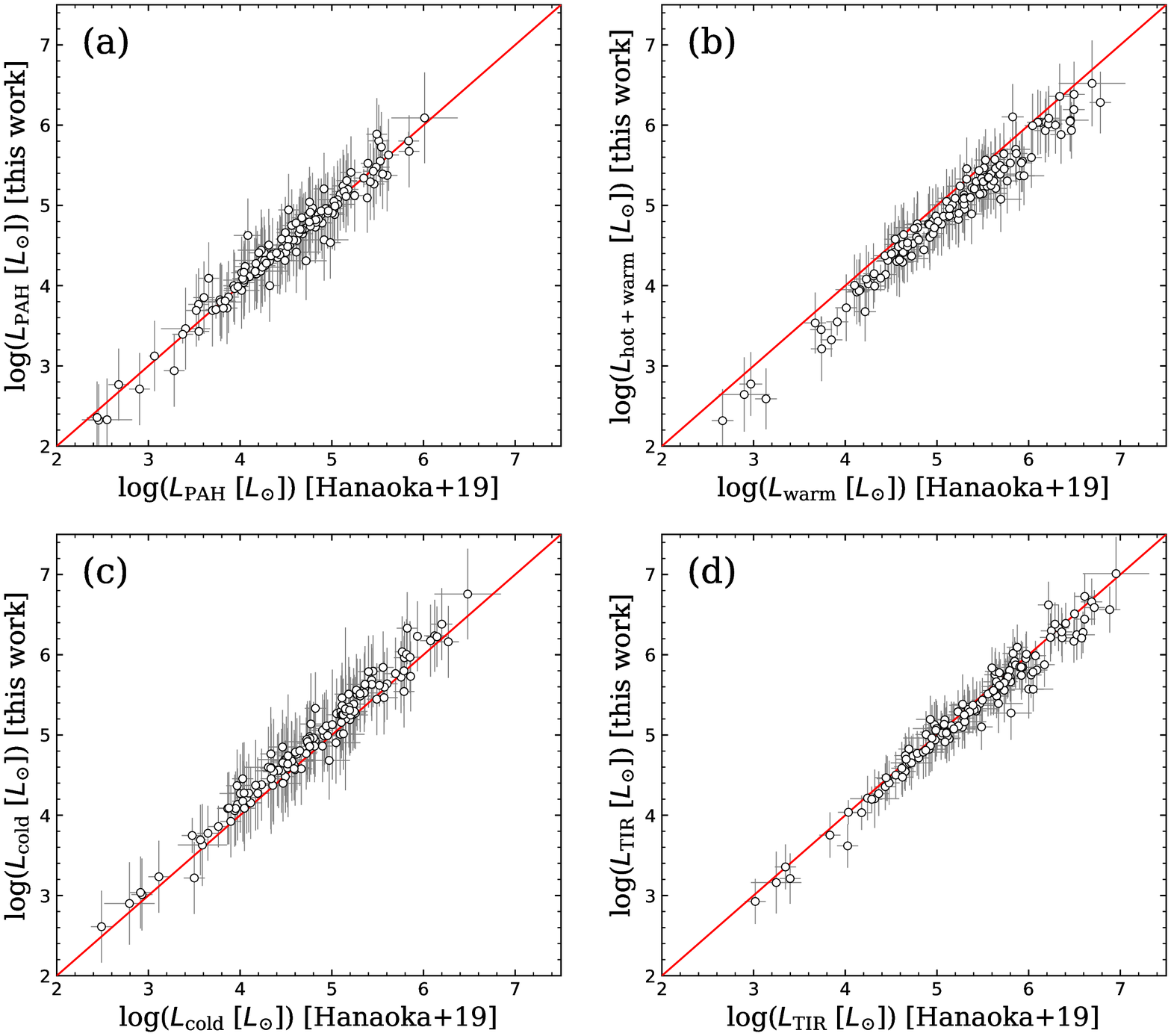}
  \end{center}
  \caption{
    Comparison between the results of the local SED fitting and those of \citet{Hanaoka2019}.
    Each panel shows the comparison of (a) PAH luminosity, (b) hot and warm dust luminosity, (c) cold dust luminosity and (d) total IR luminosity, respectively.
    The red lines in each panel correspond to $y=x$.
    }
  \label{fig:Local_Total_consis}
\end{figure*}

To investigate the spatial distributions of the dust components around the IR bubbles, model fittings are performed to local spectral energy distributions (SEDs) on a pixel-by-pixel basis for the $4R \times 4R$ region of each IR bubble, where $R$ is the radius of the IR bubble.
We smoothed and re-grided all the images with the PSF size and the pixel size of the Herschel 500~$\mu$m band images, $35''.0$ and $11''.5 \times 11''.5$, respectively.
Background intensity levels are estimated by averaging the values of the pixels within the background regions defined by the same procedure as \citet{Hattori2016} and \citet{Hanaoka2019}.
The error of the intensity in each pixel is calculated from the standard deviation of the intensities within the background region and the systematic errors of 10\% are also considered in the local SEDs for AKARI, WISE mid-IR and Herschel data (D. Ishihara et al. in preparation; \cite{Jarrett2011}; \cite{Molinari2016}).
We created the local SED of each pixel and fitted it with a dust model which includes PAHs, hot dust, warm dust and cold dust components as follows:
\begin{dmath}
  F_{\nu}(\lambda) = a I_{\nu,\rm{PAH}}(\lambda) + b \lambda^{- \beta_{\rm{h}}} B_{\nu}(\lambda,T_{\rm{h}}) + c \lambda^{- \beta_{\rm{w}}} B_{\nu}(\lambda,T_{\rm{w}}) + d \lambda^{- \beta_{\rm{c}}} B_{\nu}(\lambda,T_{\rm{c}}),
  \label{eq:SED_model}  
\end{dmath}
where $I_{\nu,\rm{PAH}}(\lambda)$ is a PAH intensity model as a function of $\lambda$ (Draine \& Li 2007), $B_{\nu}(\lambda,T)$ is the Planck function with dust temperature $T$, and $\beta$ is the emissivity power-law index.
The temperatures of the warm and cold dust components are allowed to vary around $\sim60$~K and $\sim20$~K, respectively, as the initial parameters estimated from the SEDs of each IR bubble in \citet{Hanaoka2019}.
While that of the hot dust component is fixed at 123~K which is a typical value of the IR bubbles analyzed in this study.
The emissivity power-law index generally shows a value near 2 for H\emissiontype{II} regions in our Galaxy (\cite{Anderson2012}; \cite{Hanaoka2019}), however, it can vary depending on the optical properties of dust grains.
Here, the emissivity power-law index of the cold dust component, $\beta_{\rm{c}}$, is set to be free, when the $\beta_{\rm{c}}$ value of 2 is not statistically acceptable in the SED fitting.

Figure~\ref{fig:Local_SED_fit} shows examples of the local SED-fitting results for S7 (named ``RCW 120'' by \cite{Rodgers1960}).
The pixel denoted with ``A'' in the left panel of figure~\ref{fig:Local_SED_fit} is dominated by the hot dust component, while the pixel denoted with ``B'' requires the emissivity power-law index higher than 2.
When the SED fitting is statistically rejected, a not a number (i.e., NaN) is assigned to the corresponding pixel (i.e., masked).
To remove outliers, we applied a median filter, replacing a central pixel with the median value of the surrounding $3\times3$ pixels.
If the masked region occupies more than 50\% of the entire area within $2R$, we removed the corresponding IR bubbles from our sample.
As a result, we obtain the brightness distributions of 157 IR bubbles, which is 95\% of the 165 IR bubbles.
Examples of the brightness distributions thus obtained are shown in figure~\ref{fig:intensity_map}.

In order to check the consistency between the results of \citet{Hanaoka2019} and local SED fitting results in this study, the IR luminosities integrated over the area of $<2R$ are calculated for all the components, based on the brightness distributions derived with the local SED fitting.
Figure~\ref{fig:Local_Total_consis} shows comparison between the luminosities thus obtained from the local SED fitting and the luminosities estimated from the photometry fluxes within the $2R$ circular regions in \citet{Hanaoka2019}.
Here, we compare the warm dust luminosities of \citet{Hanaoka2019} with the sum of the hot and warm dust luminosities of the local SED fitting.
Figure~\ref{fig:Local_Total_consis} shows an overall consistency between the results of \citet{Hanaoka2019} and the local SED fitting.

\begin{figure*}
  \begin{center}
    \subfigure{
      \mbox{\raisebox{0mm}{\rotatebox{90}{\small{Galactic Latitude}}}}
      \mbox{\raisebox{0mm}{\includegraphics[width=35mm, bb=0 50 260 240, clip]{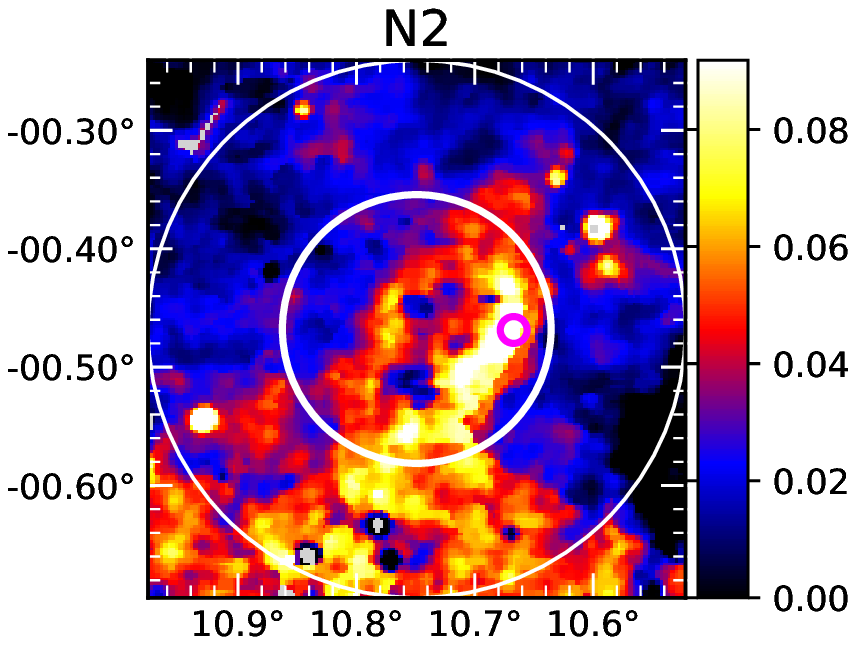}}}
      \mbox{\raisebox{0mm}{\includegraphics[width=35mm, bb=0 50 260 240, clip]{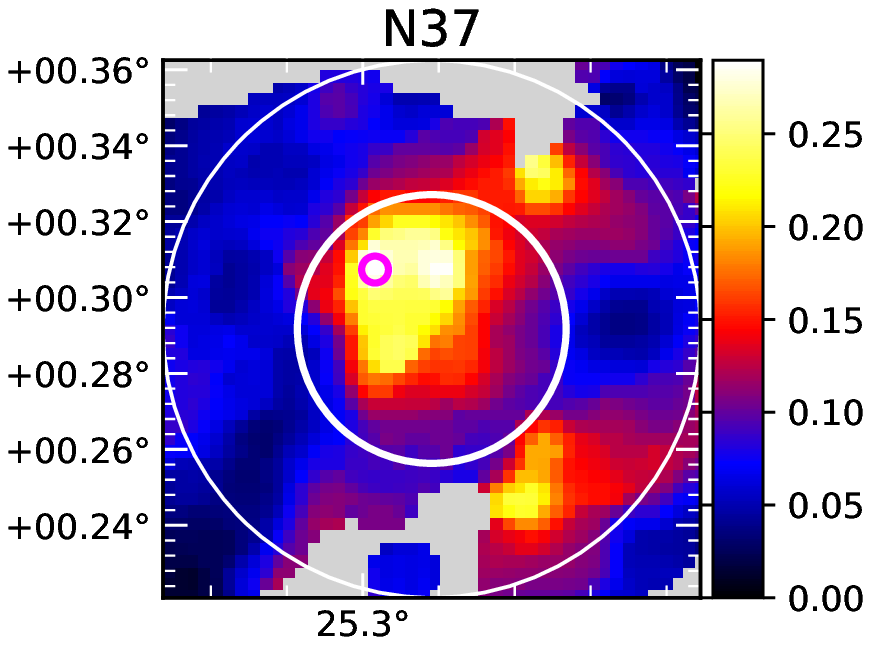}}}
      \mbox{\raisebox{0mm}{\includegraphics[width=35mm, bb=0 50 260 240, clip]{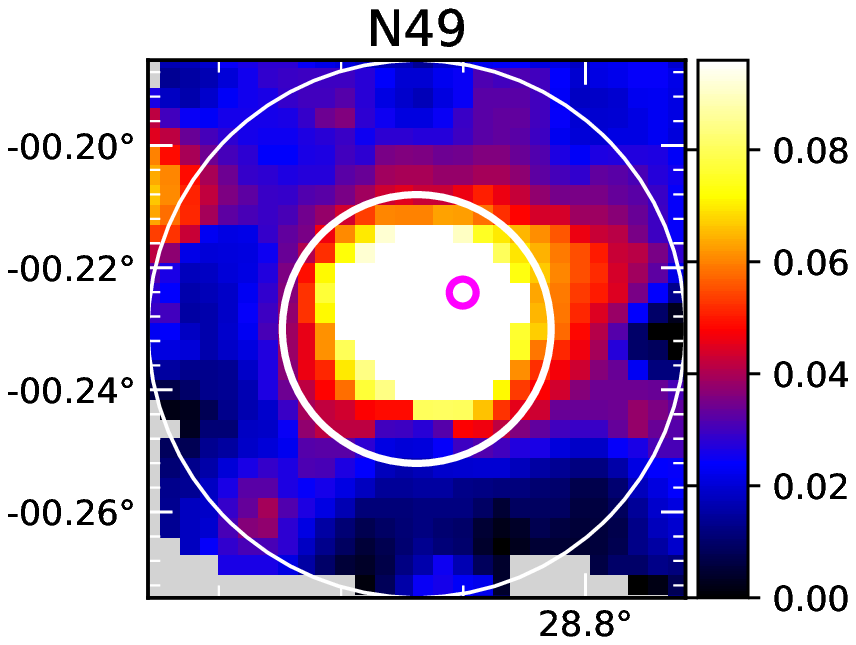}}}
      \mbox{\raisebox{0mm}{\includegraphics[width=35mm, bb=0 50 260 240, clip]{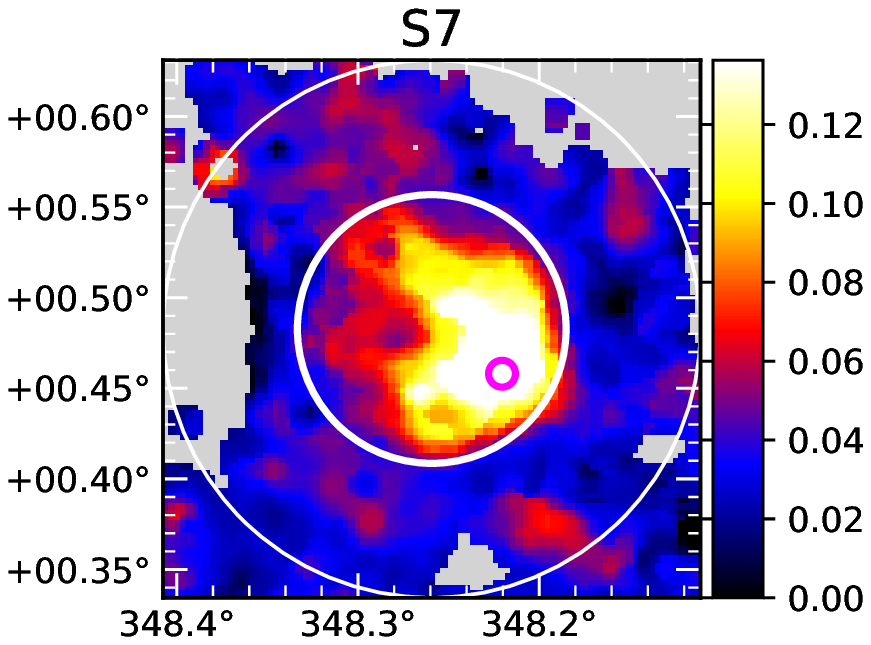}}}
    }
    \subfigure{
      \mbox{\raisebox{0mm}{\rotatebox{90}{\small{Galactic Latitude}}}}
      \mbox{\raisebox{0mm}{\includegraphics[width=35mm, bb=0 50 260 240, clip]{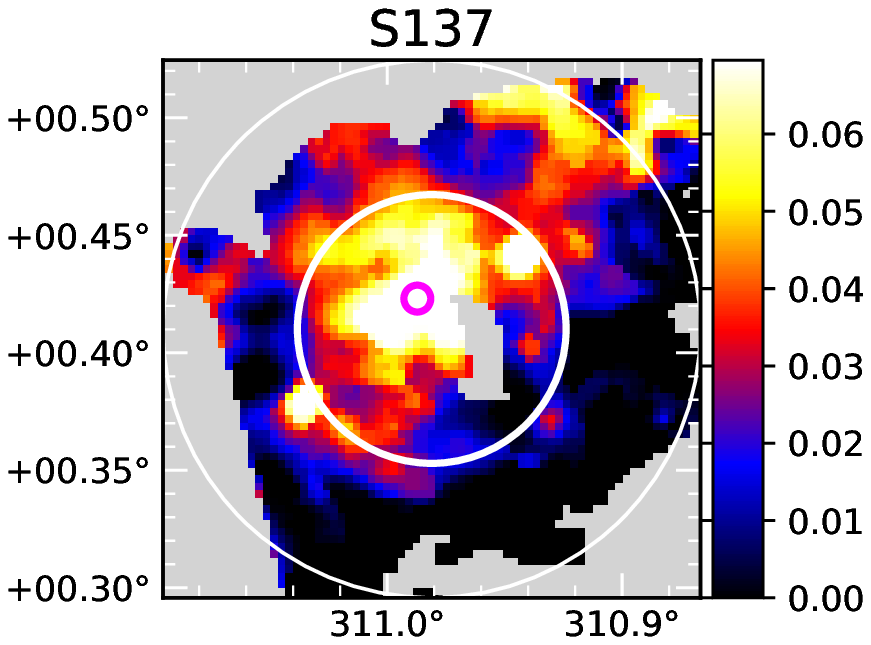}}}
      \mbox{\raisebox{0mm}{\includegraphics[width=35mm, bb=0 50 260 240, clip]{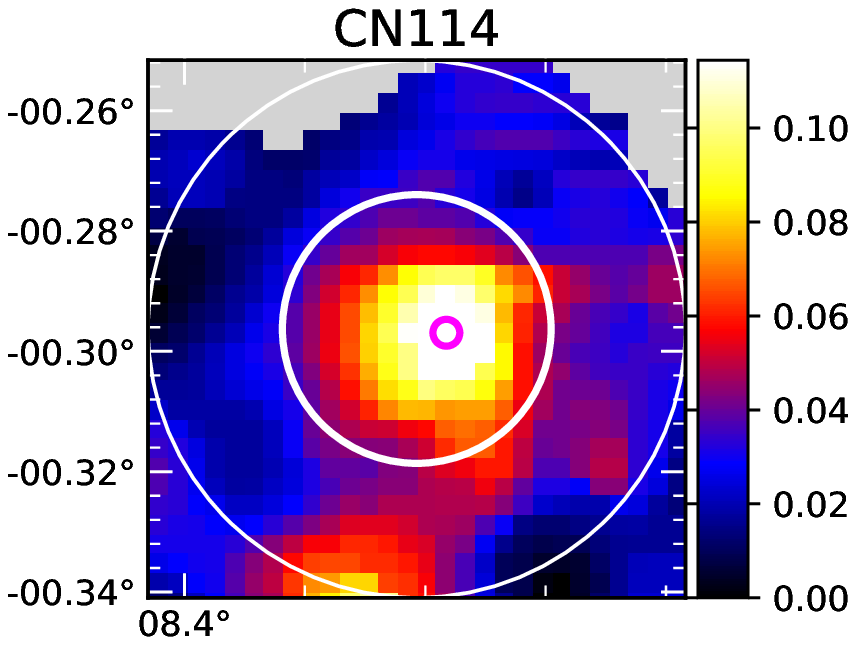}}}
      \mbox{\raisebox{0mm}{\includegraphics[width=35mm, bb=0 50 260 240, clip]{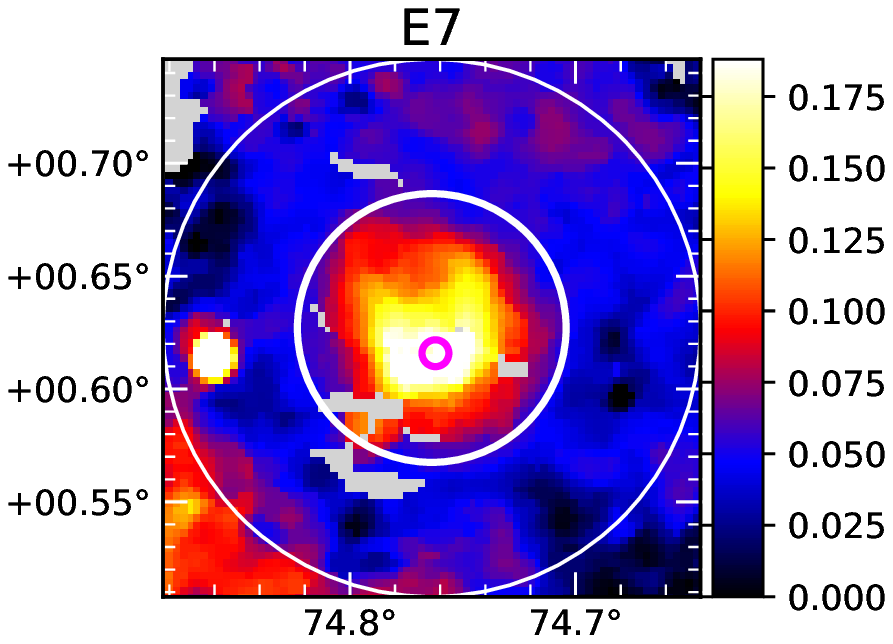}}}
      \mbox{\raisebox{0mm}{\includegraphics[width=35mm, bb=0 50 260 240, clip]{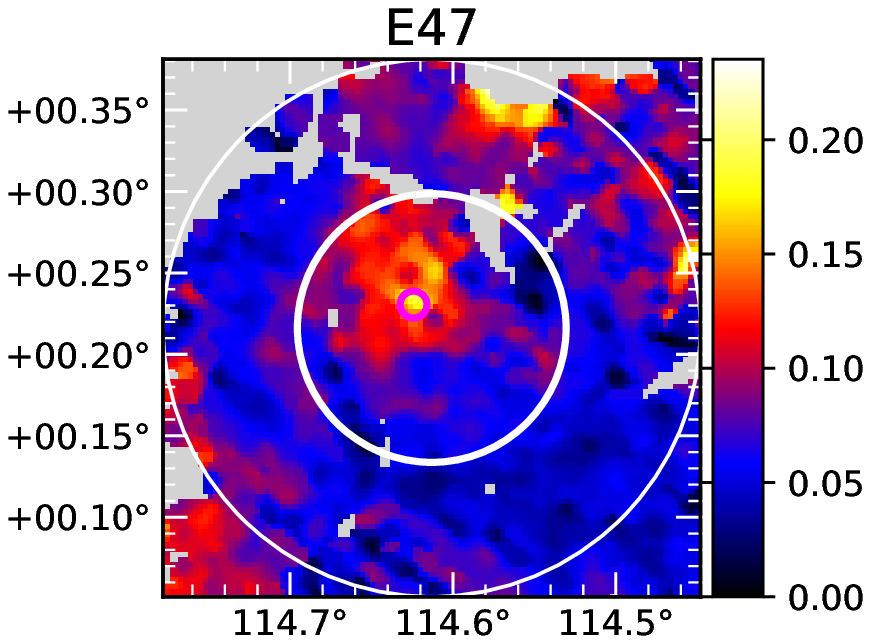}}}
    }
    \subfigure{
      \mbox{\raisebox{0mm}{\rotatebox{90}{\small{Galactic Latitude}}}}
      \mbox{\raisebox{0mm}{\includegraphics[width=35mm, bb=0 50 260 240, clip]{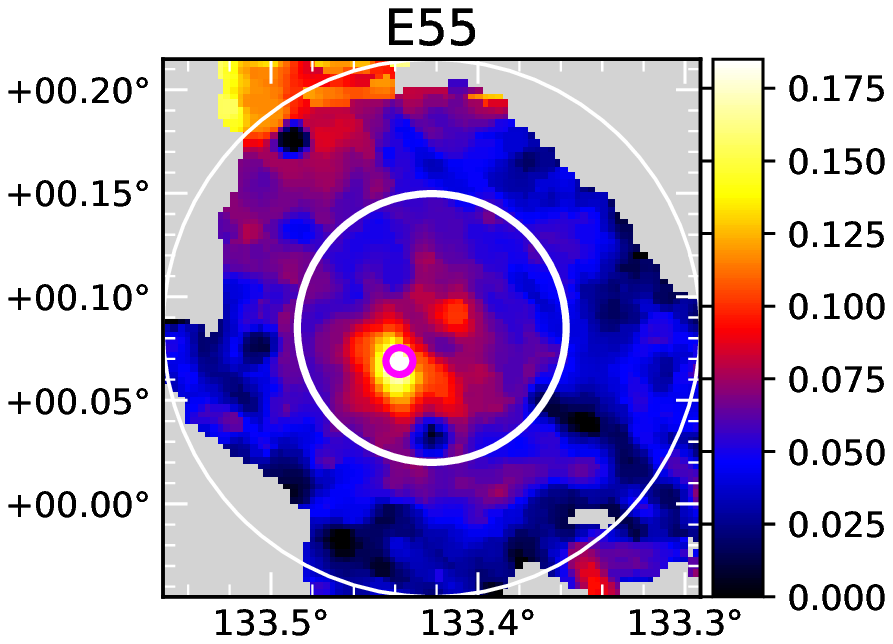}}}
      \mbox{\raisebox{0mm}{\includegraphics[width=35mm, bb=0 50 260 240, clip]{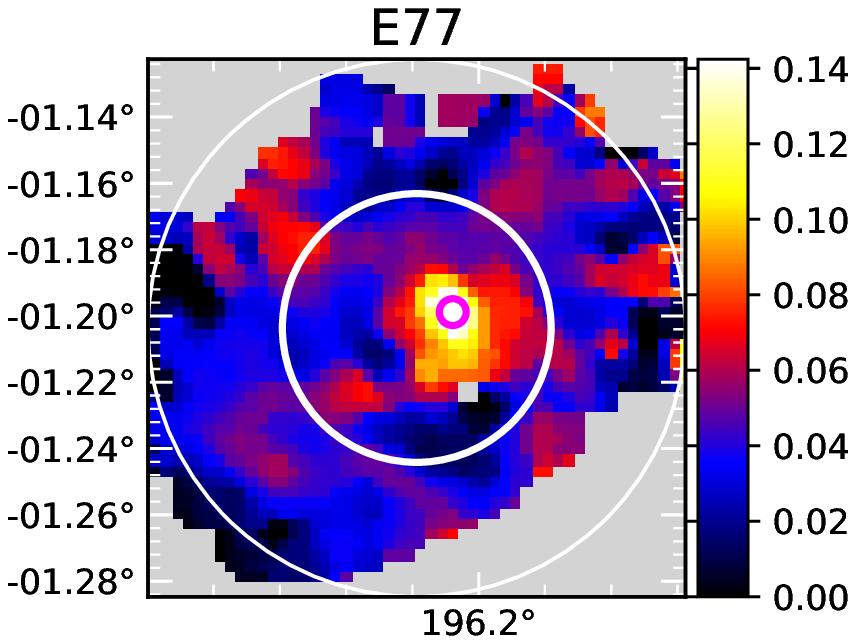}}}
      \mbox{\raisebox{0mm}{\includegraphics[width=35mm, bb=0 50 260 240, clip]{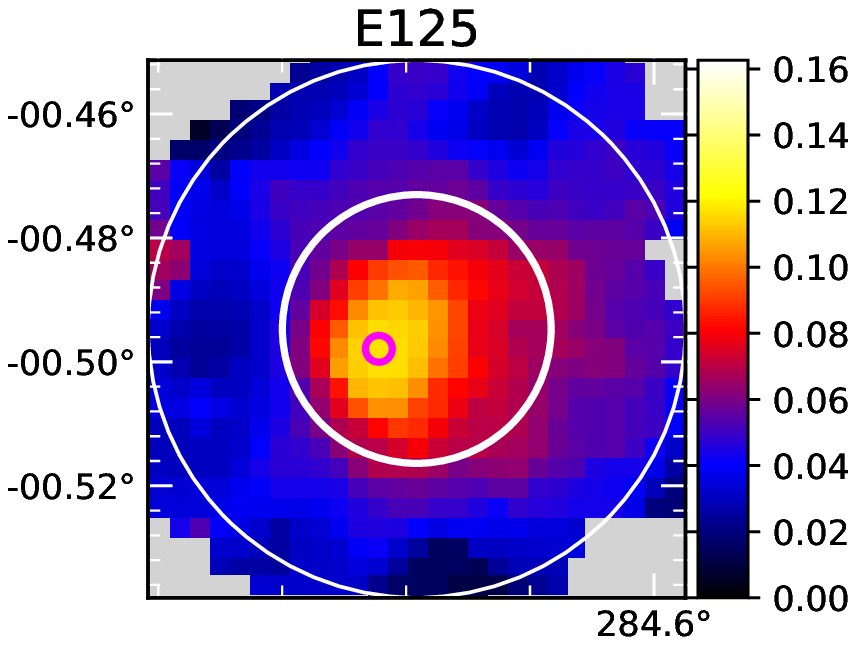}}}
      \mbox{\raisebox{0mm}{\includegraphics[width=35mm, bb=0 50 260 240, clip]{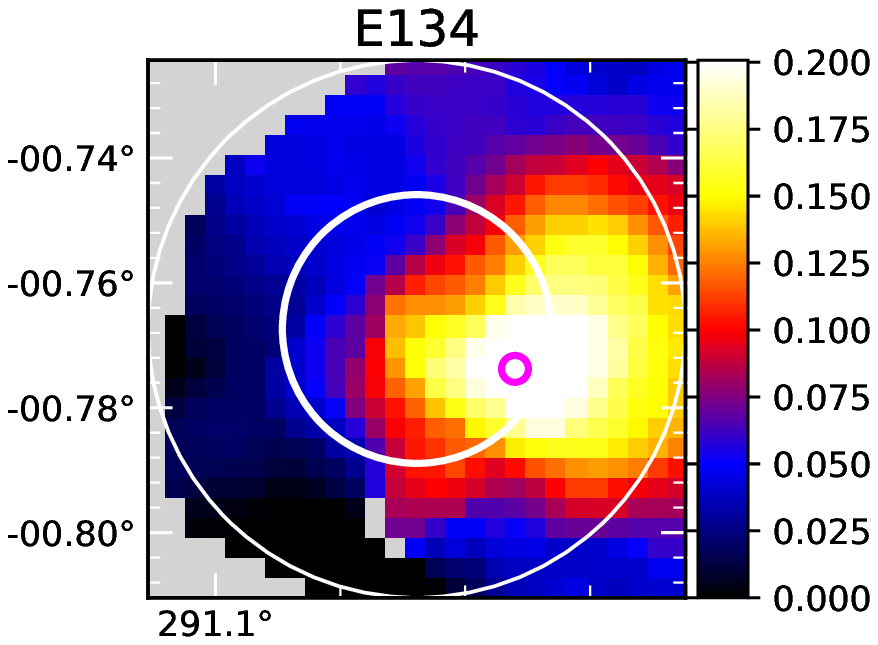}}}
    }
    \subfigure{\mbox{\raisebox{0mm}{\hspace{13mm}\small{Galactic Longitude}\hspace{8mm}\small{Galactic Longitude}\hspace{8mm}\small{Galactic Longitude}\hspace{8mm}\small{Galactic Longitude}\hspace{10mm}}}}
  \end{center}
  \caption{
    Examples of the $I_{\rm{hot}}/I_{\rm{TIR}}$ maps, together with the peak positions of the hot dust emission denoted with magenta circles.
    The thick and thin white circles correspond to the $R$ and $2R$ circular regions, respectively.
    }
  \label{fig:Ihot_map_for_pp}
\end{figure*}

\begin{figure}
  \begin{center}
    \subfigure{\includegraphics[width=0.5\linewidth,bb= -10 10 490 465,clip]{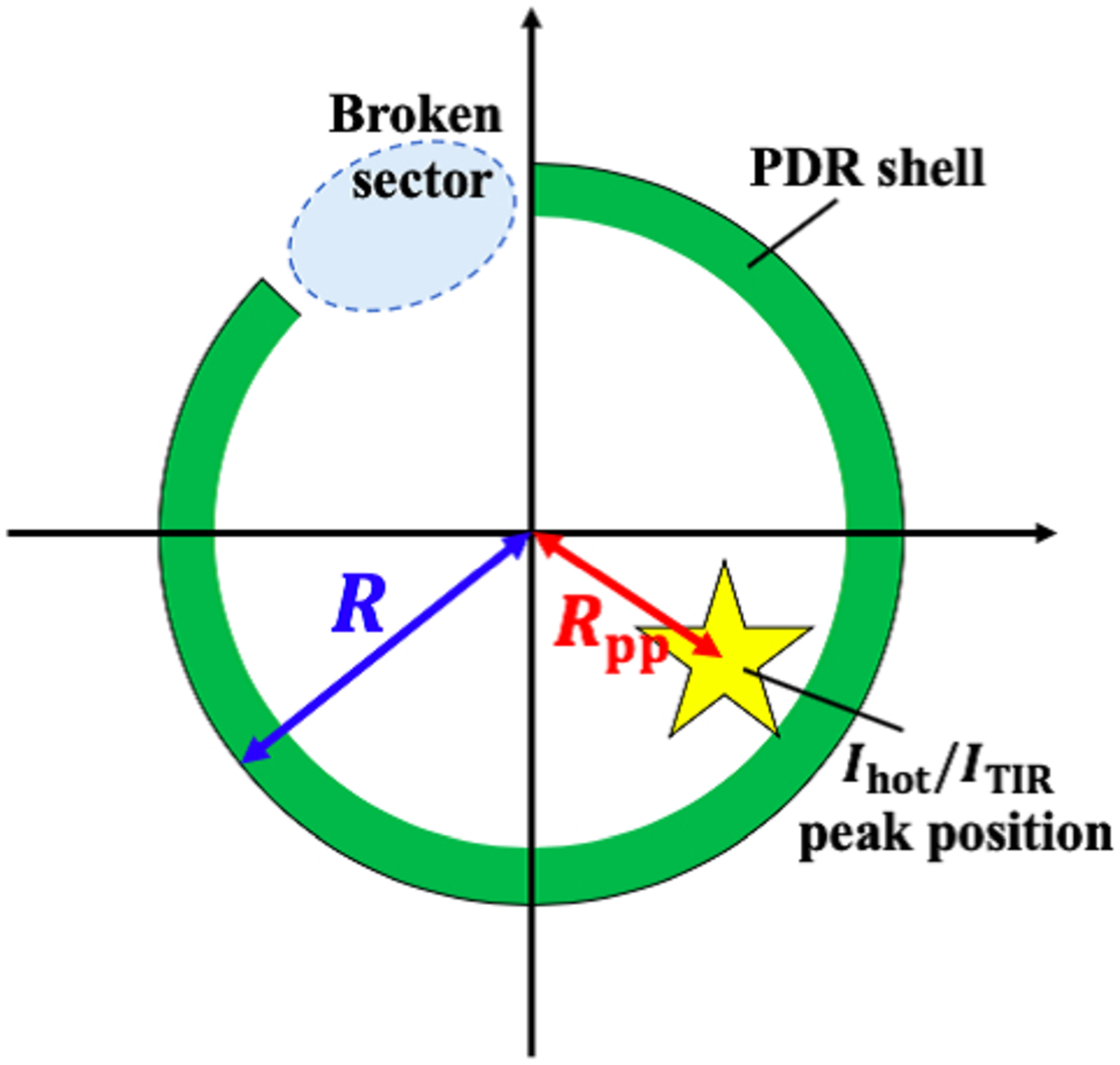}}
    \subfigure{\includegraphics[width=0.9\linewidth,clip]{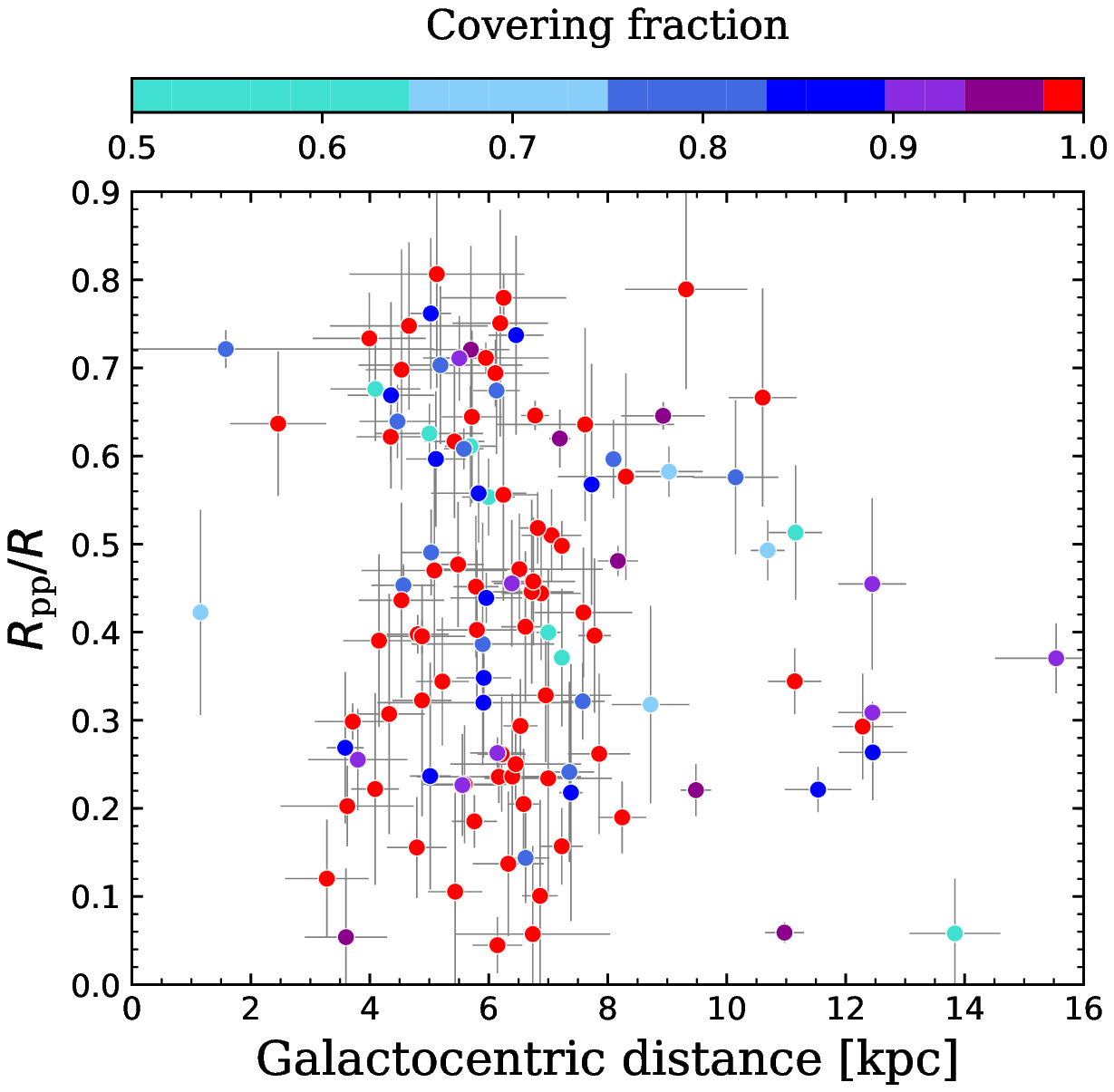}}
  \end{center}
  \caption{
    (Upper) schematic view to explain the definition of $R_{\rm{pp}}$.
    (Lower) the peak position of the hot dust emission relative to the IR bubble radius, $R_{\rm{pp}}/R$, plotted as a function of the Galactocentric distance.
    The data points are color-coded according to the CF obtained in \citet{Hanaoka2019} (the color will be discussed in Section 4).
  }
  \label{fig:rpp_GCdistance}
\end{figure}

\section{Result}
To estimate the positions of the heating sources, we investigate the brightness ratio maps of $I_{\rm{hot}}/I_{\rm{TIR}}$ for the IR bubbles.
Figure~\ref{fig:Ihot_map_for_pp} shows the $I_{\rm{hot}}/I_{\rm{TIR}}$ maps for the sample IR bubbles in figure~\ref{fig:intensity_map}.
The previous study by \citet{Hattori2016} showed that the color temperature of the dust emission peaks near the bubble centers for many of the well-defined closed bubbles, indicating the presence of heating sources (i.e., massive ionizing stars) in the centers.
  In our study, we choose only the IR bubbles which have a clear single $I_{\rm{hot}}/I_{\rm{TIR}}$ peak, so that we can mitigate the effect of mistakenly selecting a hot spot irrelevant to a massive star or confusion sources.
  Then we hereafter regard the positions of the $I_{\rm{hot}}/I_{\rm{TIR}}$ peaks as those of the heating sources.
We obtain the peak position of $I_{\rm{hot}}/I_{\rm{TIR}}$ within the inner edge of the shell (defined as 0.8$R$) for each IR bubble.
When there is no significant hot dust emission detected within the shell boundary, we obtain the peak position of the warm dust component, using the $I_{\rm{warm}}/I_{\rm{TIR}}$ map instead of the $I_{\rm{hot}}/I_{\rm{TIR}}$ map.
We then derive the peak positions for 117 IR bubbles, which have a single peak clearly identified within the shell boundary, out of the 165 IR bubbles (their peak positions are summarized in Appendix).
For 7 out of the 117 IR bubbles, we used the $I_{\rm{warm}}/I_{\rm{TIR}}$ maps to investigate the peak positions.
Then, we define the offset value from the center of each shell as the distance between the shell center and the peak position, $R_{\rm{pp}}$, relative to the angular shell radius $R$ (see the upper panel of figure~\ref{fig:rpp_GCdistance}).
The lower panel of figure~\ref{fig:rpp_GCdistance} shows the resultant $R_{\rm{pp}}/R$ plotted against the Galactocentric distance.
The values of $R$ and the Galactocentric distance of each IR bubble are cataloged in \citet{Hanaoka2019}.
The error of $R_{\rm{pp}}$ is evaluated by a quarter of the PSF size ($35''.0$) of the Herschel 500~$\mu$m band images (i.e., $9''$, similar to the pixel scale of $11''.5$).
From this panel, we find that many IR bubbles in inner Galactic regions tend to have large offsets from the centers of the PDR shells; with the Poisson statistics, $29\pm 6$\% (28/96) of the IR bubbles at $< 8$~kpc have $R_{\rm{pp}}/R > 0.6$, while $14\pm 8$\% (3/22) of the IR bubbles at $> 8$~kpc have $R_{\rm{pp}}/R > 0.6$, and thus the difference is marginally significant, although the statistics are not good enough for the latter.

\begin{figure*}
  \begin{center}
    \subfigure{
      \mbox{\raisebox{0mm}{\rotatebox{90}{\small{Galactic Latitude}}}}
      \mbox{\raisebox{0mm}{\includegraphics[width=35mm, bb=0 50 260 240, clip]{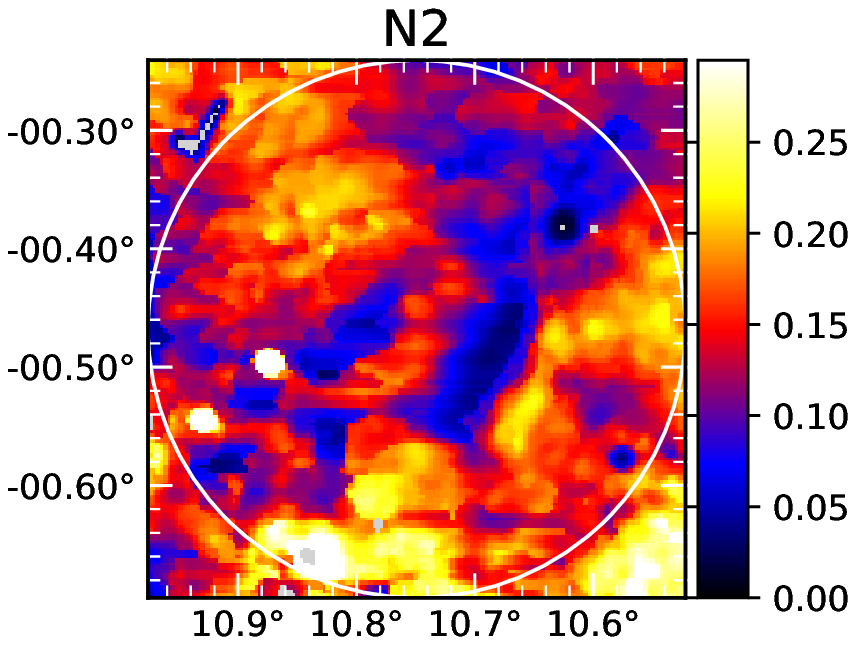}}}
      \mbox{\raisebox{0mm}{\includegraphics[width=35mm, bb=0 50 260 240, clip]{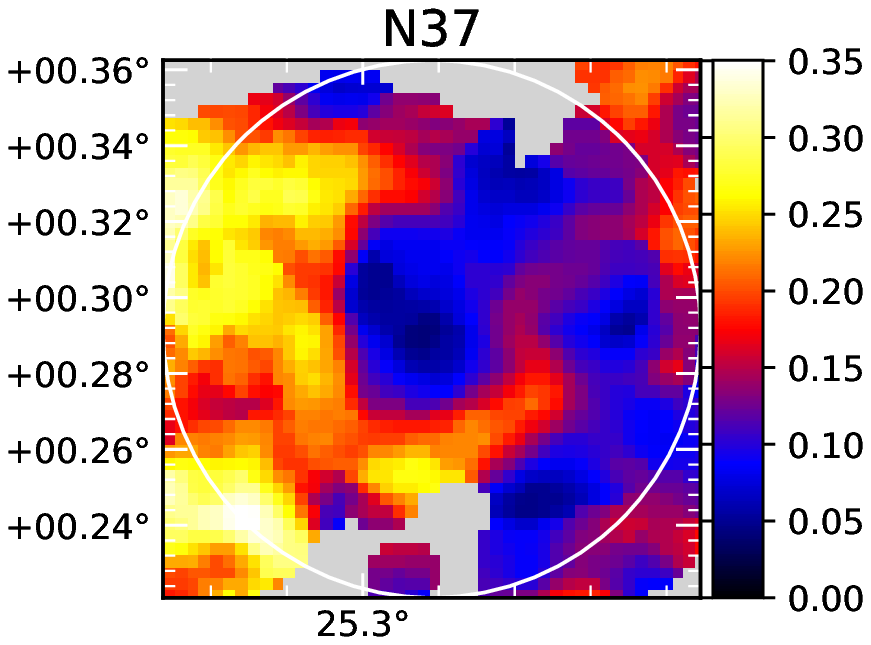}}}
      \mbox{\raisebox{0mm}{\includegraphics[width=35mm, bb=0 50 260 240, clip]{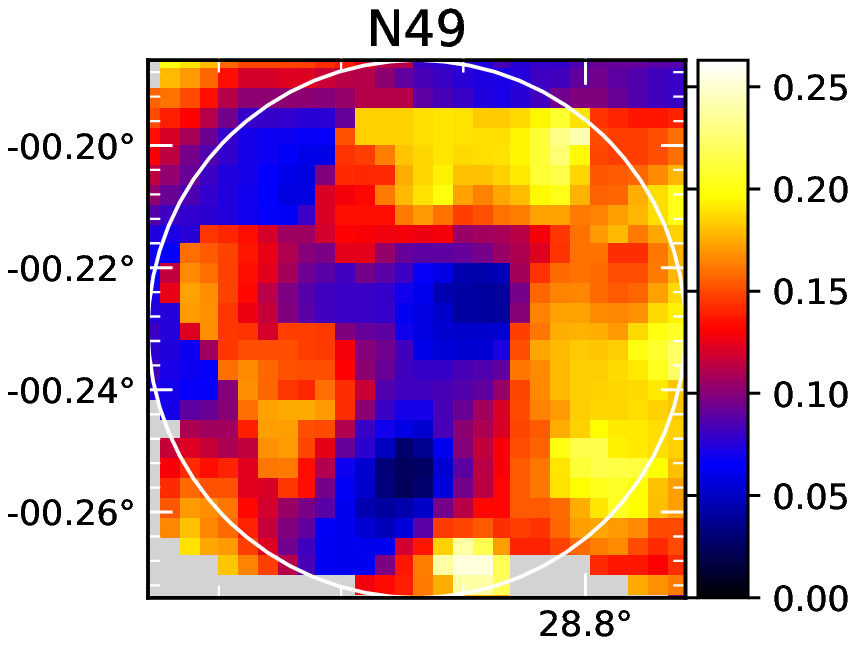}}}
      \mbox{\raisebox{0mm}{\includegraphics[width=35mm, bb=0 50 260 240, clip]{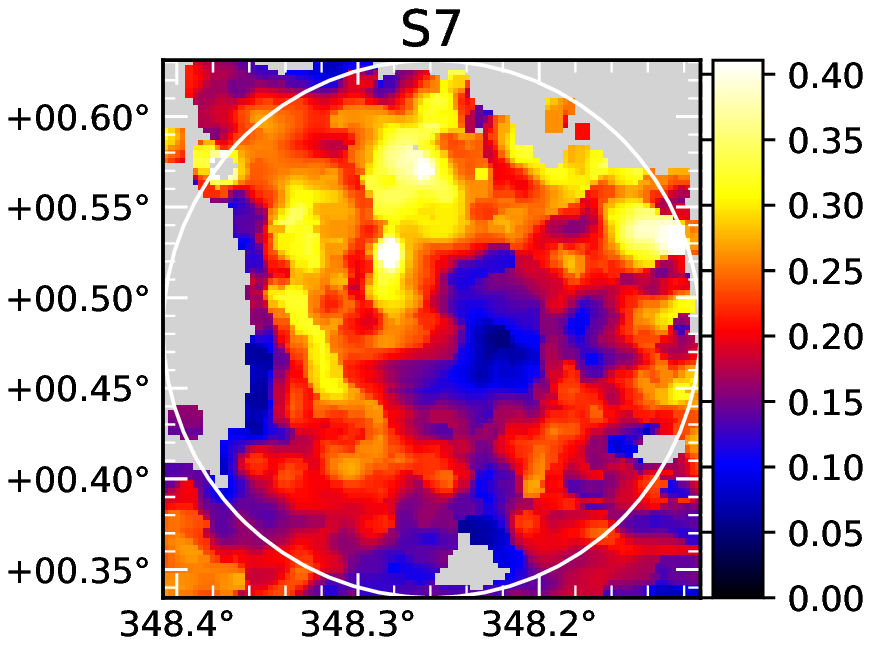}}}
    }
    \subfigure{
      \mbox{\raisebox{0mm}{\rotatebox{90}{\small{Galactic Latitude}}}}
      \mbox{\raisebox{0mm}{\includegraphics[width=35mm, bb=0 50 260 240, clip]{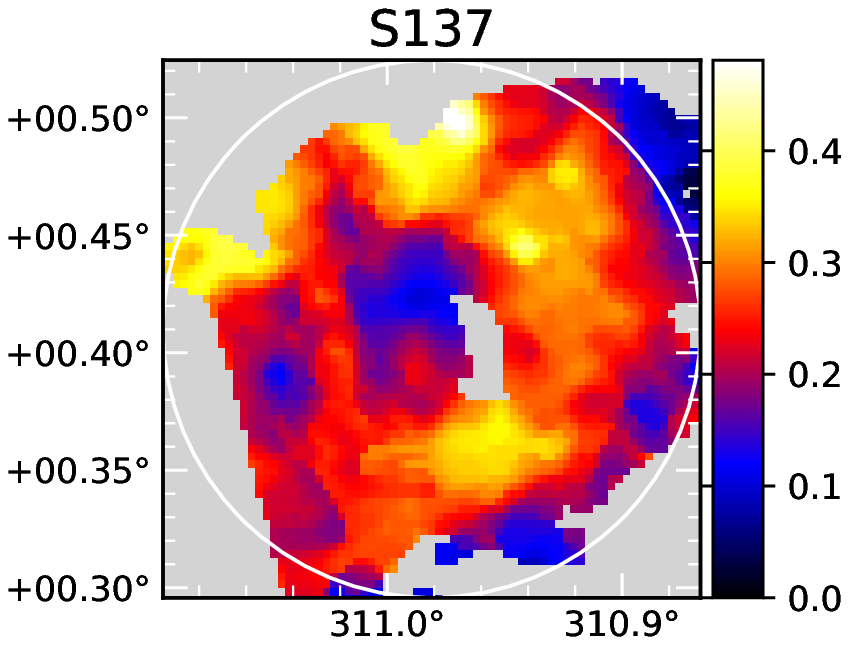}}}
      \mbox{\raisebox{0mm}{\includegraphics[width=35mm, bb=0 50 260 240, clip]{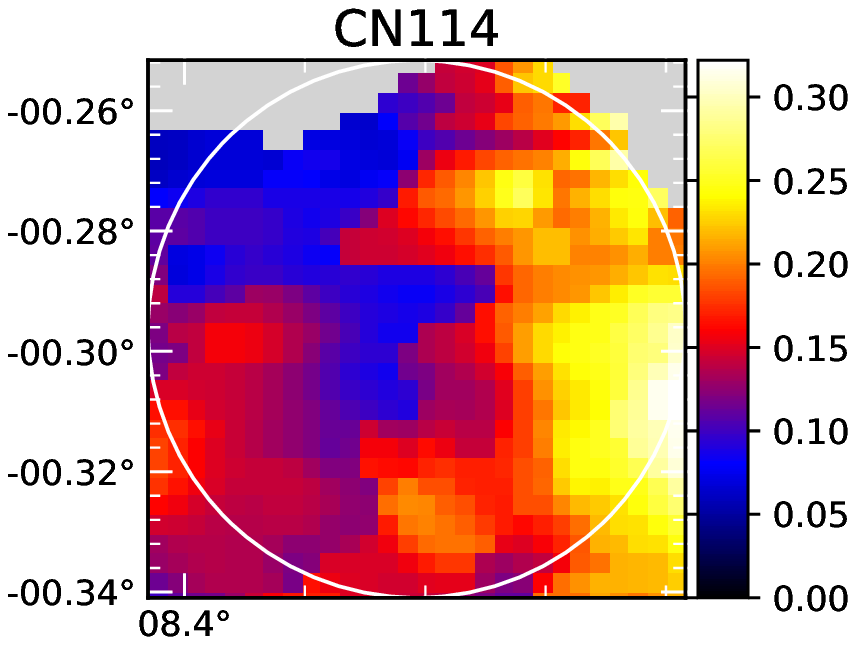}}}
      \mbox{\raisebox{0mm}{\includegraphics[width=35mm, bb=0 50 260 240, clip]{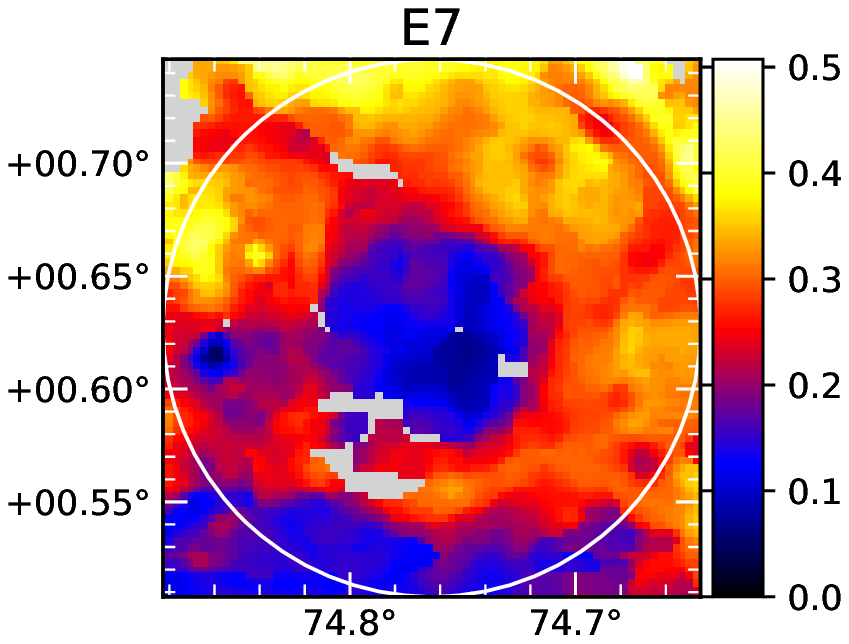}}}
      \mbox{\raisebox{0mm}{\includegraphics[width=35mm, bb=0 50 260 240, clip]{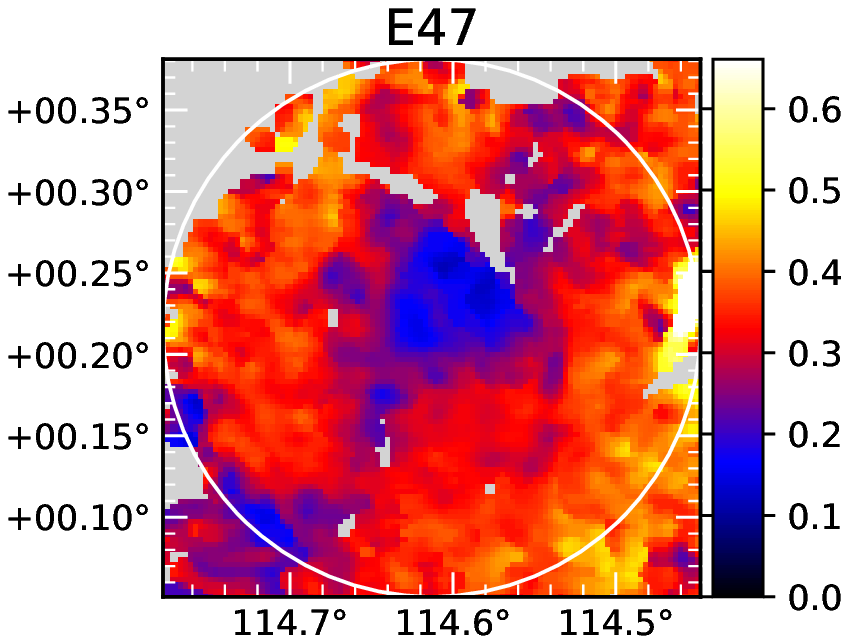}}}
    }
    \subfigure{
      \mbox{\raisebox{0mm}{\rotatebox{90}{\small{Galactic Latitude}}}}
      \mbox{\raisebox{0mm}{\includegraphics[width=35mm, bb=0 50 260 240, clip]{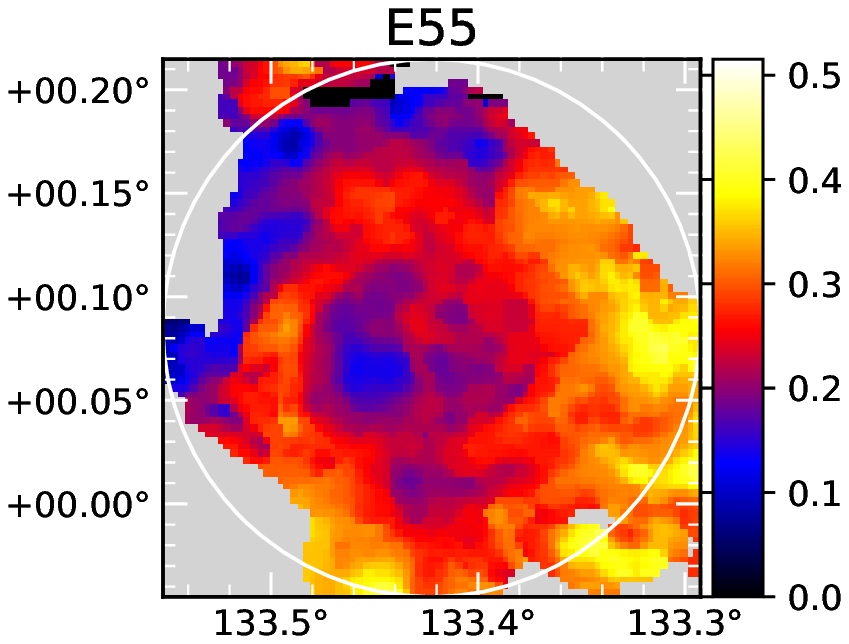}}}
      \mbox{\raisebox{0mm}{\includegraphics[width=35mm, bb=0 50 260 240, clip]{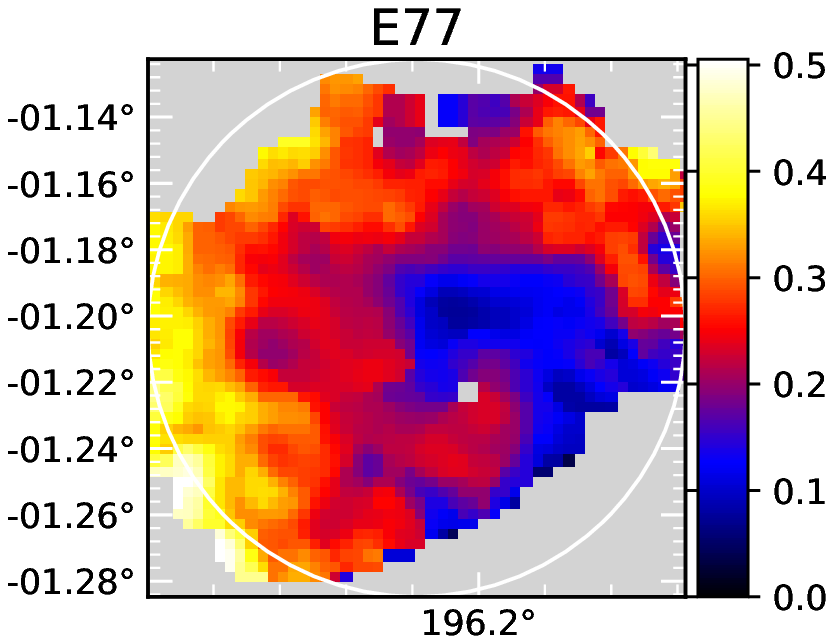}}}
      \mbox{\raisebox{0mm}{\includegraphics[width=35mm, bb=0 50 260 240, clip]{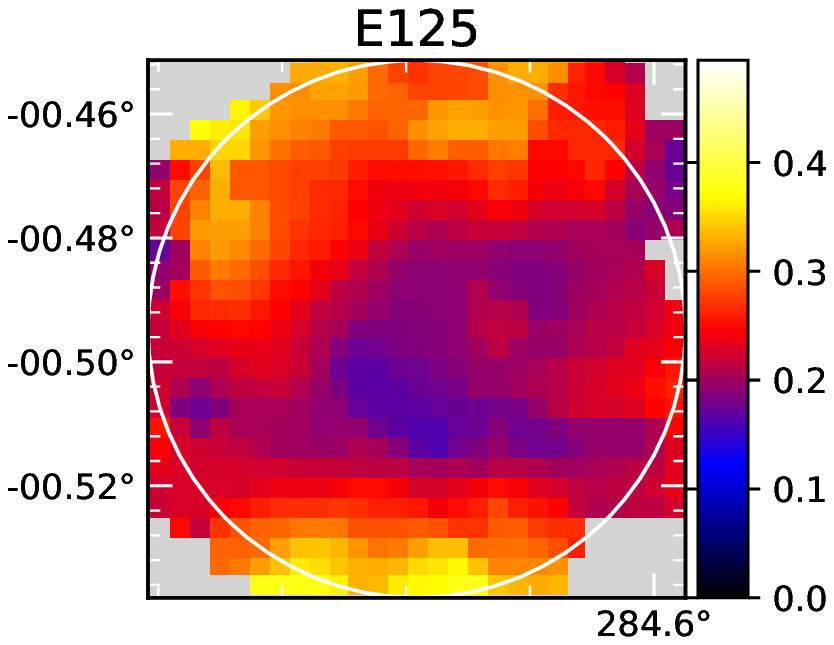}}}
      \mbox{\raisebox{0mm}{\includegraphics[width=35mm, bb=0 50 260 240, clip]{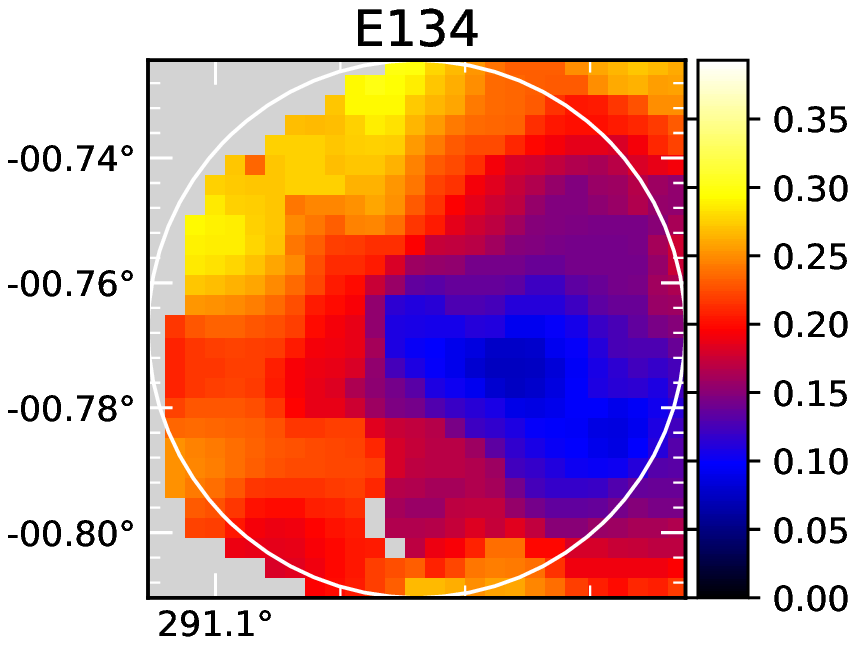}}}
    }
    \subfigure{\mbox{\raisebox{0mm}{\hspace{13mm}\small{Galactic Longitude}\hspace{8mm}\small{Galactic Longitude}\hspace{8mm}\small{Galactic Longitude}\hspace{8mm}\small{Galactic Longitude}\hspace{10mm}}}}
  \end{center}
  \caption{
  Examples of the $I_{\rm{PAH}}/I_{\rm{TIR}}$ maps.
  The white circles correspond to the $2R$ circular regions.
  }
  \label{fig:Ipah_map_for_PAHCV}
\end{figure*}

\begin{figure*}
  \begin{center}
    \includegraphics[width=0.8\linewidth,clip]{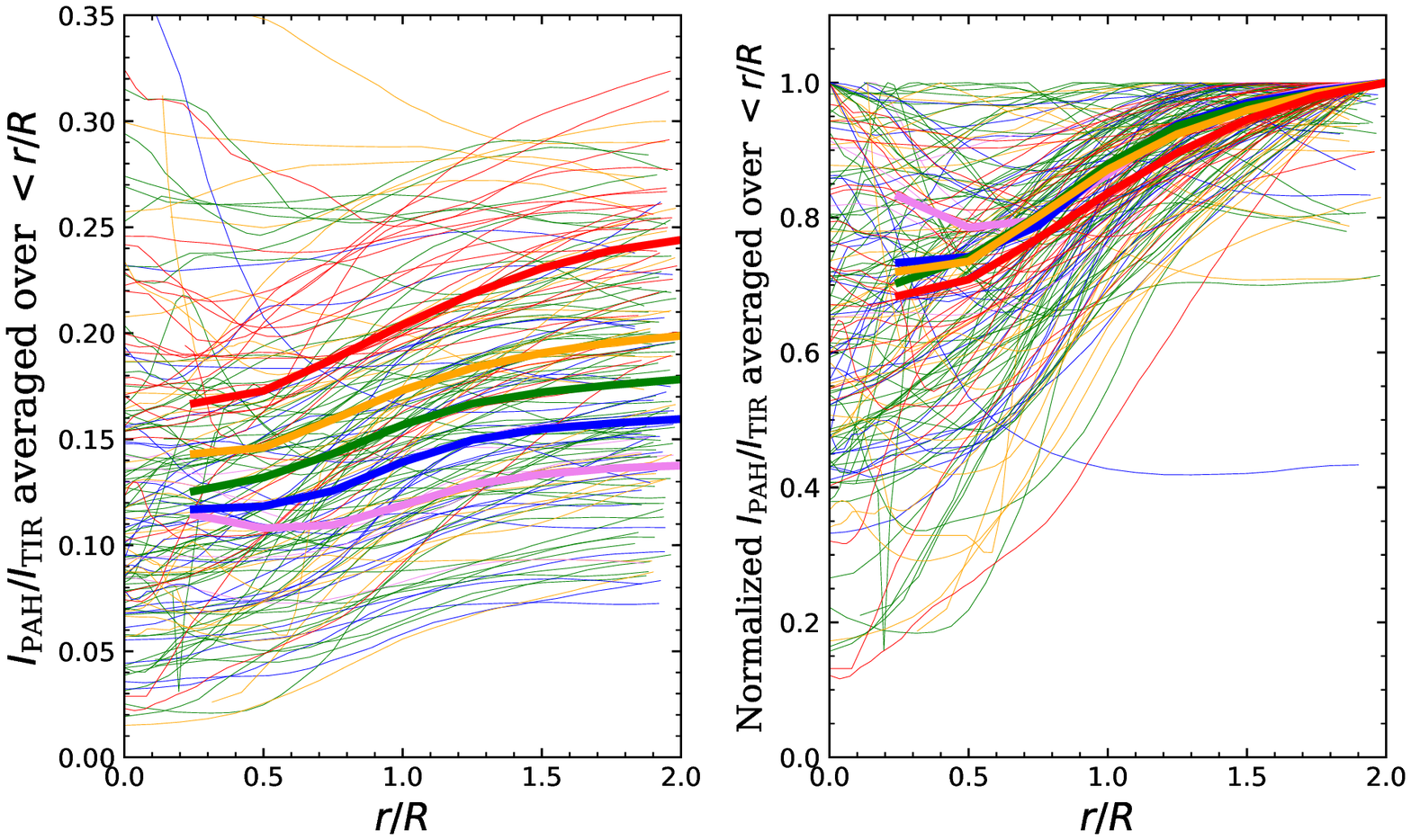}
  \end{center}
  \caption{
    (Left) radial profiles of $I_{\rm{PAH}}/I_{\rm{TIR}}$ averaged over the pixels within a circular region of a radius $r$.
    (Right) the same as the left panel, but all the profiles are normalized.
    The profiles are color-coded with the Galactocentric distance; the violet, blue, green, orange and red curves correspond to 0$-$3, 3$-$5, 5$-$7, 7$-$9 and 9$-$12~kpc from the Galactic center, respectively, while the bold curves are their averages.
  }
  \label{fig:radial_profile_GCdistance}
\end{figure*}

Figure~\ref{fig:Ipah_map_for_PAHCV} shows the brightness ratio maps of $I_{\rm{PAH}}/I_{\rm{TIR}}$ for the sample IR bubbles in figure~\ref{fig:intensity_map}.
From the $I_{\rm{PAH}}/I_{\rm{TIR}}$ distribution maps, we investigate the radial profiles of $I_{\rm{PAH}}/I_{\rm{TIR}}$ averaged over the pixels within a certain radial position $r$, which are derived by systematically reducing the resolution of $I_{\rm{PAH}}/I_{\rm{TIR}}$ maps to match the lowest resolution in our sample.
Figure~\ref{fig:radial_profile_GCdistance} shows the $I_{\rm{PAH}}/I_{\rm{TIR}}$ radial profiles plotted against $r/R$, color-coded according to the Galactocentric distance.
This result shows that $I_{\rm{PAH}}/I_{\rm{TIR}}$ monotonically increases with $r$ to $\sim R$ which is the radius of each PDR shell.
This global trend is reasonable because the PAH emission dominates in the PDR shells as pointed out by \citet{Hattori2016}.
Figure~\ref{fig:radial_profile_GCdistance} also shows that $I_{\rm{PAH}}/I_{\rm{TIR}}$ systematically increases with the Galactocentric distance, while the radial profiles are similar among those at different Galactocentric distances, as seen in the right panel of the figure.
This is consistent with the result of $L_{\rm{PAH}}/L_{\rm{TIR}}$ shown in \citet{Hanaoka2019}, further indicating that IR bubbles in outer Galactic regions have systematically (i.e., not locally but globally) high $L_{\rm{PAH}}/L_{\rm{TIR}}$ within their 2$R$ regions.

\begin{figure*}
  \begin{center}
    \subfigure{
      \mbox{\raisebox{0mm}{\rotatebox{90}{\small{Galactic Latitude}}}}
      \mbox{\raisebox{0mm}{\includegraphics[width=35mm, bb=0 50 260 240, clip]{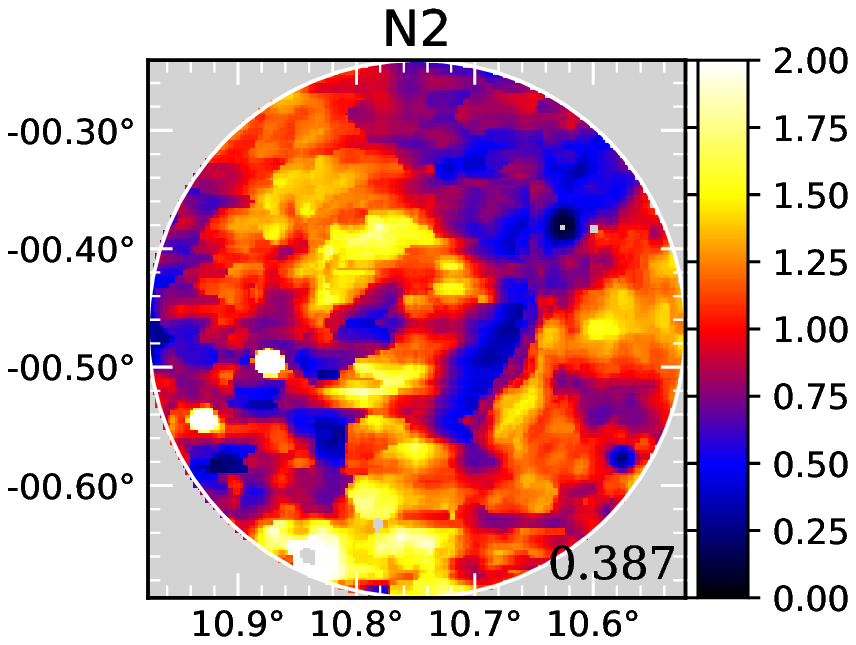}}}
      \mbox{\raisebox{0mm}{\includegraphics[width=35mm, bb=0 50 260 240, clip]{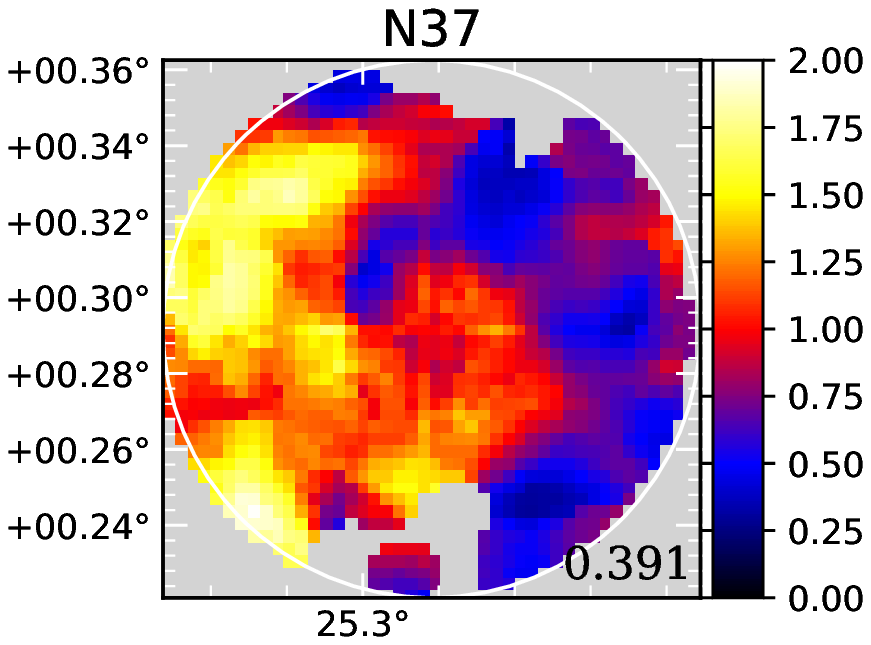}}}
      \mbox{\raisebox{0mm}{\includegraphics[width=35mm, bb=0 50 260 240, clip]{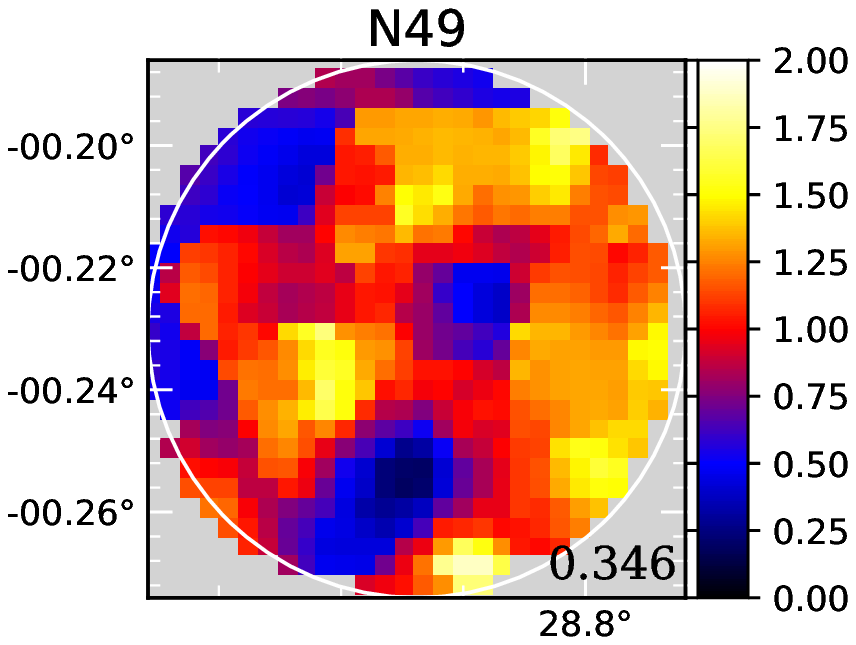}}}
      \mbox{\raisebox{0mm}{\includegraphics[width=35mm, bb=0 50 260 240, clip]{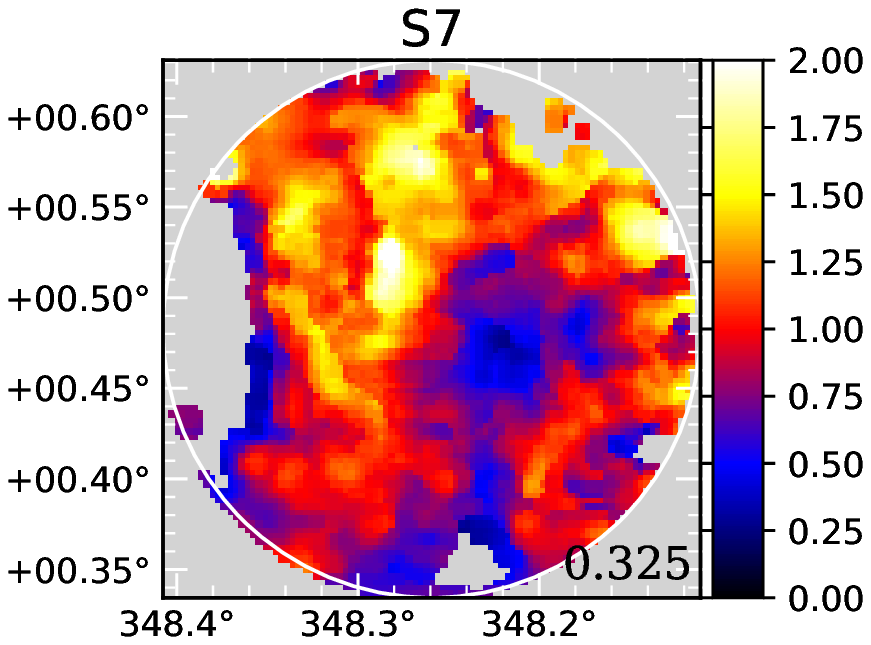}}}
    }
    \subfigure{
      \mbox{\raisebox{0mm}{\rotatebox{90}{\small{Galactic Latitude}}}}
      \mbox{\raisebox{0mm}{\includegraphics[width=35mm, bb=0 50 260 240, clip]{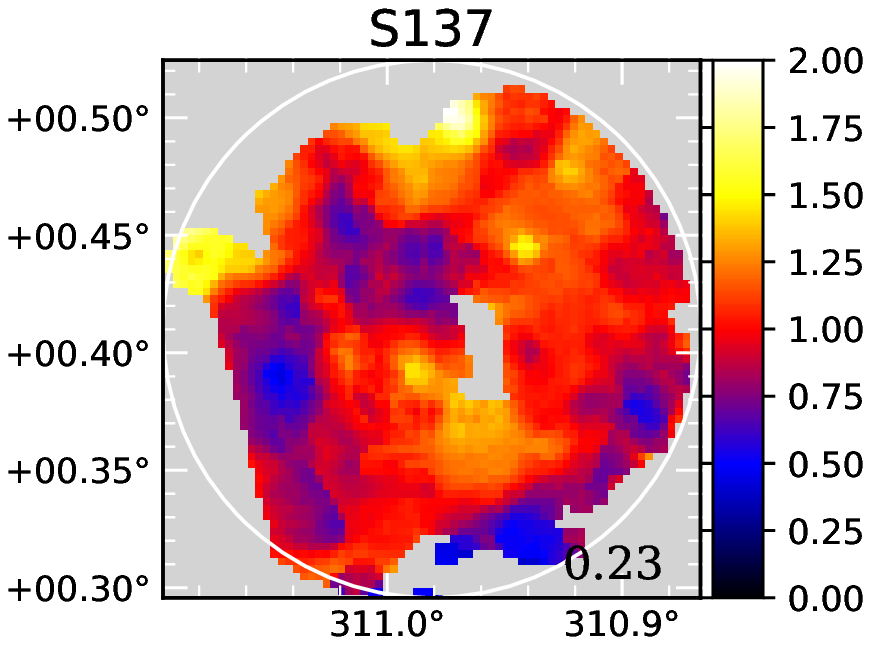}}}
      \mbox{\raisebox{0mm}{\includegraphics[width=35mm, bb=0 50 260 240, clip]{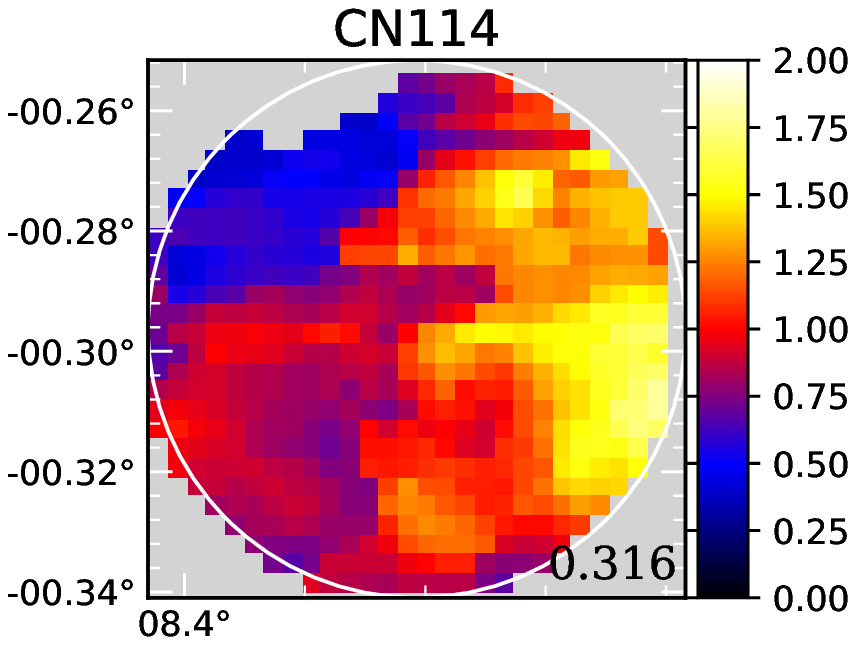}}}
      \mbox{\raisebox{0mm}{\includegraphics[width=35mm, bb=0 50 260 240, clip]{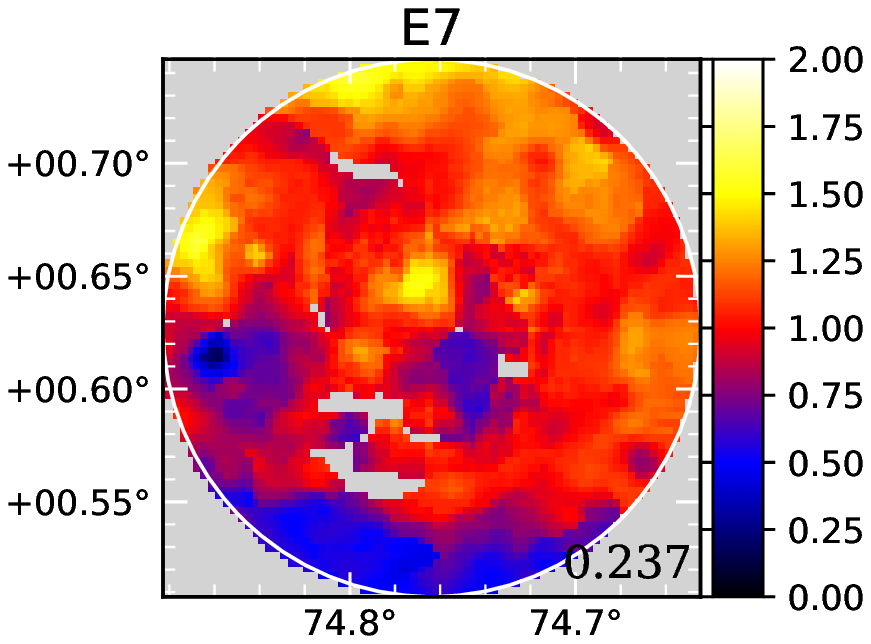}}}
      \mbox{\raisebox{0mm}{\includegraphics[width=35mm, bb=0 50 260 240, clip]{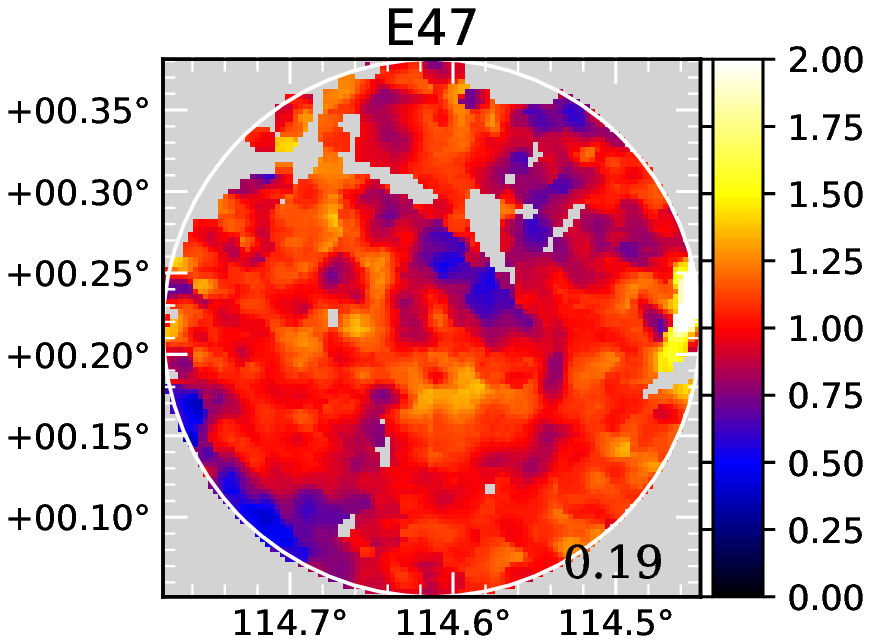}}}
    }
    \subfigure{
      \mbox{\raisebox{0mm}{\rotatebox{90}{\small{Galactic Latitude}}}}
      \mbox{\raisebox{0mm}{\includegraphics[width=35mm, bb=0 50 260 240, clip]{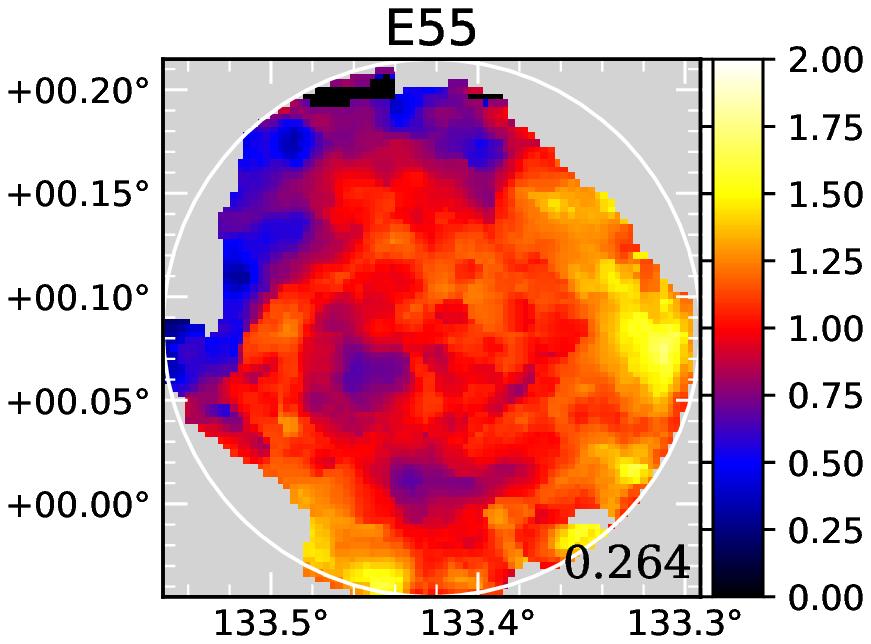}}}
      \mbox{\raisebox{0mm}{\includegraphics[width=35mm, bb=0 50 260 240, clip]{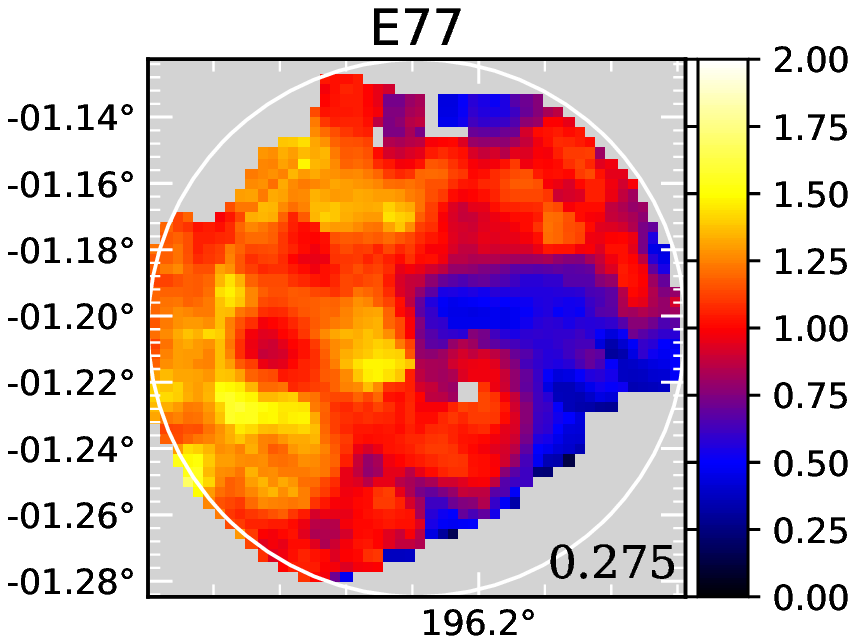}}}
      \mbox{\raisebox{0mm}{\includegraphics[width=35mm, bb=0 50 260 240, clip]{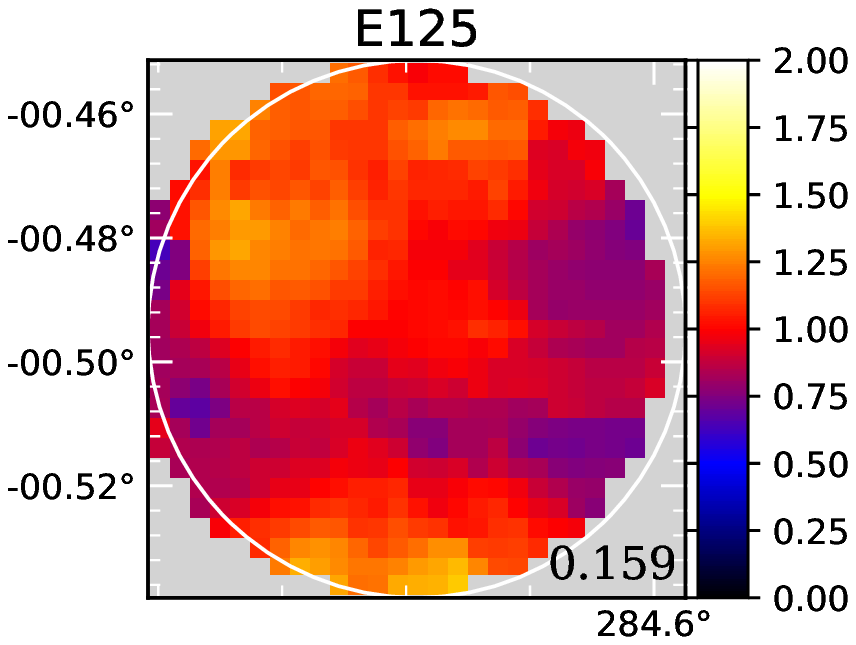}}}
      \mbox{\raisebox{0mm}{\includegraphics[width=35mm, bb=0 50 260 240, clip]{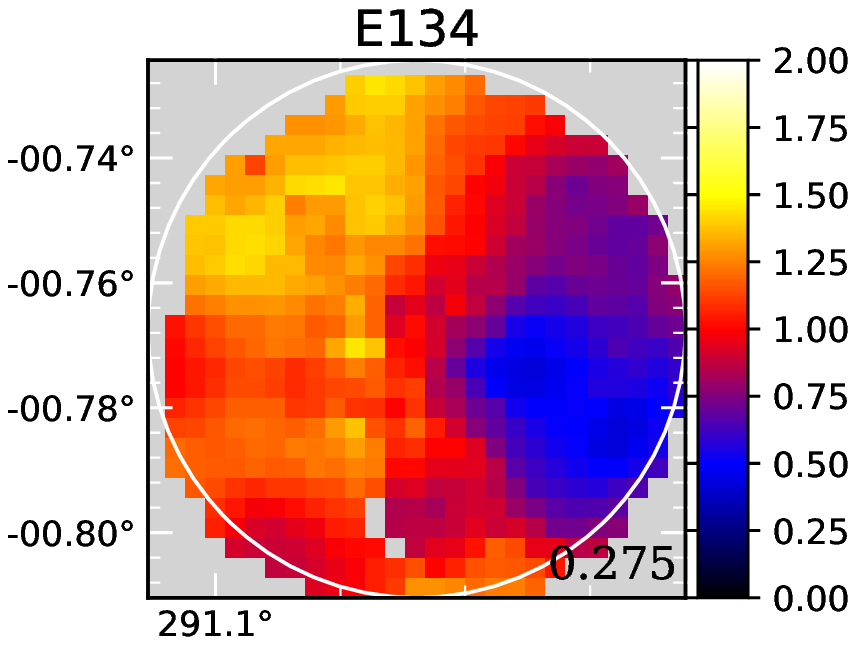}}}
    }
    \subfigure{\mbox{\raisebox{0mm}{\hspace{13mm}\small{Galactic Longitude}\hspace{8mm}\small{Galactic Longitude}\hspace{8mm}\small{Galactic Longitude}\hspace{8mm}\small{Galactic Longitude}\hspace{10mm}}}}
  \end{center}
  \caption{
    Examples of the $I_{\rm{PAH}}/I_{\rm{TIR}}$ maps after removing the radial trends of the PAH emission with the radial profiles (see figure~\ref{fig:radial_profile_GCdistance}).
    The white circles correspond to the $2R$ circular regions.
    The coefficients of variations (CVs) of the $I_{\rm{PAH}}/I_{\rm{TIR}}$ are denoted at the right bottom of each panel.
  }
  \label{fig:Ipah_flat_map}
\end{figure*}

We quantify the local spatial variation of the PAH intensity within each IR bubble.
To remove the radial global trend of the PAH intensity as seen in figure~\ref{fig:radial_profile_GCdistance}, we normalize the $I_{\rm{PAH}}/I_{\rm{TIR}}$ maps with the radial profile of each IR bubble (i.e., setting the azimuthally-averaged value to be a unity for given $r$).
Figure~\ref{fig:Ipah_flat_map} shows the resultant $I_{\rm{PAH}}/I_{\rm{TIR}}$ maps after removal of the radial trend for the same IR bubbles as in figure~\ref{fig:Ipah_map_for_PAHCV}.
We obtain the coefficients of variations (CVs) of the $I_{\rm{PAH}}/I_{\rm{TIR}}$ values within the 2$R$ circular region in each IR bubble, which are calculated as the standard deviations divided by the mean values.
We compare the $I_{\rm{PAH}}/I_{\rm{TIR}}$ CVs of the IR bubbles in inner Galactic regions with those in outer Galactic regions.

\begin{figure}
  \begin{center}
    \includegraphics[width=0.9\linewidth,clip]{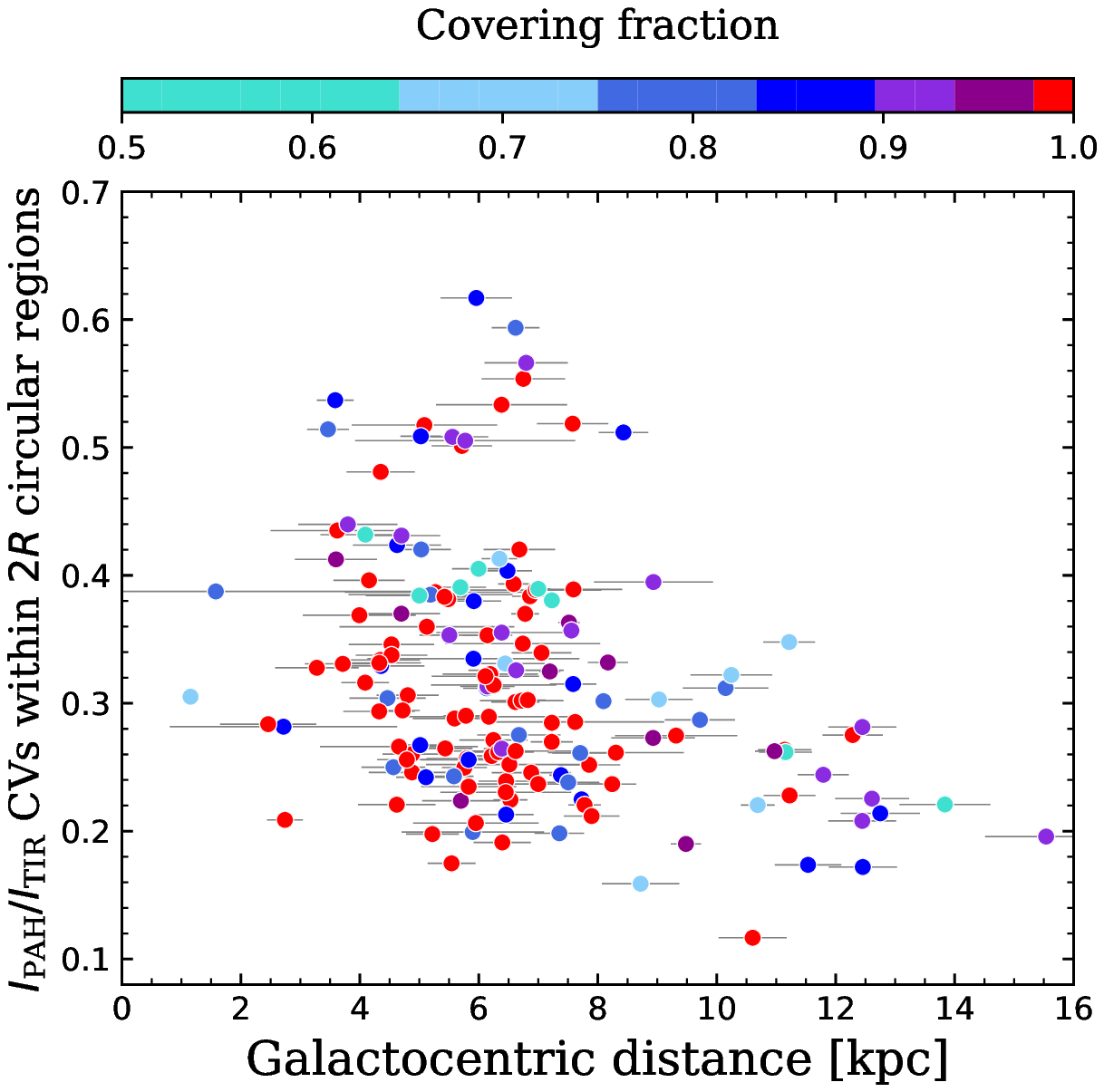}
  \end{center}
  \caption{
    Coefficients of variations (CVs), calculated as the standard deviations divided by the mean values, within the $2R$ circular regions of the $I_{\rm{PAH}}/I_{\rm{TIR}}$ maps after removing the radial profile (figure~\ref{fig:radial_profile_GCdistance}), plotted against the Galactocentric distance.
    The data points are color-coded according to the CF obtained in \citet{Hanaoka2019} (the color will be discussed in Section 4).
  }
  \label{fig:CVs_vs_GCdistance}
\end{figure}

Figure~\ref{fig:CVs_vs_GCdistance} shows that the IR bubbles in inner Galactic regions are likely to have higher $I_{\rm{PAH}}/I_{\rm{TIR}}$ CVs than the IR bubbles in outer Galactic regions, indicating that $I_{\rm{PAH}}/I_{\rm{TIR}}$ in inner Galactic regions have not only higher values on average but also larger spatial variations relative to the average than those in outer Galactic regions.

\begin{figure}
      \begin{center}
        \includegraphics[width=0.9\linewidth,clip]{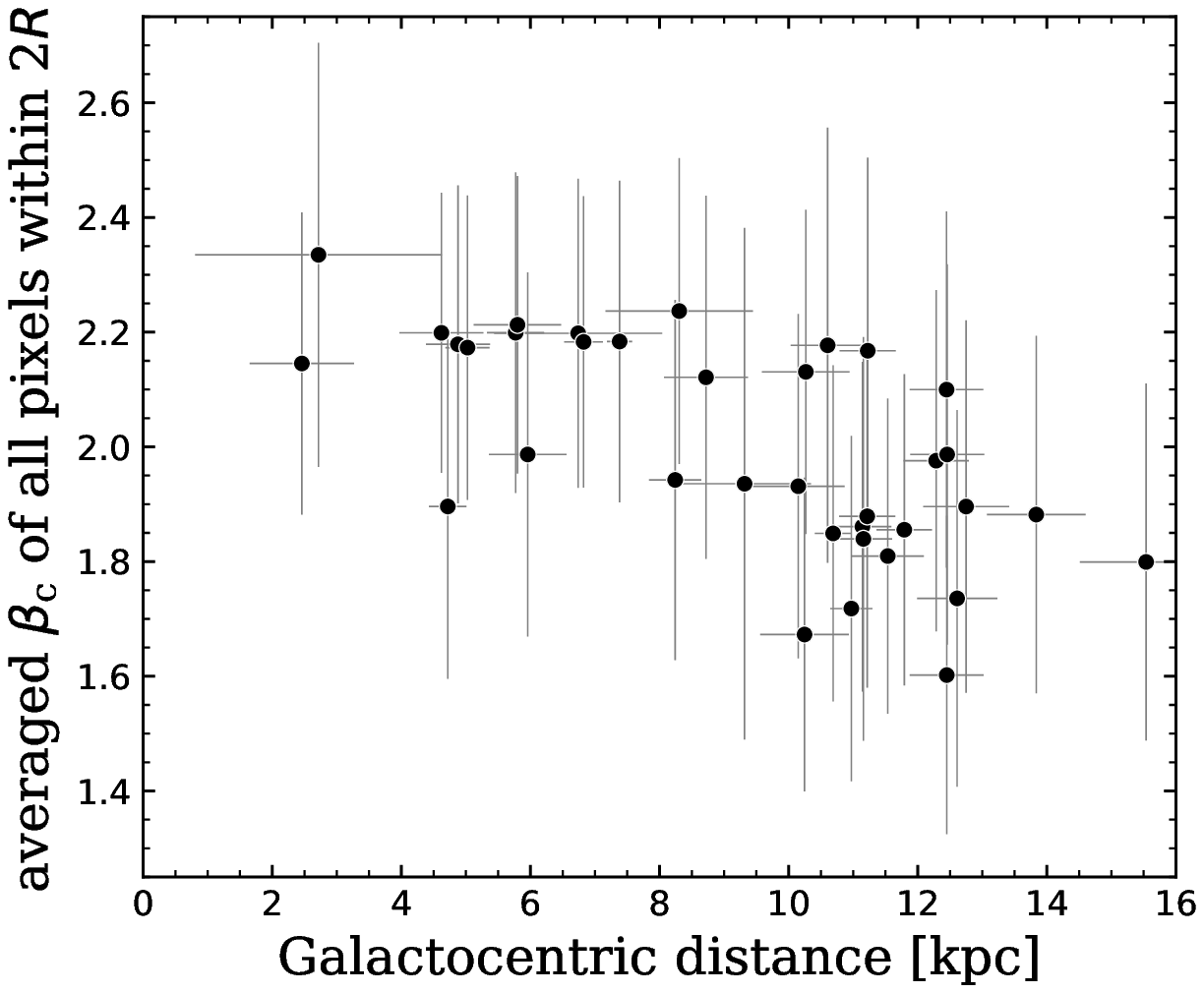}
      \end{center}
  \caption{
    Averaged emissivity power-law index of the cold dust component, $\beta_{\rm{c}}$, within the $2R$ circular region for each of the IR bubbles where 20\% of the total pixels within the $2R$ region are unacceptable for $\beta_{\rm{c}}=2$.
    The error bars correspond to the standard deviations of the $\beta_{\rm{c}}$ values in the $2R$ regions.
  }
  \label{fig:beta_ave}
\end{figure}

In fitting the SEDs, some IR bubbles show the pixels where $\beta_{\rm{c}}$ of 2 is not statistically acceptable.
To reliably identify the pixels having $\beta_{\rm{c}} \neq 2$, we selected the pixels whose $\beta_{\rm{c}}$ values are different from 2 with 5$\sigma$ significance.
For the IR bubble where such pixels occupy 20\% of the total pixels within the 2$R$ circular region, the averaged $\beta_{\rm{c}}$ in the $2R$ region is plotted against the Galactocentric distance with the standard deviation as an error bar in figure~\ref{fig:beta_ave}.
This figure shows that $\beta_{\rm{c}}$ is equal to 2 within the error bars for almost all the IR bubbles, which is consistent with the fitting results of the SEDs created from the photometry fluxes integrated within the $2R$ circular regions in \citet{Hanaoka2019} where $\beta_{\rm{c}}=2$ is acceptable for all the IR bubbles.
The averaged value decreases with the Galactocentric distance; the $\beta_{\rm{c}}$ values tend to be higher than 2 in inner Galactic regions, while they are lower than 2 in outer Galactic regions.
A Kolmogorov-Smirnov (K-S) test shows that the difference is statistically significant (p-value is 0.01), although the uncertainties of $\beta_{\rm{c}}$ are large.
Such a decreasing trend of $\beta_{\rm{c}}$ is observed in nearby galaxies as well, e.g., M31 and M33 (\cite{Smith2012}; \cite{Draine2014}; \cite{Tabatabaei2014}).
These previous studies of the nearby galaxies suggest a change in the dust properties at 3$-$4~kpc from the Galactic center based on the change of the $\beta_{\rm{c}}$ value.
The result of figure~\ref{fig:beta_ave} shows that the transition point of our Galaxy is located around $\sim$8~kpc from the Galactic center.
Also, \citet{Giannetti2017} show that the gas-to-dust ratio in our Galaxy increases with the Galactocentric distance while the metallicity decreases from IR and CO observations.

\section{Discussion}
\subsection{Variations of the IR bubble properties along the whole Galactic plane}

We examine the relation between the positions of the heating sources and the shell morphology.
We show that many IR bubbles in inner Galactic regions have large $R_{\rm{pp}}/R$, which indicates that the heating sources are likely to be closer to the dense shell regions around the IR bubbles.
For example, for the IR bubbles with offsets as large as $R_{\rm{pp}}/R > 0.6$, $28\pm5$ and $3\pm2$ IR bubbles are located at $<8$~kpc and $> 8$~kpc, respectively, with the errors from the Poisson statistics.
In figure~\ref{fig:rpp_GCdistance}, to show the dependence of the offsets on the covering fractions of the PAH shells (CFs) obtained by \citet{Hanaoka2019}, we further color-coded the data points with CF.
The figure suggests that the fraction of closed bubbles may increase at low $R_{\rm{pp}}/R$.
  For example, the IR bubbles with CF$=1.0$ occupy $70\pm 20$\% of the total IR bubbles at $R_{\rm{pp}}/R < 0.2$, while they are $52\pm 7$\% at $R_{\rm{pp}}/R > 0.2$.

\begin{figure}
  \begin{center}
    \subfigure{\includegraphics[width=0.5\linewidth, bb=-10 10 390 370, clip]{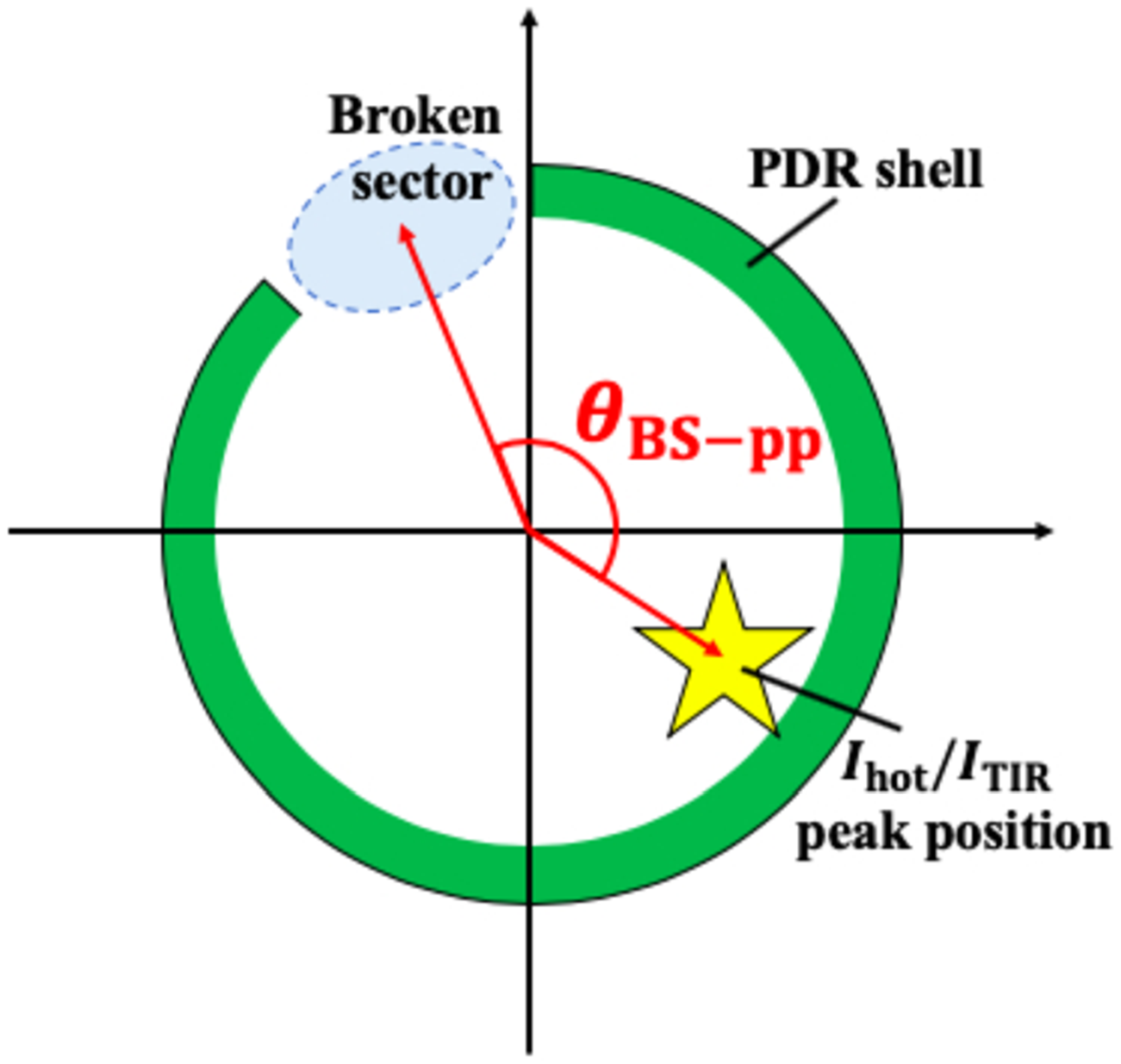}}
    \subfigure{\includegraphics[width=0.9\linewidth, clip]{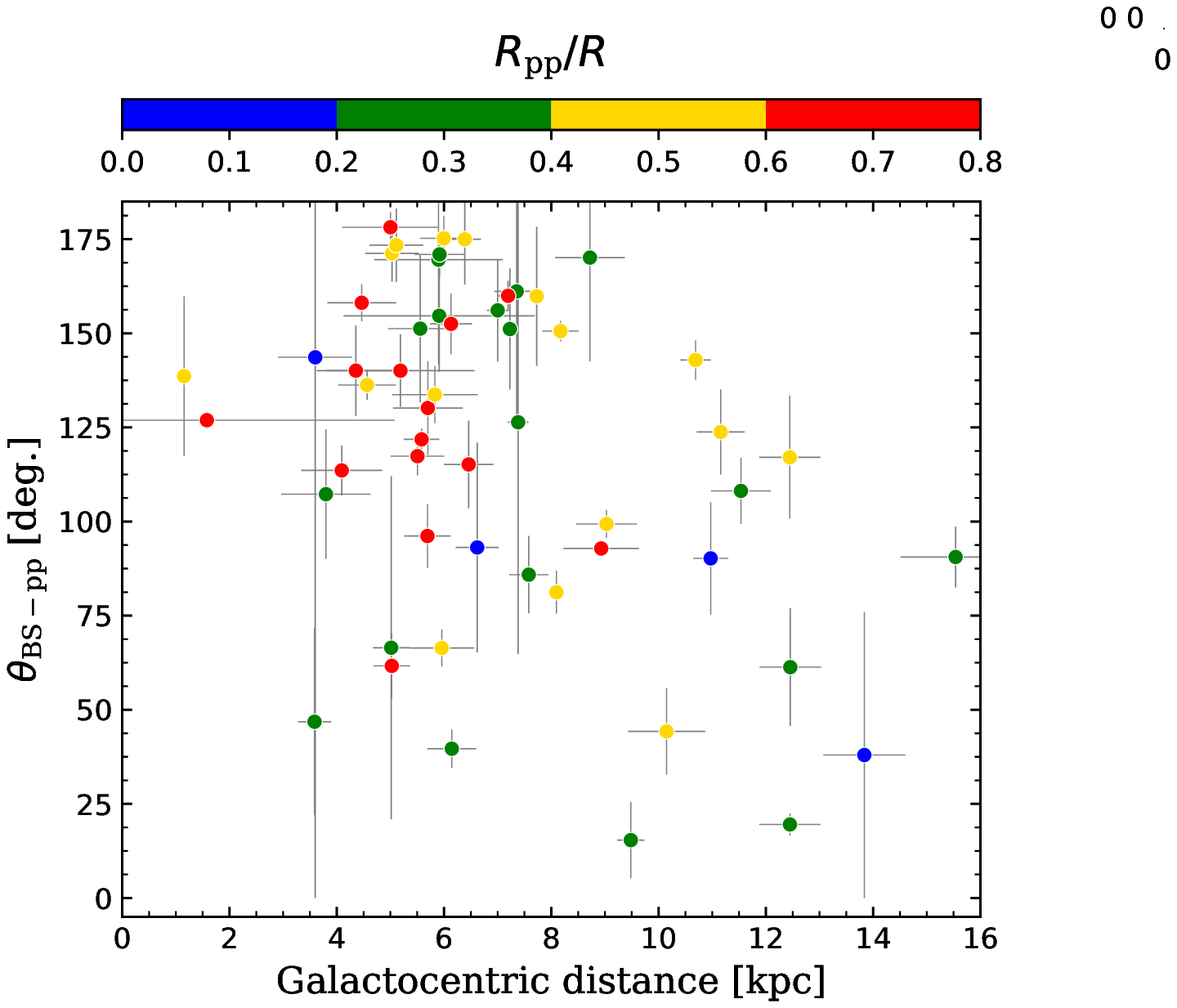}}
  \end{center}
  \caption{
    (Upper) schematic view to explain the definition of $\theta_{\rm{BS-pp}}$.
    (Lower) the angle between the positions of the peak and the broken sector, $\theta_{\rm{BS-pp}}$, plotted as a function of the Galactocentric distance.
    The data points are color-coded according to $R_{\rm{pp}}/R$.
  }
  \label{fig:theta_BS_pp_vs_GCdistance}
\end{figure}

Then, for the broken bubbles, we obtained the direction of the broken sector for each IR bubble based on the CF estimation in \citet{Hanaoka2019}. 
We investigate the difference in the direction angle, $\theta_{\rm{BS-pp}}$, between the heating source position and the broken sector direction as viewed from the shell center (see the upper panel of figure~\ref{fig:theta_BS_pp_vs_GCdistance}).
As errors of $\theta_{\rm{BS-pp}}$, we consider those associated with the positional uncertainties of the heating sources which are evaluated to be one image pixel. Then we calculate the angle which are subtended by the chord of one pixel length at Rpp from the center.
The relation thus obtained is shown in the lower panel of figure~\ref{fig:theta_BS_pp_vs_GCdistance}, from which we find that the heating sources are located nearly on the opposite side of the broken sector directions (i.e., $\theta_{\rm{BS-pp}} \sim 180^{\circ}$) for a significant fraction of the IR bubbles in inner Galactic regions.
In the figure, the color is added to the data points to show $R_{\rm{pp}}/R$, where it is found that the IR bubbles with large $R_{\rm{pp}}/R$ tend to have large $\theta_{\rm{BS-pp}}$.
Hence a significant fraction of the IR bubbles in inner Galactic regions have heating sources largely offset from the center to the shell, where the broken sector directions are related to the positions of the heating sources nearly on their opposite sides.

One possible reason for the large offset of the heating sources and the large $\theta_{\rm{BS-pp}}$ value in inner Galactic regions is local density gradients within each IR bubble, i.e., H\emissiontype{II} regions may be easier to expand toward the lower density direction, and the lower density side of the shell tends to be broken.
On the other hand, \citet{Hattori2016} mentioned that the offset of the heating sources can be explained by formation of massive stars on the boundary of the collided clouds, and in that case, the broken structure can be favorably observed.
If the CCC process is important for the broken bubbles in inner Galactic regions, it is expected that their massive stars are substantially offset to the opposite side of the broken sector direction.
This interpretation is also consistent with the result shown in the lower panel of figure~\ref{fig:theta_BS_pp_vs_GCdistance}, in which the IR bubbles with larger $\theta_{\rm{BS-pp}}$ ($\sim 180^{\circ}$) tend to have larger offsets from the center.

\begin{figure}
  \begin{center}
    \includegraphics[width=0.9\linewidth,clip]{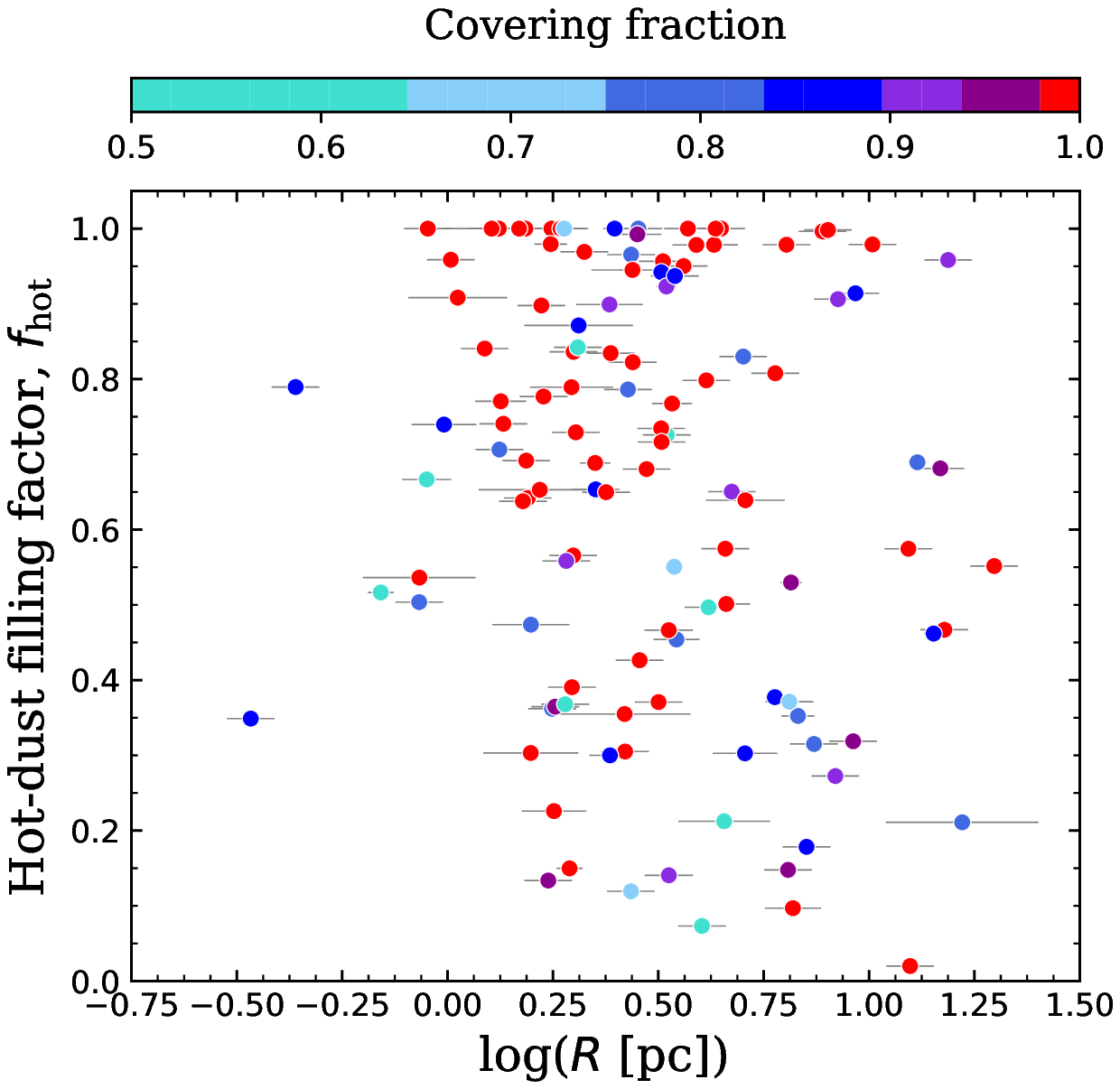}
  \end{center}
  \caption{
    Relation between the hot-dust filling factor in the H\emissiontype{II} regions and the shell radius of each IR bubble, color-coded according to the CF.
    The shell radii and CFs were obtained in \citet{Hanaoka2019}.
  }
  \label{fig:filling_R_CF}
\end{figure}

To further discuss the morphology formation mechanisms of the IR bubbles, we investigate the hot-dust filling factor ($f_{\rm{hot}}$) in the H\emissiontype{II} region of each IR bubble.
We derive ratios of the pixels brighter than 20\% of the $I_{\rm{hot}}$ intensity at the peak position of $I_{\rm{hot}}/I_{\rm{TIR}}$ within the inner edge of the shell (i.e., $<0.8R$).
For the IR bubbles without significant hot dust emission within the shell boundaries, we used the $I_{\rm{warm}}$ intensity maps.
Figure~\ref{fig:filling_R_CF} shows $f_{\rm{hot}}$ plotted against the shell radius, color-coded according to CF.
This figure shows that the large broken bubbles (i.e., log($R$ [pc])$>0.6$ and CF$<0.9$ from \cite{Hanaoka2019}) tend to have lower $f_{\rm{hot}}$.
Indeed, a K-S test shows that the distributions of the large broken bubbles and large closed bubbles (i.e., log($R$ [pc])$>0.6$ and CF$=1.0$ from \cite{Hanaoka2019}) are statistically different from each other (p-value is 0.009).
Moreover, table~\ref{table:large_broken} shows that 5 out of 10 large broken bubbles having $f_{\rm{hot}}$ lower than 0.5 also have $R_{\rm{pp}}/R>0.5$ and $\theta_{\rm{BS-pp}}>90^{\circ}$.
\citet{Hattori2016} suggested that many of the large broken bubbles may have been formed through the CCC process.
In this case, the H\emissiontype{II} gas and the hot dust are expected to be rather localized on the dense shell formed through CCC than fill the inside of the shell.
The above 5 large broken bubbles in the table may be candidate targets which have been formed by CCC.

\begin{table}
  \caption{IR properties of the large broken bubbles having low $f_{\rm{hot}}$.}
  \label{table:large_broken}
  \begin{center}
    \begin{tabular}{c|cccc}
      \hline
      & log($R$ [pc])\footnotemark[$*$] & $f_{\rm{hot}}$ & $R_{\rm{pp}}/R$ & $\theta_{\rm{BS-pp}}$\\\hline
      N2 & $1.22\pm0.18$ & 0.21 & $0.72\pm0.02$ & 127$^{\circ}$.0 \\
      N6 & $0.87\pm0.06$ & 0.32 & $0.45\pm0.02$ & 136$^{\circ}$.0 \\
      N39 & $0.62\pm0.06$ & 0.50 & $0.68\pm0.06$ & 114$^{\circ}$.0 \\
      S1 & $0.66\pm0.11$ & 0.21 & $0.62\pm0.03$ & 178$^{\circ}$.0 \\
      S27 & $0.78\pm0.01$ & 0.38 & $0.76\pm0.09$ & 62$^{\circ}$.0 \\
      S66 & $0.83\pm0.04$ & 0.35 & $0.61\pm0.02$ & 122$^{\circ}$.0 \\
      S76 & $0.71\pm0.08$ & 0.30 & $0.44\pm0.03$ & 66$^{\circ}$.0 \\
      S92 & $1.15\pm0.02$ & 0.46 & $0.35\pm0.03$ & 171$^{\circ}$.0 \\
      E21 & $0.81\pm0.06$ & 0.37 & $0.58\pm0.03$ & 99$^{\circ}$.0 \\
      E95 & $0.85\pm0.06$ & 0.18 & $0.22\pm0.03$ & 108$^{\circ}$.0 \\\hline
    \end{tabular}
  \end{center}
  \begin{tabnote}
    \footnotemark[$*$] \cite{Hanaoka2019}
  \end{tabnote}
\end{table}


In figure~\ref{fig:CVs_vs_GCdistance}, to show the dependence of the CVs of $I_{\rm{PAH}}/I_{\rm{TIR}}$ on CFs, we color-coded the data points with CF.
The figure shows that, for the IR bubbles in inner Galactic regions ($<8$~kpc), 17 out of 25 bubbles which have the $I_{\rm{PAH}}/I_{\rm{TIR}}$ CVs larger than 0.4 are observed as broken bubbles, while 38 out of 102 bubbles with the $I_{\rm{PAH}}/I_{\rm{TIR}}$ CVs smaller than 0.4 are broken bubbles.
The former ratio ($70\pm 20$\%) is significantly higher than the latter ratio ($37\pm 6$\%), based on the Poisson statistics.
Hence, the broken morphology may be the key to understanding the large spatial variations of the fractional PAH intensities.
We discuss possible processes to produce the observed large $I_{\rm{PAH}}/I_{\rm{TIR}}$ CVs in inner Galactic regions as seen in figure~\ref{fig:CVs_vs_GCdistance}. 
Considering the possibility of PAH destruction around the IR bubbles, PAHs can be destroyed by UV radiation, hot gas and interstellar shocks (e.g., \cite{Tielens2005}; \cite{Rapacioli2005}; Micelotta et al. 2010a, 2010b).
First, we only discuss the effects of interstellar shocks, because the effects of the PAH destruction by the UV radiation and hot gas in H\emissiontype{II} regions are likely to be reflected by the $I_{\rm{PAH}}/I_{\rm{TIR}}$ radial profiles (figure~\ref{fig:radial_profile_GCdistance}), which are already removed from the $I_{\rm{PAH}}/I_{\rm{TIR}}$ distribution maps in deriving the CVs (figure~\ref{fig:CVs_vs_GCdistance}).
The PAH destruction in the interstellar shocks needs a shock velocity above 100~km~s$^{-1}$ (Micelotta et al. 2010a).
The observed collisional velocities of CCC are as slow as 10$-$30~km~s$^{-1}$ (e.g., Inoue \& Fukui 2013; \cite{Fukui2016}), which are too slow to destroy the PAHs.

\begin{figure}
  \begin{center}
    \includegraphics[width=0.9\linewidth,clip]{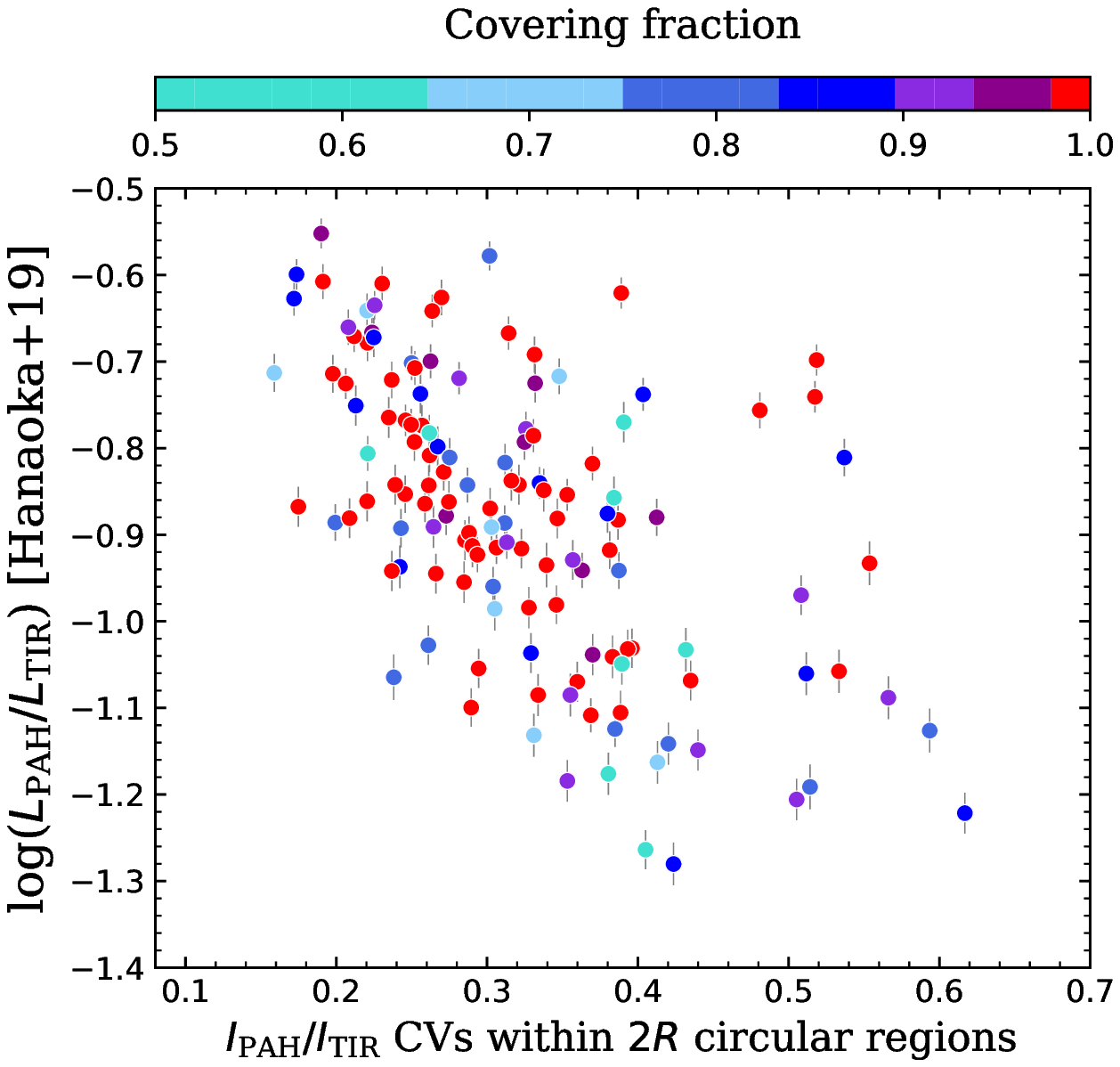}
  \end{center}
  \caption{
    Relation between $L_{\rm{PAH}}/L_{\rm{TIR}}$ obtained by \citet{Hanaoka2019} and the $I_{\rm{PAH}}/I_{\rm{TIR}}$ CVs, color-coded with the CF.
  }
  \label{fig:PAHratio_vs_PAHCV}
\end{figure}

Then, we discuss the possibility of PAH production around the IR bubbles.
PAHs can be produced by shattering carbonaceous dust grains with the collisional velocity of about 1$-$100~km~s$^{-1}$ (Jones et al. 1996); the collisional velocities of CCC are in this velocity range.
However, shortly after the production, PAHs might be destroyed by intense UV radiation from massive stars.
We calculate the PAH destruction timescale in H\emissiontype{II} regions using a C-H band photo-dissociation time scale in the diffuse ISM, $\tau_{\rm{UV,pd}}$, as described by the following equation (\cite{Jones2014}; Jones \& Habart 2015):
\begin{equation}
  \tau_{\rm{UV,pd}} = \frac{10^4}{G_0}\bigl\{ 2.7+\frac{6.5}{(a \rm{[nm]})^{1.4}} + 0.04(a \rm{[nm]})^{1.3}\bigr\} \ \ \ \rm{[yr]},
\end{equation}
where $G_0$ and $a$ are the radiation field intensity integrated for far-UV wavelengths relative to the solar neighborhood and grain radius, respectively.
We adopted the typical $G_0$ of 10$^2-$10$^4$ in H\emissiontype{II} regions whose bolometric luminosities of sources are in a range of 10$^3-$10$^5$~L$_{\rm{\odot}}$ (\cite{Decataldo2017}) and the typical PAH size of $6$~\AA~(\cite{Tielens2005}) to find that $\tau_{\rm{UV,pd}}\sim$10$-$10$^3$~yr.
This calculation implies that PAHs produced through shattering of carbonaceous grains will be destroyed in the H\emissiontype{II} region within timescales much shorter than the lifetime of IR bubbles ($\sim$10$^6$~yr).
Therefore, we can expect that CCC could decrease fractional PAH luminosities, $L_{\rm{PAH}}/L_{\rm{TIR}}$, as a whole with a large $I_{\rm{PAH}}/I_{\rm{TIR}}$ CV.
Figure~\ref{fig:PAHratio_vs_PAHCV} indeed shows that the IR bubbles with low $L_{\rm{PAH}}/L_{\rm{TIR}}$ tend to have large $I_{\rm{PAH}}/I_{\rm{TIR}}$ CVs.
Moreover, such IR bubbles tend to have lower CFs, which is in line with the CCC trend pointed out by \citet{Hattori2016}.
Hence, the large spatial variations of $I_{\rm{PAH}}/I_{\rm{TIR}}$ may also be caused by the CCC process.

We discuss the $\beta_{\rm{c}}$ values in the context of the properties of the interstellar dust.
According to the standard dust model defined in \citet{Desert1990}, $\beta_{\rm{c}} \simeq 2$ is expected for silicate grains, while $\beta_{\rm{c}} \simeq 1$ for carbonaceous grains.
Indeed, \citet{Mennella1998} obtained that amorphous carbon grains have a $\beta_{\rm{c}}$ value of $\sim$1 which is lower value than that of amorphous silicate grains in their laboratory measurement.
Among them, carbon-rich (C-rich) asymptotic giant branch (AGB) stars are considered as suppliers of carbonaceous dust including PAHs, while oxygen-rich (O-rich) AGB stars as suppliers of silicate dust (\cite{Latter1991}; \cite{Tielens2008}).
\citet{Ishihara2011} investigated the distributions of C-rich and O-rich AGB stars in our Galaxy and obtained that the C-rich AGB stars are uniformly distributed within the Galactic disk, while the O-rich AGB stars are concentrated toward the Galactic center.
Moreover, the result of $L_{\rm{PAH}}/L_{\rm{TIR}}$ of \citet{Hanaoka2019} suggests that the environments in outer Galactic regions belong to those relatively rich in PAHs.
The result shows the tendency that $\beta_{\rm{c}} \gtrsim 2$ for inner Galactic regions while $\beta_{\rm{c}} \lesssim 2$ for outer Galactic regions.  
Thus, the trend of the $\beta_{\rm{c}}$ value also supports that the IR bubbles are relatively rich in carbonaceous dust in outer Galactic regions.

\subsection{Properties of the IR bubbles in the cloud collision environments}
\begin{figure*}
  \begin{center}
    \subfigure{
      \mbox{\raisebox{5mm}{\rotatebox{90}{\small{Galactic Latitude}}}}
      \mbox{\raisebox{0mm}{\includegraphics[width=0.75\linewidth, bb=30 40 800 255, clip]{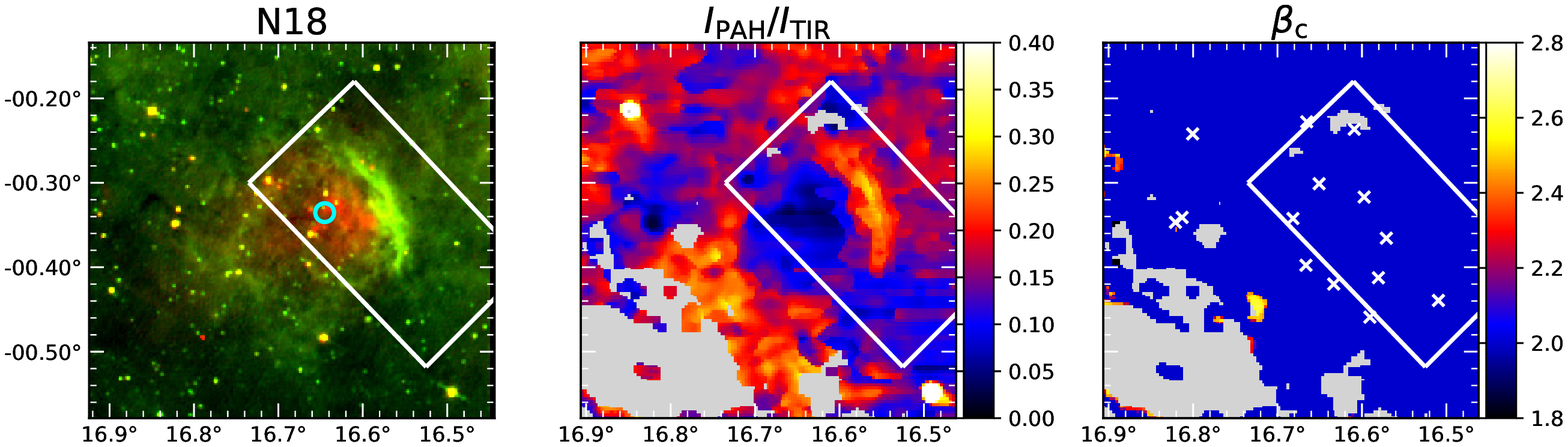}}}
    }
    \subfigure{
      \mbox{\raisebox{5mm}{\rotatebox{90}{\small{Galactic Latitude}}}}
      \mbox{\raisebox{0mm}{\includegraphics[width=0.75\linewidth, bb=30 40 800 255, clip]{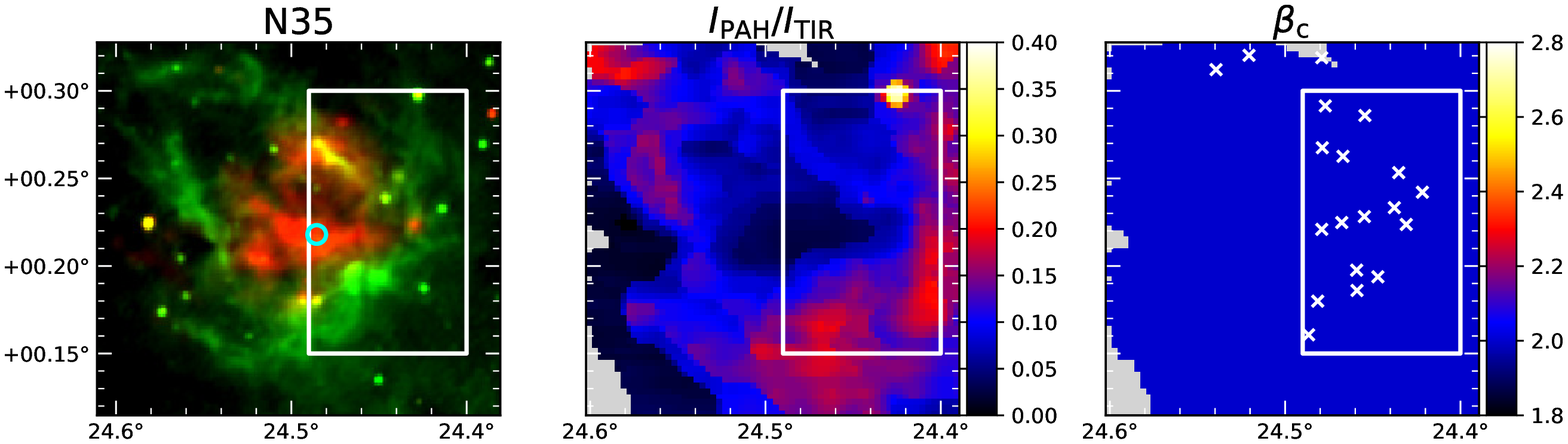}}}
    }
    \subfigure{
      \mbox{\raisebox{5mm}{\rotatebox{90}{\small{Galactic Latitude}}}}
      \mbox{\raisebox{0mm}{\includegraphics[width=0.75\linewidth, bb=30 40 800 255, clip]{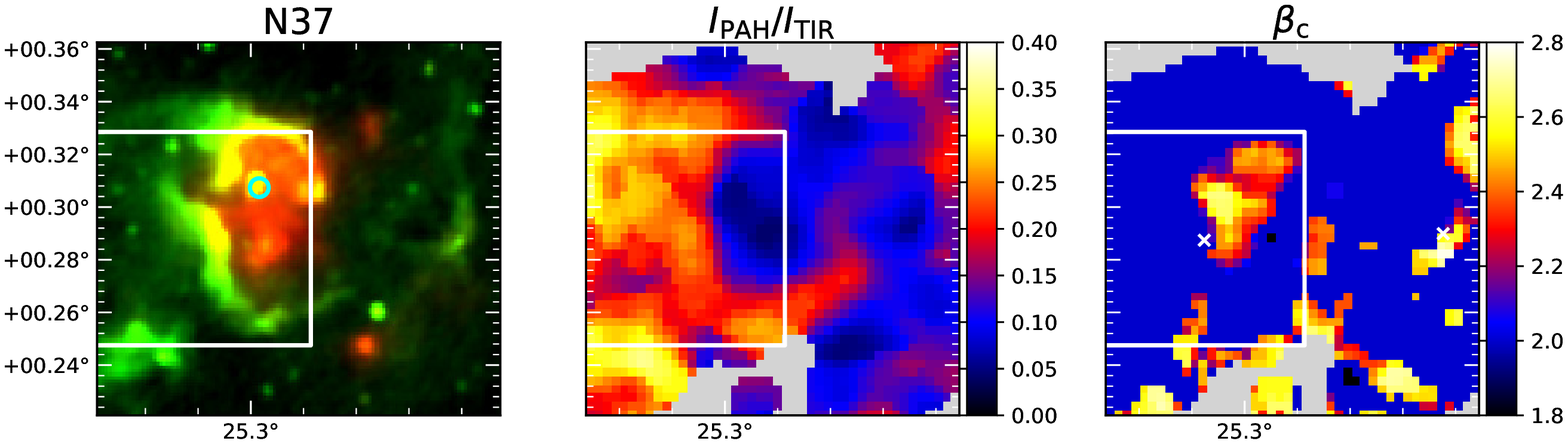}}}
    }
    \subfigure{
      \mbox{\raisebox{5mm}{\rotatebox{90}{\small{Galactic Latitude}}}}
      \mbox{\raisebox{0mm}{\includegraphics[width=0.75\linewidth, bb=30 40 800 255, clip]{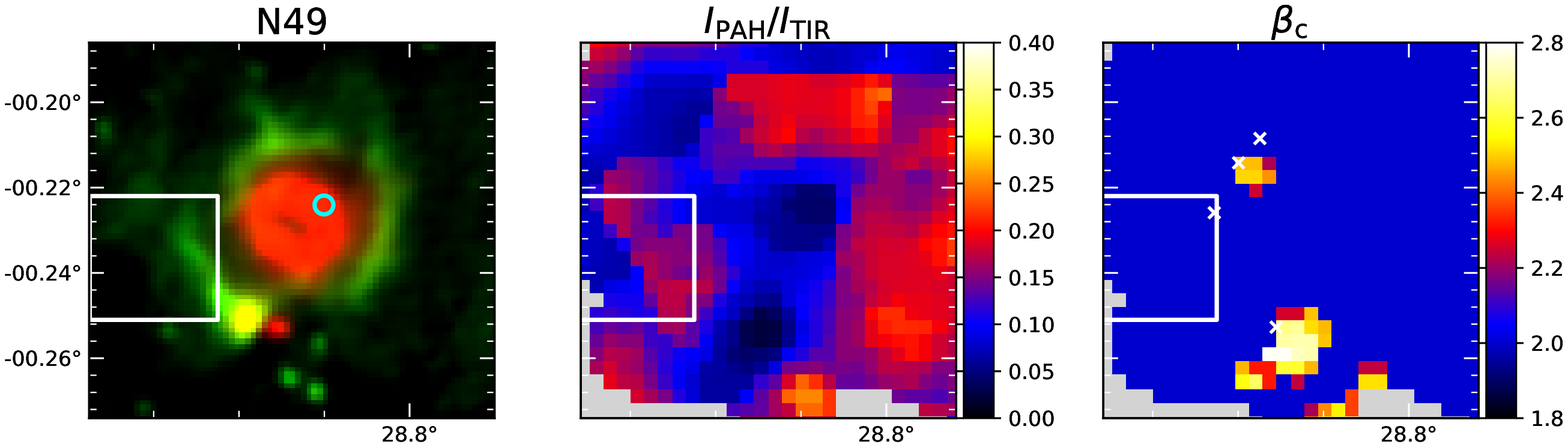}}}
    }
    \subfigure{
      \mbox{\raisebox{5mm}{\rotatebox{90}{\small{Galactic Latitude}}}}
      \mbox{\raisebox{0mm}{\includegraphics[width=0.75\linewidth, bb=30 40 800 255, clip]{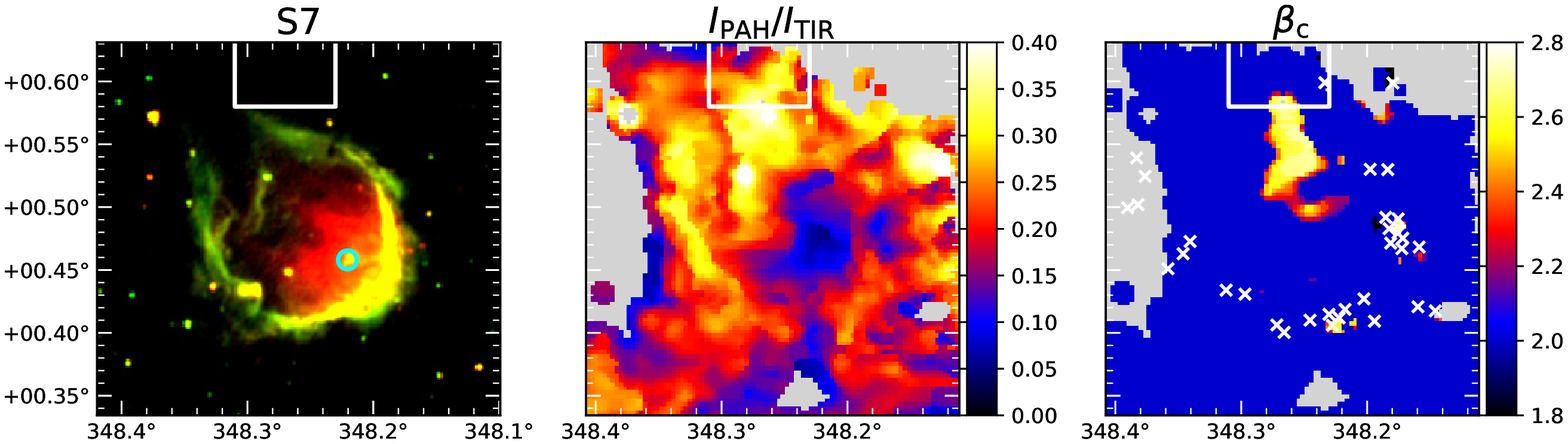}}}
    }
    \subfigure{
      \mbox{\raisebox{5mm}{\rotatebox{90}{\small{Galactic Latitude}}}}
      \mbox{\raisebox{0mm}{\includegraphics[width=0.75\linewidth, bb=30 40 800 255, clip]{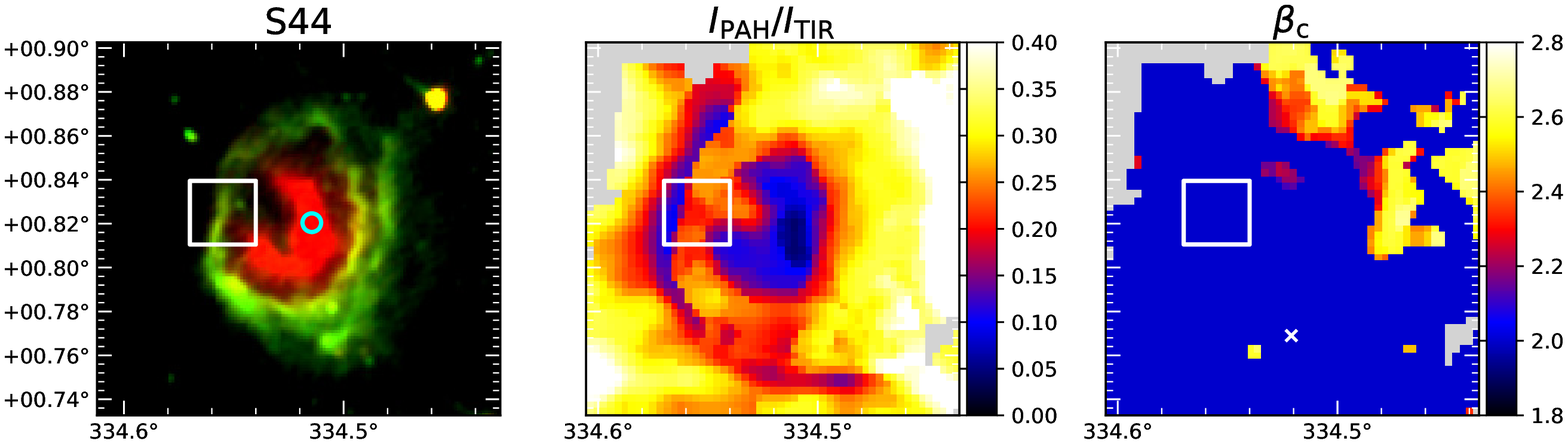}}}
    }
    \subfigure{\mbox{\raisebox{0mm}{\hspace{16mm}\small{Galactic Longitude}\hspace{12mm}\small{Galactic Longitude}\hspace{15mm}\small{Galactic Longitude}\hspace{8mm}}}}
  \end{center}
  \caption{
    PAH intensity ratio, $I_{\rm{PAH}}/I_{\rm{TIR}}$, and the emissivity power-law index of the cold dust component, $\beta_{\rm{c}}$, maps of the IR bubbles having bridge features in the CO position-velocity maps.
    The left panels show the AKARI 9~$\mu$m and 18~$\mu$m band images in green and red, respectively.
    The cyan circles in the left panels correspond to the peak positions of the hot dust emissions.
    The white squares correspond to the regions where the bridge features had been observed (N18; \cite{Torii2018N18}, N35; \cite{Torii2018N35}, N37; \cite{Baug2016}, N49; \cite{Dewangan2017}, S7; \cite{Torii2015}, S44; \cite{Kohno2018}).
    The white crosses in the right panels correspond to the positions of massive clumps identified by \citet{Csengeri2014}.
  }
  \label{fig:CCC_bubble_map}
\end{figure*}

To verify the IR properties in the cloud collision environments, we examine the spatial distributions of the dust components for the IR bubbles which are studied well with CO and most probably associated with CCC.
Among our sample IR bubbles, there are 6 IR bubbles having bridge features in the CO position-velocity maps within the $2R$ regions, which suggest that the cloud collisions are still on-going around the IR bubbles (N18; \cite{Torii2018N18}, N35; \cite{Torii2018N35}, N37; \cite{Baug2016}, N49; \cite{Dewangan2017}, S7; \cite{Torii2015}, S44; \cite{Kohno2018}).
In table~\ref{table:CCC_total_local_SED_property}, we summarize the properties of those IR bubbles obtained by the local SED fittings.
Two of them (N37 and S7) have the CCC trends in the IR properties, i.e., $R_{\rm{pp}}/R>0.5$, $\theta_{\rm{BS-pp}}>90^{\circ}$ and $I_{\rm{PAH}}/I_{\rm{TIR}}$ CV$>0.3$.
However, the other IR bubbles have small $R_{\rm{pp}}/R$ and closed shells (see table~\ref{table:CCC_total_local_SED_property} and left panels in figure~\ref{fig:CCC_bubble_map}).
We interpret that the latter are formed by nearly head-on collisions between two clouds and viewed face-on from us.
In this case, the massive star is presumably located near the shell center and the shell morphology is likely to appear ``closed''.

The center panels in figure~\ref{fig:CCC_bubble_map} show the $I_{\rm{PAH}}/I_{\rm{TIR}}$ distribution maps.
Indeed, N37 and S7, which have the CCC trends in the IR properties, have clearly high $I_{\rm{PAH}}/I_{\rm{TIR}}$ values in the white square regions, where the cloud collisions are likely to be underway, judging from the positional coincidence with the bridge features in the CO position-velocity maps.
Those trends are consistent with our hypothesis, i.e., the CCC process can produce PAHs through shattering of carbonaceous dust as discussed above.
On the other hand, the maps of N18, N35, N49 and S44 show that the $I_{\rm{PAH}}/I_{\rm{TIR}}$ values are not systematically higher in the square regions.
The effects of the PAH production process on the $I_{\rm{PAH}}/I_{\rm{TIR}}$ value might be difficult to be observed due to the overlap of multiple components along the line of sight, when CCC is viewed face-on from us.

\begin{table}
  \caption{IR properties of the IR bubbles having bridge features in the CO position-velocity maps.}
  \label{table:CCC_total_local_SED_property}
  \begin{center}
    \begin{tabular}{c|cccc}
      \hline
       & CF\footnotemark[$*$] & $R_{\rm{pp}}/R$ & $\theta_{\rm{BS-pp}}$ & $I_{\rm{PAH}}/I_{\rm{TIR}}$ CV\\\hline
      N18 & 1.00 & $0.40 \pm 0.02$ & $-$ & $0.31$ \\
      N35 & 1.00 & $0.20 \pm 0.05$ & $-$ & $0.43$ \\
      N37 & 0.67 & $0.61 \pm 0.07$ & 96$^{\circ}$.0 & $0.39$ \\
      N49 & 1.00 & $0.44 \pm 0.11$ & $-$ & $0.35$ \\
      S7 & 0.96 & $0.62 \pm 0.03$ & 160$^{\circ}$.0 & $0.33$ \\
      S44 & 1.00 & $0.16 \pm 0.06$ & $-$ & $0.26$ \\\hline
    \end{tabular}
  \end{center}
  \begin{tabnote}
    \footnotemark[$*$] \cite{Hanaoka2019}
  \end{tabnote}
\end{table}

The right panels in figure~\ref{fig:CCC_bubble_map} show that there are some clumps having $\beta_{\rm{c}} > 2.0$ within the $2R$ regions.
N37 and S7 have large $\beta_{\rm{c}}$ in the white square regions where the bridge features have been observed.
Such high $\beta_{\rm{c}}$ values have been observed in dense cold molecular clouds (e.g., Kuan et al. 1996; Lis \& Menten 1998; \cite{Planck2011}), and dust grains coated with thick icy mantles can have $\beta_{\rm{c}}$ significantly higher than 2 (e.g., \cite{Aannestad1975}; \cite{Guttler1952}; \cite{Wickramasinghe1967}).
Therefore, this result suggests that CCC may increase $\beta_{\rm{c}}$ due to the presence of cold dense clouds which are likely to be formed on the collision surface.
The coagulation on dust grains requires gas densities of $n_{\rm{H}} > 10^4$~cm$^{-3}$ (Chokshi et al. 1993), thus the high $\beta_{\rm{c}}$ regions are likely candidates for massive star formation.
To verify the hypothesis, we plot massive clumps identified by \citet{Csengeri2014} with ATLASGAL surveys in the right panels of figure~\ref{fig:CCC_bubble_map}.
  The massive clumps are located near some of the high $\beta_{\rm{c}}$ regions (e.g., those in N37, N49 and S7), but not for the other high $\beta_{\rm{c}}$ regions, which do not provide us with a very convincing result.
Hence, high $R_{\rm{pp}}/R$ and $\theta_{\rm{BS-pp}}$ as well as the local enhancements of the PAH fractional intensity ($I_{\rm{PAH}}/I_{\rm{TIR}} > 0.3$) and the cold dust emissivity power-law index ($\beta_{\rm{c}} > 2$) may be good probes to search for nearly edge-on CCC candidates, while they may not be valid for face-on CCC.
Whichever the case is, we need more statistics to verify the relationships between IR and radio CO properties for the IR bubbles indicative of CCC.

\section{Summary}
Using AKARI, Herschel and WISE data, we have studied the spatial distributions of dust components around 165 IR bubbles along the whole Galactic plane ($0^{\circ}\leq l < 360^{\circ}$, $|b|\leq 5^{\circ}$), which are decomposed into PAH, hot dust ($123$~K), warm dust ($\sim60$~K) and cold dust ($\sim20$~K).
As a result, we find that the offsets of dust heating sources from the shell centers, which are estimated with the hot or warm dust emission, in inner Galactic regions are larger than those in outer Galactic regions.
Moreover, the spatial variations of the $I_{\rm{PAH}}/I_{\rm{TIR}}$ and the $\beta_{\rm{c}}$ values around the IR bubbles in inner Galactic regions are larger than those in outer Galactic regions.
The offsets of dust heating sources suggest that a local density gradient within each IR bubble or the CCC process may make larger contribution to forming the shell morphology in inner Galactic regions.
Considering possible production and destruction processes of PAHs around the IR bubbles, we suggest that the large spatial variations of the PAH intensities in inner Galactic regions may be caused by CCC.
The $\beta_{\rm{c}}$ trend suggests that the IR bubbles may be relatively rich in carbonaceous dust in outer Galactic regions, which is aligned with the result of the higher ratios of PAH to dust in \citet{Hanaoka2019}.
Finally, we discuss the IR properties of the cloud collision environments around the IR bubbles which are studied with the previous CO observations, and our results suggest the possibility that the dust properties are influenced by the cloud collision.

\begin{ack}
This research is based on observations with AKARI, a JAXA project with the participation of ESA.
Herschel is an ESA space observatory with science instruments provided by European-led Principal Investigator consortia and with important participation from NASA.
This publication makes use of data products from the Wide-field Infrared Survey Explorer, which is a joint project of the University of California, Los Angeles, and the Jet Propulsion Laboratory/California Institute of Technology, funded by the National Aeronautics and Space Administration.
We thank all the members of the AKARI, Herschel, and WISE projects, particularly the all-sky survey, Hi-GAL, and AllWISE data reduction teams.
This research was supported by JSPS KAKENHI Grant Number 18H01252.
\end{ack}

\appendix
\section*{Summary of the IR properties derived in this study.}
\begin{longtable}{p{0.7cm} c @{\hspace{0.2cm}} c @{\hspace{0.4cm}} c @{\hspace{0.3cm}} c @{\hspace{0.4cm}} c @{\hspace{0.2cm}} c @{\hspace{0.2cm}} c @{\hspace{0.2cm}}}
  \caption{Summary of the local SED properties.}
  \label{table:local_SED_property}
  \hline
  Name & $l_{\rm{pp}}$ [$^{\circ}$] & $b_{\rm{pp}}$ [$^{\circ}$] & $R_{\rm{pp}}/R$ & $\theta_{\rm{BS-pp}}$ [$^{\circ}$] & $f_{\rm{hot}}$ & $I_{\rm{PAH}}/I_{\rm{TIR}}$ CV & $\beta_{\rm{c}}$\\
  \endfirsthead
  \hline
  Name & $l_{\rm{pp}}$ [$^{\circ}$] & $b_{\rm{pp}}$ [$^{\circ}$] & $R_{\rm{pp}}/R$ & $\theta_{\rm{BS-pp}}$ [$^{\circ}$] & $f_{\rm{hot}}$ & $I_{\rm{PAH}}/I_{\rm{TIR}}$ CV & $\beta_{\rm{c}}$\\
  \hline
  \endhead
  \hline
  \endfoot
  \hline
  \endlastfoot
  \hline
N2	&	$10.67$	&	$-0.47$	&	$0.72\pm$0.02	&	$127.0\pm2.0$	&	$0.21$	&	$0.39$	&	$-$	\\
N4	&	$11.90$	&	$0.74$	&	$0.34\pm$0.07	&	$-$	&	$0.57$	&	$0.20$	&	$-$	\\
N6	&	$12.48$	&	$-0.61$	&	$0.45\pm$0.02	&	$136.0\pm4.0$	&	$0.32$	&	$0.25$	&	$-$	\\
N10	&	$13.18$	&	$0.04$	&	$0.39\pm$0.10	&	$-$	&	$0.84$	&	$0.40$	&	$-$	\\
N11	&	$13.23$	&	$0.08$	&	$0.32\pm$0.13	&	$-$	&	$0.84$	&	$0.26$	&	$2.18\pm$0.28	\\
N12	&	$13.77$	&	$-0.03$	&	$0.62\pm$0.03	&	$-$	&	$0.98$	&	$0.48$	&	$-$	\\
N14	&	$-$	&	$-$	&	$-$	&	$-$	&	$-$	&	$0.17$	&	$-$	\\
N15	&	$-$	&	$-$	&	$-$	&	$-$	&	$-$	&	$0.42$	&	$-$	\\
N16	&	$15.00$	&	$0.05$	&	$0.32\pm$0.06	&	$155.0\pm15.0$	&	$0.91$	&	$0.34$	&	$-$	\\
N18	&	$16.65$	&	$-0.34$	&	$0.40\pm$0.02	&	$-$	&	$1.00$	&	$0.31$	&	$-$	\\
N29	&	$23.06$	&	$0.55$	&	$0.29\pm$0.05	&	$-$	&	$1.00$	&	$0.22$	&	$-$	\\
N30	&	$23.12$	&	$0.55$	&	$0.75\pm$0.13	&	$-$	&	$0.54$	&	$0.32$	&	$-$	\\
N34	&	$-$	&	$-$	&	$-$	&	$-$	&	$-$	&	$0.39$	&	$-$	\\
N35	&	$24.49$	&	$0.22$	&	$0.20\pm$0.05	&	$-$	&	$1.00$	&	$0.43$	&	$-$	\\
N36	&	$24.84$	&	$0.09$	&	$0.26\pm$0.06	&	$107.0\pm17.0$	&	$0.65$	&	$0.44$	&	$-$	\\
N37	&	$25.30$	&	$0.31$	&	$0.61\pm$0.07	&	$96.0\pm8.0$	&	$0.84$	&	$0.39$	&	$-$	\\
N39	&	$25.36$	&	$-0.18$	&	$0.68\pm$0.06	&	$114.0\pm7.0$	&	$0.50$	&	$0.43$	&	$-$	\\
N40	&	$25.38$	&	$-0.36$	&	$0.81\pm$0.13	&	$-$	&	$1.00$	&	$0.36$	&	$-$	\\
N44	&	$-$	&	$-$	&	$-$	&	$-$	&	$-$	&	$0.22$	&	$2.20\pm$0.24	\\
N45	&	$26.98$	&	$-0.07$	&	$0.75\pm$0.09	&	$-$	&	$0.57$	&	$0.27$	&	$-$	\\
N46	&	$27.32$	&	$-0.12$	&	$0.67\pm$0.11	&	$140.0\pm12.0$	&	$0.65$	&	$0.33$	&	$-$	\\
N47	&	$28.01$	&	$-0.02$	&	$0.73\pm$0.05	&	$-$	&	$0.81$	&	$0.37$	&	$-$	\\
N49	&	$28.82$	&	$-0.22$	&	$0.44\pm$0.11	&	$-$	&	$0.97$	&	$0.35$	&	$-$	\\
N50	&	$29.01$	&	$0.08$	&	$0.70\pm$0.09	&	$140.0\pm10.0$	&	$0.83$	&	$0.39$	&	$-$	\\
N52	&	$-$	&	$-$	&	$-$	&	$-$	&	$-$	&	$0.42$	&	$-$	\\
N54	&	$31.14$	&	$0.29$	&	$0.67\pm$0.07	&	$153.0\pm8.0$	&	$0.36$	&	$0.31$	&	$-$	\\
N56	&	$32.59$	&	$0.00$	&	$0.47\pm$0.12	&	$-$	&	$0.96$	&	$0.52$	&	$-$	\\
N59	&	$33.09$	&	$-0.12$	&	$0.40\pm$0.02	&	$-$	&	$0.57$	&	$0.25$	&	$-$	\\
N61	&	$34.17$	&	$0.17$	&	$0.55\pm$0.04	&	$175.0\pm6.0$	&	$0.73$	&	$0.41$	&	$-$	\\
N62	&	$34.33$	&	$0.20$	&	$0.64\pm$0.10	&	$-$	&	$0.78$	&	$0.50$	&	$-$	\\
N64	&	$34.76$	&	$-0.67$	&	$0.04\pm$0.03	&	$-$	&	$0.98$	&	$0.35$	&	$-$	\\
N65	&	$35.01$	&	$0.33$	&	$0.26\pm$0.07	&	$-$	&	$0.73$	&	$0.26$	&	$-$	\\
N68	&	$35.66$	&	$-0.04$	&	$0.24\pm$0.03	&	$-$	&	$0.47$	&	$0.29$	&	$-$	\\
N71	&	$38.31$	&	$-0.03$	&	$0.26\pm$0.02	&	$40.0\pm5.0$	&	$0.27$	&	$0.31$	&	$-$	\\
N72	&	$38.35$	&	$-0.13$	&	$0.22\pm$0.15	&	$126.0\pm62.0$	&	$0.79$	&	$0.24$	&	$2.18\pm$0.28	\\
N73	&	$38.74$	&	$-0.14$	&	$0.39\pm$0.14	&	$170.0\pm28.0$	&	$1.00$	&	$0.20$	&	$-$	\\
N74	&	$38.91$	&	$-0.44$	&	$0.10\pm$0.11	&	$-$	&	$1.00$	&	$0.38$	&	$-$	\\
N77	&	$40.43$	&	$-0.04$	&	$0.72\pm$0.12	&	$130.0\pm12.0$	&	$0.36$	&	$0.22$	&	$-$	\\
N79	&	$41.52$	&	$0.04$	&	$0.64\pm$0.11	&	$-$	&	$1.00$	&	$0.28$	&	$-$	\\
N80	&	$-$	&	$-$	&	$-$	&	$-$	&	$-$	&	$0.36$	&	$-$	\\
N81	&	$42.06$	&	$-0.59$	&	$0.71\pm$0.02	&	$-$	&	$0.55$	&	$0.21$	&	$-$	\\
N82	&	$42.11$	&	$-0.63$	&	$0.40\pm$0.09	&	$-$	&	$0.83$	&	$0.26$	&	$2.21\pm$0.26	\\
N84	&	$42.83$	&	$-0.15$	&	$0.57\pm$0.14	&	$160.0\pm18.0$	&	$0.35$	&	$0.23$	&	$-$	\\
N90	&	$43.77$	&	$0.07$	&	$0.41\pm$0.09	&	$-$	&	$0.69$	&	$0.30$	&	$-$	\\
N91	&	$44.26$	&	$0.09$	&	$0.78\pm$0.03	&	$-$	&	$0.02$	&	$0.31$	&	$-$	\\
N92	&	$44.34$	&	$-0.82$	&	$0.24\pm$0.09	&	$-$	&	$0.90$	&	$0.19$	&	$-$	\\
N95	&	$45.39$	&	$-0.72$	&	$0.56\pm$0.12	&	$-$	&	$0.64$	&	$0.27$	&	$-$	\\
N98	&	$47.03$	&	$0.23$	&	$0.14\pm$0.08	&	$-$	&	$0.65$	&	$0.26$	&	$-$	\\
N101	&	$-$	&	$-$	&	$-$	&	$-$	&	$-$	&	$0.33$	&	$-$	\\
N114	&	$52.25$	&	$0.71$	&	$0.26\pm$0.09	&	$-$	&	$1.00$	&	$0.25$	&	$-$	\\
N115	&	$53.55$	&	$-0.00$	&	$0.16\pm$0.04	&	$-$	&	$0.31$	&	$0.27$	&	$-$	\\
N117	&	$54.10$	&	$-0.06$	&	$0.44\pm$0.08	&	$-$	&	$0.43$	&	$0.25$	&	$-$	\\
N120	&	$-$	&	$-$	&	$-$	&	$-$	&	$-$	&	$0.52$	&	$-$	\\
N123	&	$57.54$	&	$-0.28$	&	$0.58\pm$0.12	&	$-$	&	$0.37$	&	$0.26$	&	$2.24\pm$0.27	\\
N124	&	$58.61$	&	$0.63$	&	$0.24\pm$0.10	&	$161.0\pm34.0$	&	$0.71$	&	$0.20$	&	$-$	\\
N126	&	$59.61$	&	$0.31$	&	$0.42\pm$0.07	&	$-$	&	$0.95$	&	$0.39$	&	$-$	\\
N127	&	$60.66$	&	$-0.03$	&	$0.60\pm$0.04	&	$81.0\pm6.0$	&	$0.50$	&	$0.30$	&	$-$	\\
N130	&	$-$	&	$-$	&	$-$	&	$-$	&	$-$	&	$0.32$	&	$-$	\\
N131	&	$-$	&	$-$	&	$-$	&	$-$	&	$-$	&	$0.40$	&	$-$	\\
N133	&	$63.17$	&	$0.45$	&	$0.40\pm$0.09	&	$-$	&	$0.96$	&	$0.22$	&	$-$	\\
S1	&	$349.81$	&	$-0.56$	&	$0.62\pm$0.03	&	$178.0\pm4.0$	&	$0.21$	&	$0.38$	&	$-$	\\
S7	&	$348.22$	&	$0.46$	&	$0.62\pm$0.03	&	$160.0\pm4.0$	&	$0.13$	&	$0.33$	&	$-$	\\
S8	&	$-$	&	$-$	&	$-$	&	$-$	&	$-$	&	$0.21$	&	$-$	\\
S11	&	$345.49$	&	$0.40$	&	$0.20\pm$0.06	&	$-$	&	$0.74$	&	$0.39$	&	$-$	\\
S13	&	$345.13$	&	$-0.75$	&	$0.65\pm$0.02	&	$-$	&	$0.50$	&	$0.37$	&	$-$	\\
S14	&	$344.76$	&	$-0.54$	&	$0.18\pm$0.03	&	$-$	&	$0.80$	&	$0.25$	&	$-$	\\
S15	&	$343.91$	&	$-0.65$	&	$0.23\pm$0.07	&	$-$	&	$0.79$	&	$0.29$	&	$-$	\\
S17	&	$343.47$	&	$-0.04$	&	$0.45\pm$0.08	&	$-$	&	$1.00$	&	$0.29$	&	$-$	\\
S18	&	$-$	&	$-$	&	$-$	&	$-$	&	$-$	&	$0.29$	&	$1.90\pm$0.30	\\
S20	&	$-$	&	$-$	&	$-$	&	$-$	&	$-$	&	$0.36$	&	$-$	\\
S23	&	$341.26$	&	$-0.35$	&	$0.48\pm$0.07	&	$-$	&	$0.39$	&	$0.38$	&	$-$	\\
S27	&	$340.06$	&	$-0.14$	&	$0.76\pm$0.09	&	$62.0\pm8.0$	&	$0.38$	&	$0.51$	&	$2.17\pm$0.27	\\
S29	&	$338.91$	&	$0.62$	&	$0.23\pm$0.06	&	$151.0\pm20.0$	&	$0.90$	&	$0.51$	&	$-$	\\
S36	&	$337.94$	&	$-0.48$	&	$0.49\pm$0.05	&	$171.0\pm7.0$	&	$0.45$	&	$0.42$	&	$-$	\\
S37	&	$337.69$	&	$-0.35$	&	$0.10\pm$0.11	&	$-$	&	$1.00$	&	$0.27$	&	$-$	\\
S41	&	$-$	&	$-$	&	$-$	&	$-$	&	$-$	&	$0.33$	&	$-$	\\
S44	&	$334.51$	&	$0.82$	&	$0.16\pm$0.06	&	$-$	&	$0.77$	&	$0.26$	&	$-$	\\
S51	&	$332.67$	&	$-0.63$	&	$0.62\pm$0.09	&	$-$	&	$1.00$	&	$0.38$	&	$-$	\\
S62	&	$331.33$	&	$-0.34$	&	$0.60\pm$0.08	&	$173.0\pm10.0$	&	$0.30$	&	$0.24$	&	$-$	\\
S64	&	$331.04$	&	$-0.14$	&	$0.71\pm$0.05	&	$117.0\pm5.0$	&	$0.14$	&	$0.35$	&	$-$	\\
S66	&	$330.84$	&	$-0.39$	&	$0.61\pm$0.02	&	$122.0\pm3.0$	&	$0.35$	&	$0.24$	&	$-$	\\
S70	&	$329.27$	&	$0.11$	&	$0.24\pm$0.13	&	$66.0\pm46.0$	&	$0.94$	&	$0.27$	&	$-$	\\
S71	&	$327.99$	&	$-0.10$	&	$0.70\pm$0.14	&	$-$	&	$0.69$	&	$0.34$	&	$-$	\\
S73	&	$-$	&	$-$	&	$-$	&	$-$	&	$-$	&	$0.24$	&	$-$	\\
S74	&	$327.53$	&	$-0.86$	&	$0.74\pm$0.11	&	$115.0\pm12.0$	&	$0.74$	&	$0.21$	&	$-$	\\
S76	&	$326.95$	&	$0.01$	&	$0.44\pm$0.03	&	$66.0\pm5.0$	&	$0.30$	&	$0.62$	&	$1.99\pm$0.32	\\
S79	&	$326.68$	&	$0.53$	&	$0.51\pm$0.05	&	$-$	&	$1.00$	&	$0.34$	&	$-$	\\
S91	&	$-$	&	$-$	&	$-$	&	$-$	&	$-$	&	$0.40$	&	$-$	\\
S92	&	$320.60$	&	$0.13$	&	$0.35\pm$0.03	&	$171.0\pm6.0$	&	$0.46$	&	$0.38$	&	$-$	\\
S96	&	$320.17$	&	$0.80$	&	$0.45\pm$0.10	&	$-$	&	$0.91$	&	$0.30$	&	$-$	\\
S97	&	$-$	&	$-$	&	$-$	&	$-$	&	$-$	&	$0.28$	&	$-$	\\
S104	&	$317.98$	&	$-0.76$	&	$0.46\pm$0.07	&	$-$	&	$0.30$	&	$0.55$	&	$-$	\\
S109	&	$-$	&	$-$	&	$-$	&	$-$	&	$-$	&	$0.53$	&	$-$	\\
S110	&	$-$	&	$-$	&	$-$	&	$-$	&	$-$	&	$0.57$	&	$-$	\\
S111	&	$-$	&	$-$	&	$-$	&	$-$	&	$-$	&	$0.35$	&	$-$	\\
S116	&	$314.24$	&	$0.43$	&	$0.69\pm$0.04	&	$-$	&	$0.10$	&	$0.32$	&	$-$	\\
S123	&	$312.96$	&	$-0.44$	&	$0.47\pm$0.06	&	$-$	&	$0.35$	&	$0.25$	&	$-$	\\
S133	&	$311.47$	&	$0.40$	&	$0.46\pm$0.07	&	$175.0\pm12.0$	&	$0.92$	&	$0.26$	&	$-$	\\
S137	&	$310.99$	&	$0.42$	&	$0.25\pm$0.04	&	$-$	&	$0.64$	&	$0.23$	&	$-$	\\
S141	&	$309.55$	&	$-0.72$	&	$0.06\pm$0.10	&	$-$	&	$0.65$	&	$0.35$	&	$2.20\pm$0.27	\\
S143	&	$-$	&	$-$	&	$-$	&	$-$	&	$-$	&	$0.26$	&	$-$	\\
S150	&	$305.54$	&	$0.35$	&	$0.23\pm$0.13	&	$-$	&	$1.00$	&	$0.24$	&	$-$	\\
S156	&	$305.26$	&	$0.23$	&	$0.33\pm$0.08	&	$-$	&	$0.95$	&	$0.39$	&	$-$	\\
S163	&	$-$	&	$-$	&	$-$	&	$-$	&	$-$	&	$0.29$	&	$-$	\\
S181	&	$-$	&	$-$	&	$-$	&	$-$	&	$-$	&	$0.24$	&	$-$	\\
S186	&	$-$	&	$-$	&	$-$	&	$-$	&	$-$	&	$0.26$	&	$-$	\\
CN24	&	$-$	&	$-$	&	$-$	&	$-$	&	$-$	&	$0.28$	&	$2.33\pm$0.37	\\
CN60	&	$-$	&	$-$	&	$-$	&	$-$	&	$-$	&	$0.51$	&	$-$	\\
CN63	&	$4.56$	&	$-0.13$	&	$0.05\pm$0.08	&	$144.0\pm144.0$	&	$0.53$	&	$0.41$	&	$-$	\\
CN71	&	$5.89$	&	$-0.43$	&	$0.50\pm$0.03	&	$-$	&	$0.15$	&	$0.28$	&	$-$	\\
CN73	&	$6.07$	&	$-0.13$	&	$0.37\pm$0.08	&	$151.0\pm16.0$	&	$0.52$	&	$0.38$	&	$-$	\\
CN88	&	$6.99$	&	$-0.28$	&	$0.56\pm$0.06	&	$134.0\pm8.0$	&	$0.87$	&	$0.26$	&	$-$	\\
CN90	&	$-$	&	$-$	&	$-$	&	$-$	&	$-$	&	$0.23$	&	$-$	\\
CN99	&	$7.38$	&	$0.70$	&	$0.64\pm$0.04	&	$158.0\pm5.0$	&	$0.69$	&	$0.30$	&	$-$	\\
CN107	&	$-$	&	$-$	&	$-$	&	$-$	&	$-$	&	$0.37$	&	$-$	\\
CN108	&	$8.12$	&	$-0.52$	&	$0.30\pm$0.02	&	$-$	&	$0.98$	&	$0.33$	&	$-$	\\
CN109	&	$-$	&	$-$	&	$-$	&	$-$	&	$-$	&	$0.43$	&	$-$	\\
CN111	&	$8.32$	&	$-0.08$	&	$0.27\pm$0.09	&	$47.0\pm25.0$	&	$1.00$	&	$0.54$	&	$-$	\\
CN114	&	$8.36$	&	$-0.30$	&	$0.22\pm$0.11	&	$-$	&	$0.98$	&	$0.32$	&	$-$	\\
CN138	&	$9.83$	&	$-0.71$	&	$0.31\pm$0.14	&	$-$	&	$0.77$	&	$0.29$	&	$-$	\\
CN139	&	$-$	&	$-$	&	$-$	&	$-$	&	$-$	&	$0.33$	&	$-$	\\
CN148	&	$-$	&	$-$	&	$-$	&	$-$	&	$-$	&	$0.41$	&	$-$	\\
CS2	&	$359.73$	&	$-0.41$	&	$0.40\pm$0.07	&	$156.0\pm14.0$	&	$0.67$	&	$0.39$	&	$-$	\\
CS33	&	$356.23$	&	$0.68$	&	$0.42\pm$0.12	&	$139.0\pm21.0$	&	$0.55$	&	$0.30$	&	$-$	\\
CS51	&	$354.19$	&	$-0.05$	&	$0.12\pm$0.07	&	$-$	&	$0.47$	&	$0.33$	&	$-$	\\
CS57	&	$353.36$	&	$-0.16$	&	$0.64\pm$0.08	&	$-$	&	$0.73$	&	$0.28$	&	$2.15\pm$0.26	\\
CS62	&	$353.09$	&	$0.33$	&	$0.14\pm$0.05	&	$93.0\pm28.0$	&	$0.47$	&	$0.59$	&	$-$	\\
CS79	&	$351.65$	&	$0.51$	&	$0.52\pm$0.04	&	$-$	&	$0.23$	&	$0.30$	&	$2.18\pm$0.25	\\
CS81	&	$-$	&	$-$	&	$-$	&	$-$	&	$-$	&	$0.51$	&	$-$	\\
E3	&	$-$	&	$-$	&	$-$	&	$-$	&	$-$	&	$0.21$	&	$-$	\\
E7	&	$74.76$	&	$0.62$	&	$0.19\pm$0.04	&	$-$	&	$0.72$	&	$0.24$	&	$1.94\pm$0.31	\\
E21	&	$82.59$	&	$0.38$	&	$0.58\pm$0.03	&	$99.0\pm4.0$	&	$0.37$	&	$0.30$	&	$-$	\\
E27	&	$90.23$	&	$1.74$	&	$0.58\pm$0.09	&	$44.0\pm12.0$	&	$0.79$	&	$0.31$	&	$1.93\pm$0.30	\\
E30	&	$-$	&	$-$	&	$-$	&	$-$	&	$-$	&	$0.32$	&	$1.67\pm$0.27	\\
E39	&	$106.25$	&	$0.97$	&	$0.67\pm$0.12	&	$-$	&	$0.64$	&	$0.12$	&	$2.18\pm$0.38	\\
E47	&	$114.62$	&	$0.23$	&	$0.22\pm$0.03	&	$15.0\pm10.0$	&	$0.99$	&	$0.19$	&	$-$	\\
E55	&	$133.44$	&	$0.07$	&	$0.34\pm$0.04	&	$-$	&	$0.98$	&	$0.26$	&	$1.86\pm$0.29	\\
E56	&	$134.20$	&	$0.81$	&	$0.51\pm$0.08	&	$124.0\pm11.0$	&	$0.37$	&	$0.26$	&	$1.84\pm$0.35	\\
E58	&	$-$	&	$-$	&	$-$	&	$-$	&	$-$	&	$0.35$	&	$1.88\pm$0.30	\\
E59	&	$-$	&	$-$	&	$-$	&	$-$	&	$-$	&	$0.23$	&	$2.17\pm$0.34	\\
E70	&	$-$	&	$-$	&	$-$	&	$-$	&	$-$	&	$0.24$	&	$1.86\pm$0.27	\\
E72	&	$182.35$	&	$0.21$	&	$0.49\pm$0.03	&	$143.0\pm5.0$	&	$0.12$	&	$0.22$	&	$1.85\pm$0.29	\\
E76	&	$190.05$	&	$0.48$	&	$0.06\pm$0.01	&	$90.0\pm15.0$	&	$0.32$	&	$0.26$	&	$1.72\pm$0.30	\\
E77	&	$196.21$	&	$-1.20$	&	$0.29\pm$0.06	&	$-$	&	$0.82$	&	$0.28$	&	$1.98\pm$0.30	\\
E80	&	$211.87$	&	$-1.33$	&	$0.46\pm$0.10	&	$117.0\pm16.0$	&	$0.56$	&	$0.21$	&	$2.10\pm$0.31	\\
E82	&	$212.02$	&	$-1.30$	&	$0.31\pm$0.01	&	$20.0\pm3.0$	&	$0.96$	&	$0.28$	&	$1.60\pm$0.28	\\
E83	&	$212.05$	&	$-0.74$	&	$0.06\pm$0.06	&	$38.0\pm38.0$	&	$0.07$	&	$0.22$	&	$1.88\pm$0.31	\\
E84	&	$212.41$	&	$-1.15$	&	$0.26\pm$0.05	&	$61.0\pm16.0$	&	$0.94$	&	$0.17$	&	$1.99\pm$0.33	\\
E85	&	$217.32$	&	$-1.40$	&	$0.37\pm$0.04	&	$91.0\pm8.0$	&	$0.91$	&	$0.20$	&	$1.80\pm$0.31	\\
E86	&	$-$	&	$-$	&	$-$	&	$-$	&	$-$	&	$0.23$	&	$1.74\pm$0.33	\\
E89	&	$-$	&	$-$	&	$-$	&	$-$	&	$-$	&	$0.21$	&	$1.90\pm$0.32	\\
E95	&	$234.77$	&	$-0.22$	&	$0.22\pm$0.03	&	$108.0\pm9.0$	&	$0.18$	&	$0.17$	&	$1.81\pm$0.27	\\
E122	&	$-$	&	$-$	&	$-$	&	$-$	&	$-$	&	$0.51$	&	$-$	\\
E124	&	$283.94$	&	$-0.83$	&	$0.65\pm$0.02	&	$93.0\pm2.0$	&	$0.68$	&	$0.27$	&	$-$	\\
E125	&	$284.64$	&	$-0.50$	&	$0.32\pm$0.11	&	$170.0\pm28.0$	&	$1.00$	&	$0.16$	&	$2.12\pm$0.32	\\
E127	&	$286.21$	&	$-0.16$	&	$0.48\pm$0.02	&	$151.0\pm3.0$	&	$0.15$	&	$0.33$	&	$-$	\\
E134	&	$291.05$	&	$-0.77$	&	$0.79\pm$0.11	&	$-$	&	$0.68$	&	$0.28$	&	$1.94\pm$0.45	\\
\end{longtable}

\clearpage


\begin{thebibliography}{}
\bibitem[Aannestad(1975)]{Aannestad1975}
  Aannestad, P.A. 1975, \apj, 200, 30
\bibitem[Anderson et al.(2012)]{Anderson2012}
  Anderson, L.D., et al. 2012, A\&A, 542, A10

\bibitem[Baug et al.(2016)]{Baug2016}
  Baug, T., Dewangan, L.K., Ojha, D.K., \& Ninan, J.P. 2016, \apj, 833, 85
\bibitem[Benjamin et al.(2003)]{Benjamin2003}  
  Benjamin, R.A., et al. 2003, \pasp, 115, 953

\bibitem[Chokshi, Tielens and Hollenbach(1993)]{Chokshi1993}
  Chokshi, A., Tielens, A.G.G.M., \& Hollenbach, D. 1993, \apj, 407, 806
\bibitem[Churchwell et al.(2006)]{Churchwell2006}
  Churchwell, E., et al. 2006, \apj, 649, 759
\bibitem[Churchwell et al.(2007)]{Churchwell2007}
  Churchwell, E., et al. 2007, \apj, 670, 428
\bibitem[Churchwell et al.(2009)]{Churchwell2009}
  Churchwell, E., et al. 2009, \pasp, 121, 213
\bibitem[Csengeri et al.(2014)]{Csengeri2014}
  Csengeri, T., et al. 2014, \aap, 565, A75
\bibitem[Cutri et al.(2013)]{Cutri2013}
  Cutri, R.M., et al. 2013, Explanatory Supplement to the AllWISE Data Release Products, Tech. rep.
  
\bibitem[Dale, Bonnell and Whitworth(2007)]{Dale2007CC}
  Dale, J.E., Bonnell, I.A., \& Whitworth, A.P. 2007, \mnras, 375, 1291
\bibitem[Decataldo et al.(2017)]{Decataldo2017}
  Decataldo, D., Ferrara, A., Pallottini, A., Gallerani, S., \& Vallini, L. 2017, \mnras, 471, 4476
\bibitem[Deharveng et al.(2010)]{Deharveng2010}  
  Deharveng, L., et al. 2010, \aap, 523, A6
\bibitem[Desert, Boulanger and Puget(1990)]{Desert1990}
  Desert, F.-X., Boulanger, F., \& Puget, J.L. 1990, \aap, 237, 215
\bibitem[Dewangan et al.(2017)]{Dewangan2017}
  Dewangan, L.K., Ojha, D.K., \& Zinchenko, I. 2017, \apj, 851, 140
\bibitem[Draine and Li(2007)]{Draine2007}
  Draine, B.T., \& Li, A. 2007, \apj, 657, 810
\bibitem[Draine et al.(2014)]{Draine2014}
  Draine, B.T., et al. 2014, \apj, 780, 172
  
\bibitem[Elmegreen(1998)]{Elmegreen1998}
  Elmegreen, B.G. 1998, in \asp, 148, Origins, ed. C.E. Woodward et al. (San Francisco: ASP), 150

\bibitem[Fukui et al.(2016)]{Fukui2016}
  Fukui, Y., et al. 2016, \apj, 820, 26
  
\bibitem[Giannetti et al.(2017)]{Giannetti2017}
  Giannetti, A., et al. 2017, \aap, 606, L12
\bibitem[Griffin et al.(2010)]{Griffin2010}
  Griffin, M.J., et al. 2010, \aap, 518, L3
\bibitem[Gritschneder et al.(2009)]{Gritschneder2009}
  Gritschneder, M., Naab, T., Walch, S., Burkert, A., \& Heitsch, F. 2009, \apj, 694, L26
\bibitem[G$\ddot{\rm{u}}$ttler(1952)]{Guttler1952}
  G$\ddot{\rm{u}}$ttler, A. 1952, Annalen der Physik, 446, 65
  
\bibitem[Habe and Ohta(1992)]{Habe1992}
  Habe, A., \& Ohta, K. 1992, \pasj, 44, 203
\bibitem[Hattori et al.(2016)]{Hattori2016}
  Hattori, Y., et al. 2016, \pasj, 68, 37
\bibitem[Hanaoka et al.(2019)]{Hanaoka2019}
  Hanaoka, M., et al. 2019, \pasj, 71, 6
\bibitem[Hawarth et al.(2015)]{Hawarth2015}
  Hawarth, T.J., et al. 2015, \mnras, 450, 10

\bibitem[Inoue and Fukui(2013)]{Inoue2013}
  Inoue, T., \& Fukui, Y. 2013, \apj, 774, L31
\bibitem[Ishihara et al.(2010)]{Ishihara2010}
  Ishihara, D., et al. 2010, \aap, 514, A1
\bibitem[Ishihara et al.(2011)]{Ishihara2011}
  Ishihara, D., Kaneda, H., Onaka, T., Ita, Y., Matsuura, M., \& Matsunaga, N. 2011, \aap, 534, A79

\bibitem[Jarrett et al.(2011)]{Jarrett2011}
  Jarrett, T.H., et al. 2011, \apj, 735, 112
\bibitem[Jones, Tielens and Hollenbach(1996)]{Jones1996}
  Jones, A.P., Tielens, A.G.G.M., \& Hollenbach, D.J. 1996, \apj, 469, 740
\bibitem[Jones and Habart(2015)]{Jones2015}
  Jones, A.P., \& Habart, E. 2015, \aap, 581, A92
\bibitem[Jones et al.(2014)]{Jones2014}
  Jones, A.P., Ysard, N., K$\ddot{\rm{o}}$hler, M., Fanciullo, L., Bocchino, M., Micelotta, E., Verstraete, L., \& Guillet, V. 2014, Faraday Discussions, 168, 313
  
\bibitem[Kohno et al.(2018)]{Kohno2018}
  Kohno, M., et al. 2018, \pasj, arXiv:1809.00118
\bibitem[Kuan, Mehringer and Snyder(1996)]{Kuan1996}
  Kuan, Y.-J., Mehringer, D.M., \& Snyder, L.E. 1996, \apj, 459, 619

\bibitem[Latter(1991)]{Latter1991}
  Latter, W.B. 1991, \apj, 377, 187
\bibitem[Lis and Menten(1998)]{Lis1998}
  Lis, D.C., \& Menten, K.M. 1998, \apj, 507, 794
  
\bibitem[Mennella et al.(1998)]{Mennella1998}
  Mennella, V., Brucato, J.R., Colangeli, L., Palumbo, P., Rotundi, A., \& Bussoletti, E. 1998, \apj, 496, 1058
\bibitem[Micelotta, Jones and Tielens(2010a)]{Micelotta2010a}
  Micelotta, E.R., Jones, A.P., \& Tielens, A.G.G.M. 2010a, \aap, 510, A37 
\bibitem[Micelotta, Jones and Tielens(2010b)]{Micelotta2010b}
  Micelotta, E.R., Jones, A.P., \& Tielens, A.G.G.M. 2010b, \aap, 510, A36 
\bibitem[Molinari et al.(2010)]{Molinari2010}
  Molinari, S., et al. 2010, \pasp, 122, 314
\bibitem[Molinari et al.(2016)]{Molinari2016}
  Molinari, S., et al. 2016, \aap, 591, A149
\bibitem[Murakami et al.(2007)]{Murakami2007}
  Murakami, H., et al. 2007, \pasj, 59, S369

\bibitem[Onaka et al.(2007)]{Onaka2007}
  Onaka, T., et al. 2007, \pasj, 59, S401

\bibitem[Pilbratt et al.(2010)]{Pilbratt2010}
  Pilbratt, G.L., et al. 2010, \aap, 518, L1
\bibitem[Planck Collaboration et al.(2011)]{Planck2011}
  Planck Collaboration, et al. 2011, \aap, 536, A23
\bibitem[Poglitsch et al.(2010)]{Poglitsch2010}
  Poglitsch, A., et al. 2010, \aap, 518, L2
  
\bibitem[Rapacioli et al.(2005)]{Rapacioli2005}
  Rapacioli, M., Calvo, F., Spiegelman, F., Joblin, C., \& Wales, D.J. 2005, Journal of Physical Chemistry A, 109, 2487
\bibitem[Rodgers et al.(1960)]{Rodgers1960}
  Rodgers, A.W., Campbell, C.T., \& Whiteoak, J.B 1960, \mnras, 121, 103
  
\bibitem[Smith et al.(2012)]{Smith2012}
  Smith, M.W.L., et al. 2012, \apj, 756, 40
  
\bibitem[Tabatabaei et al.(2014)]{Tabatabaei2014}
  Tabatabaei, F.S., et al. 2014, \aap, 561, A95
\bibitem[Takahira et al.(2018)]{Takahira2018}
  Takahira, K., Shima, K., Habe, A., \& Tasker, E.J. 2018, \pasj, 70, S58
\bibitem[Tielens(2005)]{Tielens2005}
  Tielens, A.G.G.M. 2005, The Physics and Chemistry of the Interstellar Medium
\bibitem[Tielens(2008)]{Tielens2008}
  Tielens, A.G.G.M. 2008, \araa, 46, 289
\bibitem[Torii et al.(2015)]{Torii2015}
  Torii, K., et al. 2015, \apj, 806, 7
\bibitem[Torii et al.(2018a)]{Torii2018N18}
  Torii, K., et al. 2018a, \pasj, arXiv:1706.07164
\bibitem[Torii et al.(2018b)]{Torii2018N35}
  Torii, K., et al. 2018b, \pasj, 70, S51

\bibitem[Wickramasinghe(1967)]{Wickramasinghe1967}
  Wickramasinghe, N.C. 1967, Interstellar grains
\bibitem[Wright et al.(2010)]{Wright2010}
  Wright, E.L., et al. 2010, \aj, 140, 1868
  
\bibitem[Zinnecker and Yorke(2007)]{Zinnecker2007}
  Zinnecker, H., \& Yorke, H.W. 2007, \araa, 45, 481

\end{thebibliography}
\end{document}